\documentclass[fleqn,usenatbib]{mnras}

\usepackage{newtxtext,newtxmath}

\usepackage[T1]{fontenc}

\DeclareRobustCommand{\VAN}[3]{#2}
\let\VANthebibliography\thebibliography
\def\thebibliography{\DeclareRobustCommand{\VAN}[3]{##3}\VANthebibliography}

\usepackage{graphicx}	
\graphicspath{{Figures/}}
\usepackage{color}
\usepackage[dvipsnames]{xcolor}
\usepackage{comment}
\usepackage{threeparttable}
\usepackage{subfig}

\title[MRI turbulence in stratified discs at large magnetic Prandtl numbers]{MRI turbulence in vertically stratified accretion discs at large magnetic Prandtl numbers}

\author[L. E. Held et al.]{
Loren E. Held$^{1,2}$\thanks{E-mail: leh50@cam.ac.uk (LEH)}, George Mamatsashvili$^{3,4}$ and Martin E. Pessah$^{5}$
\\
$^{1}$Max Planck Institute for Gravitational Physics (Albert Einstein Institute), Am M{\"u}hlenberg 1, Potsdam 14476, Germany\\
$^{2}$Department of Applied Mathematics and Theoretical Physics, University of Cambridge, Centre for Mathematical Sciences, Wilberforce Road,\\ 
Cambridge CB3 0WA, United Kingdom\\
$^{3}$Helmholtz-Zentrum Dresden-Rossendorf, Bautzner Landstra{\ss}e 400, Dresden 01328, Germany\\
$^{4}$E. Kharadze Georgian National Astrophysical Observatory, Abastumani 0301, Georgia\\
$^{5}$Niels Bohr International Academy, Niels Bohr Institute, Blegdamsvej 17, DK-2100 Copenhagen \O, Denmark}

\date{Accepted XXX. Received YYY; in original form ZZZ}

\pubyear{2024}

\begin{document}
\label{firstpage}
\pagerange{\pageref{firstpage}--\pageref{lastpage}}
\maketitle

\begin{abstract}
The discovery of the first binary neutron star merger, GW170817, has spawned a plethora of global numerical relativity simulations. These simulations are often ideal (with dissipation determined by the grid) and/or axisymmetric (invoking ad hoc mean-field dynamos). However, binary neutron star mergers (similar to X-ray binaries and active galactic nuclei inner discs) are characterised by large magnetic Prandtl numbers, $\rm Pm$, (the ratio of viscosity to resistivity). $\rm Pm$ is a key parameter determining dynamo action and dissipation but it is ill-defined (and likely of order unity) in ideal simulations.  To bridge this gap, we investigate the magnetorotational instability (MRI) and associated dynamo at large magnetic Prandtl numbers using fully compressible, three-dimensional, vertically stratified, isothermal simulations of a local patch of a disc. We find that, within the bulk of the disc ($z\lesssim2H$, where $H$ is the scale-height), the turbulent intensity (parameterized by the stress-to-thermal-pressure ratio $\alpha$), and the saturated magnetic field energy density, $E_\text{mag}$, produced by the MRI dynamo, both scale as a power with Pm at moderate Pm ($4\lesssim \text{Pm} \lesssim 32$): $E_\text{mag} \sim \text{Pm}^{0.74}$ and $\alpha \sim \text{Pm}^{0.71}$, respectively. At larger Pm ($\gtrsim 32$) we find deviations from power-law scaling and the onset of a plateau.  Compared to our recent unstratified study, this scaling with Pm becomes weaker further away from the disc mid-plane, where the Parker instability dominates. We perform a thorough spectral analysis to understand the underlying dynamics of small-scale MRI-driven turbulence in the mid-plane and of large-scale Parker-unstable structures in the atmosphere.
\end{abstract}




\section{Introduction}
\label{INTRO}

The transport of angular momentum is the key mechanism driving accretion in discs. Turbulence - thought to be driven by the magnetorotational instability (MRI), and potentially also magnetically-driven outflows, facilitate this angular momentum transport. Complicating matters, a characteristic of magnetohydrodynamic (MHD) turbulence is the existence of not one, but two dissipation mechanisms (viscosity and resistivity) and associated length-scales that can drastically alter the nature of MHD turbulence compared to its more familiar hydrodynamic (i.e. Kolmogorov-type) cousin. The ratio of these two diffusivities, known as the \textit{magnetic Prandtl number} (Pm), is a key parameter in MHD turbulence and in dynamo theory  \citep{rincon2019}, influencing not only the saturated state of the turbulence (with magnetic field growth and  associated angular momentum transport), but also the thermal stability of the disc \citep{balbus2008, potter2017, kawanaka2019}. 

Different discs are believed to lie in different Pm regimes: the hot, dense, partially neutrino-cooled discs from binary neutron star mergers, and the inner regions of X-ray binaries and active galactic nuclei are thought to  lie in the $\text{Pm} \gg 1$ regime \citep{balbus2008, rossi2008}, while cooler, protoplanetary, discs are generally thought to lie in the $\text{Pm} \ll 1$ regime \citep{lesur2021magnetohydrodynamics}. Global simulations however sophisticated, are almost always either ideal \citep{kiuchi2022implementation}, and thus
characterized by a purely \textit{numerical} magnetic Prandtl number which is likely of order unity \citep{minoshima2015}, or otherwise employ mean-field dynamo or viscosity prescriptions \citep{fujibayashi2020viscous,shibata2021long}, prescriptions which, while very informative in their own right, are nevertheless ad hoc. Sub-grid modeling attempts to bridge the gap between ideal simulations and those with fully explicit dissipation coefficients, but comes with its own uncertainties \citep{meheut2015, miravet2022assessment}.
Though it has been known for some time that MRI turbulence is particularly sensitive to the ratio of dissipation scales \citep{lesur2007, pessah2008, simon2011, nauman2016sustained, nauman2018}, the MRI in the regime $\text{Pm} \gg 1$ has only been sparsely investigated in the literature. Recent work \citep{heldmamatsashvili2022,Guilet2022}, however, has begun to forge inroads into this physically interesting, but numerically challenging \citep{reboul2022mri} regime, though it has so far been limited to unstratified and isothermal disc models.

\subsection{The need for explicit dissipation coefficients}
The magnetorotational instability requires a weak seed field in order to operate, though this need not be a mean (external) field. In fact, the MRI can sustain itself without the presence of a mean field by means of a dynamo, as demonstrated in both unstratified local models \citep[e.g.,][]{hawley1996local,fromang2007,riols2017magnetorotational,Walker2016,mamatsashvili2020zero}, and vertically stratified ones \citep[e.g.,][]{brandenburg1995dynamo, davisstonepesssah2010,shi2009numerically,gressel2010,salvesen2016a}. This set-up (referred to as `zero-net-flux' or ZNF) is particularly attractive because it is agnostic to the strength and geometry of the magnetic field of the central accretor or external medium, both of which can vary considerably from system to system.

Complicating matters, in the absence of explicit dissipation coefficients (viscosity and resistivity) both unstratified and stratified models exhibit the `convergence problem', a numerical artifact that very likely afflicts ideal global simulations, too. In essence, in ideal MHD simulations, turbulent transport \textit{decreases} with increasing numerical resolution \citep{pessah2007angular, fromang2007}. When vertical stratification is included, which gives the domain a `global' character in the vertical direction, one encounters the same problem \citep{bodo2014,ryan2017}. Convergence with resolution is restored when dissipation coefficients are explicitly taken into account, however, and also leads to new and interesting physics: in the regime $\text{Pm} \lesssim 1$ the zero-net-magnetic flux MRI struggles even to sustain itself \citep{Walker2016, riols2015dissipative, nauman2016sustained, mamatsashvili2020zero}, while in the opposite regime turbulent intensity is remarkably sensitive to Pm, exhibiting power-law scaling with Pm, at least up moderately large values of $\text{Pm} \gtrsim 1$ \citep{potter2017, heldmamatsashvili2022, Guilet2022}. Although a few authors have carried out stratified ZNF simulations at $\text{Pm}\gtrsim 1$ \citep{gressel2010, davisstonepesssah2010}, the only work we are aware of that investigated the $\text{Pm}\gg 1$ regime with vertical stratification is \cite{simon2011}, although even they only went up to ${\rm Pm} = 8$. In this work, we aim to explore the behavior of the MRI dynamo in stratified discs at much larger magnetic Prandtl numbers (up to \text{Pm=90}) than has been previously done.

\subsection{Magnetic buoyancy instabilities in accretion discs}
Real accretion discs are stratified in the direction perpendicular to the disc, and interesting new physics arises when stratification is taken into account. In this case magnetic fields can become buoyant, rising through the disc \citep{millerstone2000, blackman2009,shi2009numerically,  Uzdensky2013,Dudorov2019}, which is most easily revealed in the well-known 'spacetime diagrams' \citep{rincon2019, lesur2021magnetohydrodynamics}. These buoyant fields do not only play a passive role: the zero-net-flux MRI dynamo in stratified discs has been shown to behave much like an $\alpha-\Omega$ dynamo \citep{gressel2010, gressel2015, rincon2019, dhang2023shedding}. Finally, the addition of thermodynamics in stratified models further complicates things, leading to scenarios such as the interplay between the MRI and convection \citep{bodo2013fully, hirose2014, scepi2018a, held2021magnetohydrodynamic}, but comes with its own numerical challenges \citep{gressel2013,held2021magnetohydrodynamic}.

Yet another interesting effect that can occur in stratified discs is that of \textit{Parker instability} (and associated dynamo). The Parker mode (also known as the undulatory mode) is characterized by perturbations whose wavenumbers are \textit{parallel} to the magnetic field ($\mathbf{k} \parallel \mathbf{B}$) \citep{pringle2007astrophysical,stone2007magnetic, hillier2016nature}. The instability involves bending of field lines and plasma slides down these field lines forming overdense regions in the troughs between the crests. Assuming the gravitational acceleration points in the $z$-direction, the linear instability criterion is given by $dB/dz < 0$ \citep{pringle2007astrophysical}. 

The idea that magnetic fields in accretion discs might undergo Parker instability, leading to the generation of vertical field from toroidal and radial fields has been around for a long time \citep{shu1974parker,Tout_Pringle1992}. In early 3D, vertically stratified, zero-net-flux, isothermal simulations the disc was found to be only marginally unstable to Parker instability \citep{stonehawley1996}, however this was likely due to the restricted vertical size of the domain which encompassed only $\pm 2$ scale-heights either side of the mid-plane. Later, simulations in taller boxes (spanning 6-9 scale-heights on either side of the mid-plane) initialized with a relatively strong initial toroidal magnetic field (i.e. with a gas-to-magnetic pressure ratio $\beta \sim$ 1-25) did observe the development of Parker instability (in addition to the MRI) both in isothermal discs \citep{johansen2008high, kadowaki2018mhd}, and also in non-isothermal simulations with radiative transfer \citep{blaes2007surface, shi2009numerically, blaes2011}. As we report in this work, we find the emergence of Parker instability even when the disc is initialized with relatively weak ($\beta \sim 10^3)$ zero-net-magnetic-flux.

\subsection{Motivation and outline}
Our aim in this paper is to build on our earlier work \citep{heldmamatsashvili2022} on the saturation and energetics of zero-net-flux MRI turbulence (alternatively referred to as the `MRI dynamo') in the regime of large magnetic Prandtl number, mainly by taking into account more realistic disc physics, in particular the effect of vertical stratification. To facilitate comparison to our unstratified, isothermal, simulations, here we also adopt an isothermal equation of state. 

The structure of the paper is as follows. We outline our governing equations, numerical algorithms, and key parameters and diagnostics in Section \ref{METHODS}. In Section \ref{RESULTS_RealSpace} we present various results in physical space such as the scaling of turbulent intensity and magnetic field strength with magnetic Prandtl number, and also the vertical structure of the disc, including evidence for the Parker instability in the disc atmosphere. In Section \ref{RESULTS_SpectralSpace} we perform a detailed spectral analysis of the turbulence, focusing on how energy transfers in spectral space contribute to the dominant dynamics in different vertical parts of the disc. Finally, we present our conclusion in Section \ref{CONCLUSIONS}. As some of our key results related to magnetic buoyancy (Parker instability) occur predominantly in the atmosphere of the disc, we also investigated the effects of changing the vertical boundary conditions and of increasing the vertical box size in Appendices \ref{APPENDIX_VerticalBoundaryConditions} and \ref{APPENDIX_VerticalBoxSize}, respectively.

Readers primarily interested in how turbulent transport and magnetic field strength scale with Pm should jump straight to Section \ref{RESULTS_RealSpace}, in particular Figure \ref{FIGURE_EmagAndAlphaPmScaling}. Readers primarily interested in the key results of our spectral analysis and how these relate to the self-sustenance mechanism of the dynamo at different heights in the disc should jump to Section \ref{RESULTS_SummarySustenanceSchemesAtDifferentHeights}.

\section{Methods}
\label{METHODS}

\subsection{Governing equations}
\label{METHODS_GoverningEquations}
We work in the shearing box approximation
\citep{hawley1995,latter2017local},
which treats a local region of a disc as a Cartesian box located at some fiducial radius $r = r_0$ and orbiting with the angular frequency of the disc at that radius $\Omega_0 \equiv \Omega(r_0)$. A point in the box has
Cartesian coordinates $(x, y, z)$ along the radial, azimuthal/toroidal, and vertical directions, respectively. In this rotating frame, the equations of non-ideal MHD are

\begin{equation}
\partial_t \rho + \nabla \cdot (\rho \mathbf{u}) = 0, \label{SB1}
\end{equation}
\begin{multline}
\partial_t \mathbf{u} + \mathbf{u}\cdot\nabla \mathbf{u} = -\frac{1}{\rho} \nabla P - 2\Omega_0 \mathbf{e}_z \times \mathbf{u} + \mathbf{g_\text{eff}}+ \\\frac{1}{\mu_0 \rho}(\nabla\times\mathbf{B})\times\mathbf{B}+\frac{1}{\rho}\nabla \cdot \mathbf{T}, \label{SB2}
\end{multline} 
\begin{equation}
\partial_t \mathbf{B} = \nabla\times(\mathbf{u}\times\mathbf{B})+\eta\nabla^2\mathbf{B}, \label{SB4}
\end{equation}
with the symbols taking their usual meanings. We close the system with the equation of state for an isothermal gas $P = c_s^2 \rho$ where $c_s^2$ is the constant sound speed.
 
All our simulations are vertically stratified and the effective gravitational potential is embodied in the tidal acceleration $\mathbf{g_\text{eff}}=2q\Omega_0^2x\mathbf{e}_x-\Omega_0^2 z \mathbf{e}_z$ (third term on the right-hand side of Equation \ref{SB2}), where $q$ is the dimensionless shear parameter $q \equiv -\left.d\ln{\Omega}/d\ln{r}\right\vert_{r=r_0}$. For Keplerian discs $q=3/2$, a value we adopt throughout this paper.

To control the magnetic Prandtl number (see Section \ref{Methods_Parameters}) we employ explicit diffusion coefficients. The viscous stress tensor is given by $\mathbf{T} \equiv 2\rho \nu \mathbf{S}$, where $\nu$ is the kinematic viscosity, and $\mathbf{S} \equiv (1/2)[\nabla \mathbf{u} + (\nabla \mathbf{u})^\text{T}] - (1/3)(\nabla\cdot\mathbf{u})\mathbf{I}$ is the traceless shear tensor \citep{landau1987}. The explicit magnetic diffusivity is denoted by $\eta$: it is related to the resistivity $\xi $ via $\eta \equiv \xi/\mu_0$, where $\mu_0$ is the permeability of free space (note that from now on we will use the terms resistivity and magnetic diffusivity interchangeably). Note that in a real disc the microscopic viscosity and resistivity (and thus the magnetic Prandtl number) will depend on temperature and density (see \cite{rossi2008} and \cite{kawanaka2019} for the explicit dependence of Pm on temperature and density in a neutrino-cooled disc). However, as the simulations discussed in this paper are isothermal, we keep the viscosity and resistivity fixed in space and time in any given simulation.

\subsection{Important parameters}
\label{Methods_Parameters}
The magnetic Reynolds number compares inductive to resistive effects and is given by

\begin{equation}
\text{Rm} = \frac{c_s H}{\eta},
\end{equation}
where $c_s$ is the isothermal speed of sound, $H=c_s/\Omega_0$ is the scale-height (see Section \ref{METHODS_Units} for definitions), and $\eta$ is the magnetic diffusivity. 

The Reynolds number compares inertial to viscous forces and is given by

\begin{equation}
\text{Re} = \frac{c_s H}{\nu},
\end{equation}
where $\nu$ is the kinematic viscosity. 

Finally, the ratio of Rm to Re defines the magnetic Prandtl number,

\begin{equation}
\text{Pm} \equiv \frac{\rm Rm}{\rm Re} =  \frac{\nu}{\eta},
\end{equation}
which serves as the key control parameter in our simulations.

\subsection{Numerical set-up}
\label{METHODS_NumericalSetUp}
\subsubsection{Code}
\label{Methods_Codes}
For our simulations we use the conservative, finite-volume code \textsc{PLUTO} \citep{mignone2007}. We employ the HLLD Riemann solver, 2nd-order-in-space linear interpolation, and the 2nd-order-in-time Runge-Kutta algorithm. In addition, in order to enforce the condition that $\nabla\cdot\mathbf{B}=0$, we employ Constrained Transport (CT), and use the UCT-Contact algorithm to calculate the EMF at cell edges \citep{gardiner2005unsplit}. To allow for longer time-steps, we take advantage of the \textsc{FARGO} scheme \citep{mignone2012}. When explicit resistivity $\eta$ and viscosity $\nu$ are included, we further reduce the computational time via the Super-Time-Stepping (STS) scheme \citep{alexiades1996super}. Ghost zones are used to implement the boundary conditions.

We use the built-in shearing box module in \textsc{PLUTO}
\citep{mignone2012}. Rather than solving Equations \eqref{SB1}-\eqref{SB4} (primitive
form), \textsc{PLUTO} solves the governing equations in conservative form.

\subsubsection{Initial conditions}
\label{METHODS_InitialConditions}
All our simulations are initialized from an equilibrium exhibiting a Gaussian density profile:

\begin{equation}
\rho = \rho_0 \exp{(-z^2/(2H_0^2))},
\label{EQUN_densityprofile}
\end{equation}
where $\rho_0$ is the mid-plane density at initialization, and $H_0$ is the scale-height at the mid-plane at initialization (formally defined below).

The background velocity is given by $\mathbf{u}_0 = -q \Omega_0 x \,\mathbf{e}_y$. At initialization we usually perturb all the
velocity components with random noise exhibiting a flat power
spectrum. The perturbations $\delta \mathbf{u}$ have maximum
amplitude of about $5\times10^{-2}\,c_{s0}$, unless stated otherwise. Here $c_{s0}$ is the sound speed at initialization. All simulations are initialized with commonly used \textit{zero-net-flux} (ZNF) magnetic field configuration $\mathbf{B}_0 = B_0\sin{(2\pi x/L_x)}\mathbf{e}_z$, where $L_x$ is the radial box size. We define the field strength at initialization $B_0$ through the plasma beta parameter $\beta_0 \equiv 2\mu_0 P_0 /B_0^2$, where $P_0=\rho_0c_{s0}^2$ is the pressure at the mid-plane. We set $\beta_0 \equiv 1000$ in all our simulations.

\subsubsection{Units}
\label{METHODS_Units}
Note that from this point onwards, all quantities are given in terms of dimensionless (code) units. Time units are selected so that $\Omega_0 = 1$. The length unit is chosen so that the initial sound speed $c_{s0} = 1$, which in turn defines a reference scale-height $H_0\equiv c_{s0} / \Omega_0=1$. Finally the mass unit is set by the initial mid-plane density, which is $\rho_0 = 1$.  Magnetic field is expressed in units of $c_{s0}\sqrt{\mu_0 \rho_0}$. Pressure, stresses, and energy densities are expressed in units of $c_{s0}^2 \rho_0$. Note that we will occasionally drop the subscript on $\Omega_0$, $c_{s0}$, and $H_0$ in the text from this point onwards.

\subsubsection{Box size and resolution}
\label{METHODS_BoxSizeAndResolution}
The majority of our simulations are run at a resolution of $N_x\times N_y\times N_z=512\times512\times1024$ in a box of size $[L_x, L_y,L_z] = [4H,4H,8H]$ (i.e. $128$ cells per scale-height). In our unstratified paper we found that a resolution of 128 cells/$H$ was sufficient to resolve the resistive scale at $\text{Rm} = 18750$ (see Appendix A of \cite{heldmamatsashvili2022}). We have also run select simulations at lower resolutions of 32 cells/$H$ and 64 cells/$H$. To investigate the dependence on the vertical box size, we have repeated our fiducial run ($
\text{Pm}=4$, box size $[4H,4H,8H]$) in a taller box of size $[4H,4H,10H]$ (keeping the resolution per scale-height fixed at $128$ cells per $H$). We discuss this box size study in Appendix \ref{APPENDIX_VerticalBoxSize}. All the simulations described in this paper are listed in Tables \ref{TABLE_PmComparison}-\ref{TABLE_zBCComparison} in Appendix \ref{APPENDIX_TablesOfSimulations}.

\subsubsection{Boundary conditions and mass source term}
\label{METHODS_BoundaryConditions}
We use standard shear-periodic boundary conditions (BCs) in the $x$-direction \cite[see][]{hawley1995}, and periodic boundary conditions in the $y$-direction. In the vertical direction, we keep the ghost zones associated with the thermal variables in isothermal hydrostatic equilibrium, in the manner described in \cite{zingale2002mapping}. For the velocity components we use standard \textit{outflow} boundary conditions in the vertical direction, whereby the vertical gradients of all velocity components are zero (numerically, we set variables in the ghost zones equal to those in the active cells bordering the ghost zones). For the magnetic field we employ `vertical field' boundary conditions, also known as `pseudo-vacuum' boundary conditions. Explicitly these are defined as setting $B_x = 0$, $B_y = 0$, and $\partial B_z/\partial z = 0$ at the vertical boundaries. 

As we are interested in the dynamics in the disc atmosphere $|z| > 2H$, we also investigate the effect of using different vertical boundary conditions for the magnetic field. Altogether we explored three different types of boundary condition in the vertical direction: vertical field zBCs, outflow zBCs, and perfect conductor zBCs. These results are described in Appendix \ref{APPENDIX_VerticalBoundaryConditions} and the corresponding simulations are listed in Table \ref{TABLE_zBCComparison} in Appendix \ref{APPENDIX_TablesOfSimulations}.

Finally, to prevent mass-loss through the vertical boundary from depleting the mass in the box, we employ a simple mass source term. (This mimics what occurs in a real disc or in a global disc simulation, where mass lost from any given annulus through outflows is replenished by accretion of material from a neighboring annulus.) At the end of the $n$th step, we subtract the total mass in the box at the end of that step $M_n$ from the total mass in the box at initialization $M_0$. This mass difference $\Delta M_n \equiv M_0 - M_n$ is added back into the box with the same profile used to initialize the density (cf. Equation \ref{EQUN_densityprofile}). Thus the total mass in the box remains constant in any given simulation.

\subsection{Diagnostics}
\label{METHODS_Diagnostics}
Below we define various diagnostics in physical space. For diagnostics in Fourier space the reader should refer to Section \ref{RESULTS_SpectralGoverningEquations}.

\subsubsection{Averaged quantities}
\label{METHODS_AveragedQuantities}
The volume-average of a quantity $X$ is denoted $\langle X \rangle$ and is defined as 
\begin{equation}
\langle X \rangle(t) \equiv \frac{1}{V} \int_V X(x, y, z, t) dV,
\end{equation}
where $V$ is the volume of the box. Note that occasionally we average only over a part of the box instead of the entire domain, for example $|z| < 2H$ to capture diagnostics in the `bulk' of the disc, or $|z| > 2H$ to capture diagnostics in the `atmosphere' of the disc. In this case the region we average over is stated in the text.

We are also interested in averaging certain quantities (e.g. magnetic energy density or turbulent stresses) over time. The temporal average of a quantity $X$ is denoted $\langle{X}\rangle_t$ and is defined as
\begin{equation}
\langle X \rangle_t (x, y, z) \equiv \frac{1}{\Delta t} \int_{t_i}^{t_f} X(x, y, z, t) dt,
\end{equation}
where we integrate from some initial time $t_i$ to some final time $t_f$ and $\Delta t \equiv t_f - t_i$.

The horizontal average of a quantity $X$ is denoted $\langle{X}\rangle_{xy}$ and is defined as

\begin{equation}
    \langle X \rangle_{xy}(z,t) \equiv \frac{1}{A} \int_A X(x,y,z,t) dxdy.
\end{equation}
Horizontal averages over different coordinate directions (e.g. over the $y$- and $z$-directions) are defined in a similar manner.

\subsubsection{Reynolds and magnetic stresses and transport $\alpha$-parameter}
\label{METHODS_ReynoldsAndMagneticStressesAndAlpha}
In accretion discs, the radial transport of angular momentum is
related to the $xy$-component of the total stress
\begin{equation}
\Pi_{xy} \equiv R_{xy} + M_{xy},
\label{totalstress}
\end{equation}
in which $R_{xy} \equiv \rho u_x \delta u_y$ is the Reynolds stress, where $\delta u_y \equiv u_y + q\Omega x$ is the perturbation of the y-component of the total velocity $u_y$ about the background Keplerian flow $u_{0y} = -q \Omega_0 x$ and $M_{xy} \equiv -B_x B_y$ is the magnetic (Maxwell) stress. Note that since we will exclusively refer to fluctuating part of velocity $\delta\mathbf{u}$ below, from now on we will drop the $\delta$ and simply refer to the perturbed as $\textbf{u}$, which should not be confused with the total velocity.

The total stress is related to the classical dimensionless angular momentum transport parameter $\alpha$. This can be defined either by normalizing the total stress by the volume-averaged gas pressure $\langle P \rangle$
\begin{equation}
\alpha \equiv \frac{\langle \Pi_{xy} \rangle}{\langle P \rangle},
\label{alpha1}
\end{equation}
or alternatively by normalizing by the mid-plane pressure at initialization $P_0 = c_{s0}^2 \rho_0$
\begin{equation}
\alpha_0 \equiv \frac{\langle \Pi_{xy} \rangle}{c_{s0}^2 \rho_0},
\label{alpha2}
\end{equation}
where we remind the reader that, in code units, the mid-plane sound speed and mid-plane density at initialization are simply $c_{s0}=1$ and $\rho_0 = 1$. Note that due to the effects of vertical stratification (for which the pressure, and therefore density, decreases monotonically from the mid-plane), averaging alpha over the entire box can result in alpha defined by Equation \ref{alpha1} being as much as 1.5-3 times that defined by Equation \ref{alpha2}. Different authors use different definitions for $\alpha$ in the literature, and so caution is needed when comparing values from different sources \citep[e.g.,][]{pessah2008fundamental, heldlatter2018}.

\begin{figure}
\centering
\includegraphics[scale=0.33]
{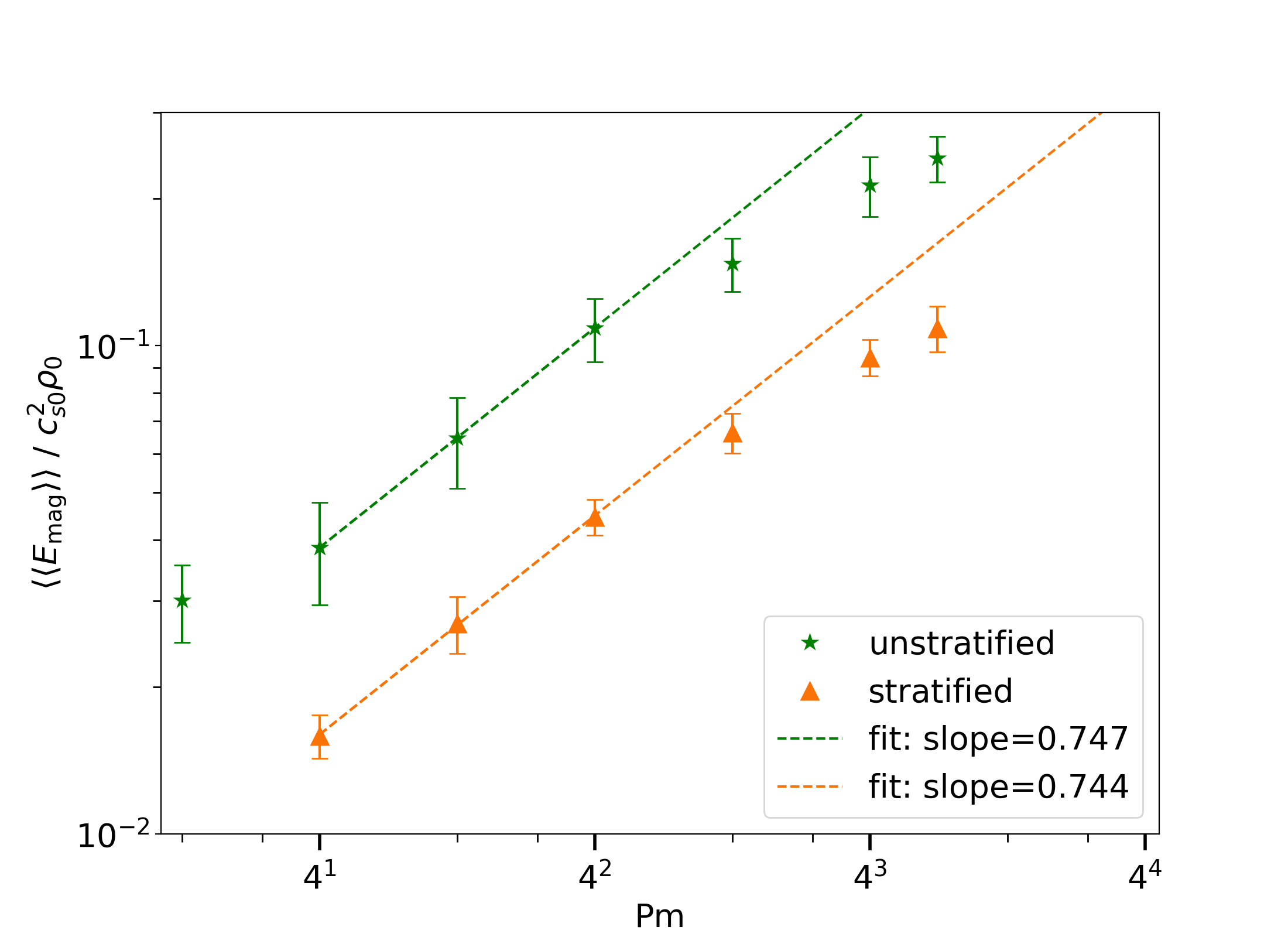}
\includegraphics[scale=0.33]{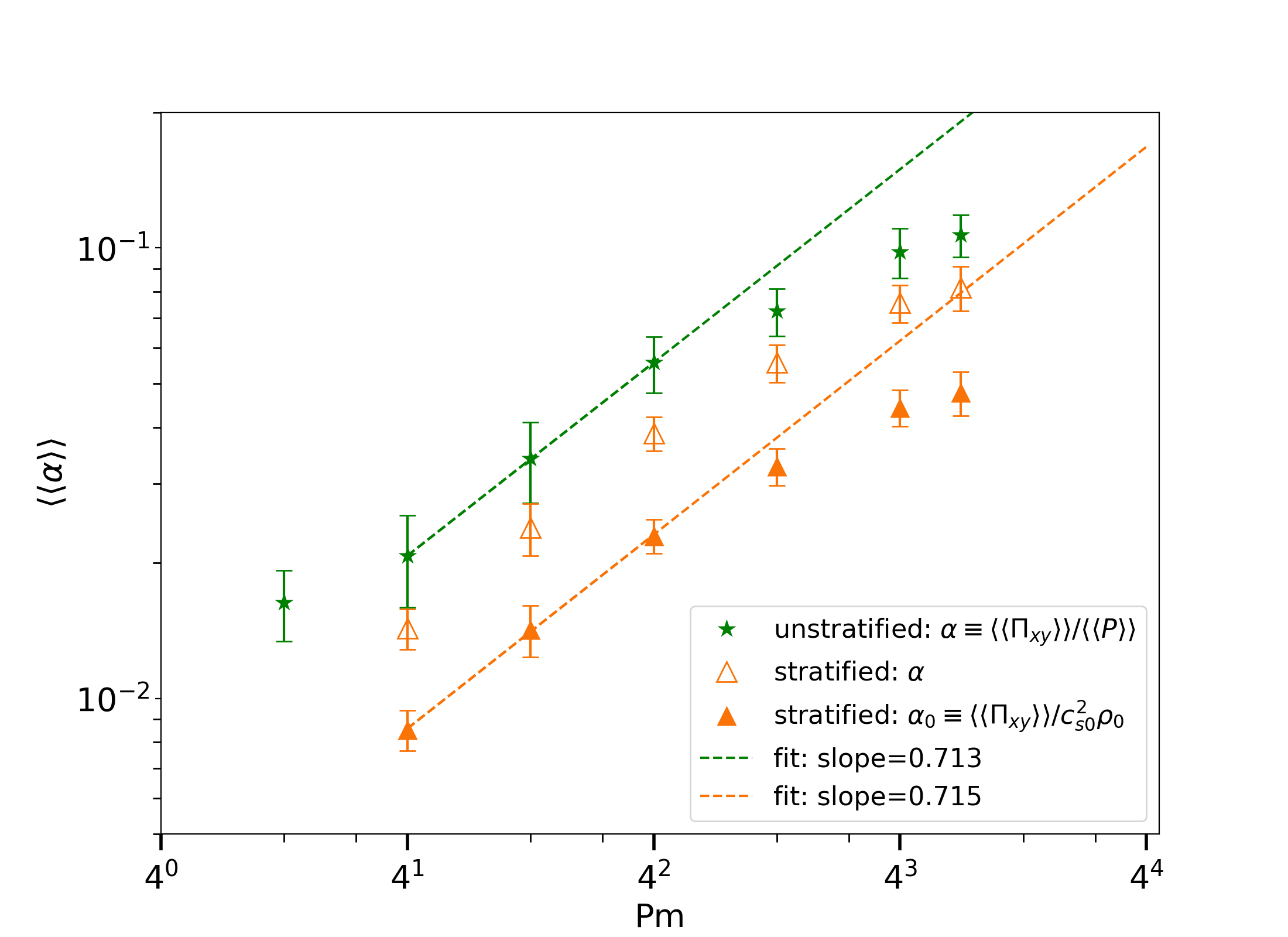}
\caption{Top: scaling of time-and-spatially averaged magnetic energy density with magnetic Prandtl number Pm in isothermal shearing box simulations of the MRI with zero-net-magnetic flux. Bottom: scaling of alpha viscosity parameter with Pm. The symbols have the following meanings: green stars denote unstratified simulations from \citep{heldmamatsashvili2022}, orange triangles denote vertically stratified simulations from this work. In the bottom panel we distinguish between $\alpha$ (solid triangles, cf. Equation \ref{alpha1}). and $\alpha_0$ (empty triangles, cf. Equation \ref{alpha2}). Note that in the stratified simulations, spatial averages were taken over the bulk of the disc ($|z|<2H$), only. Unstratified (stratified) simulations were run in boxes of size $4H\times4H\times4H$ and $4H\times4H\times8H$, respectively. All simulations were run at a resolution of $128$ cells-per-scale-height $H$ and at at fixed magnetic Reynolds number of $\text{Rm}=18750$. The abscissa is in log scale to base 4.}
\label{FIGURE_EmagAndAlphaPmScaling}
\end{figure}

\begin{figure}
\centering
\includegraphics[scale=0.24]{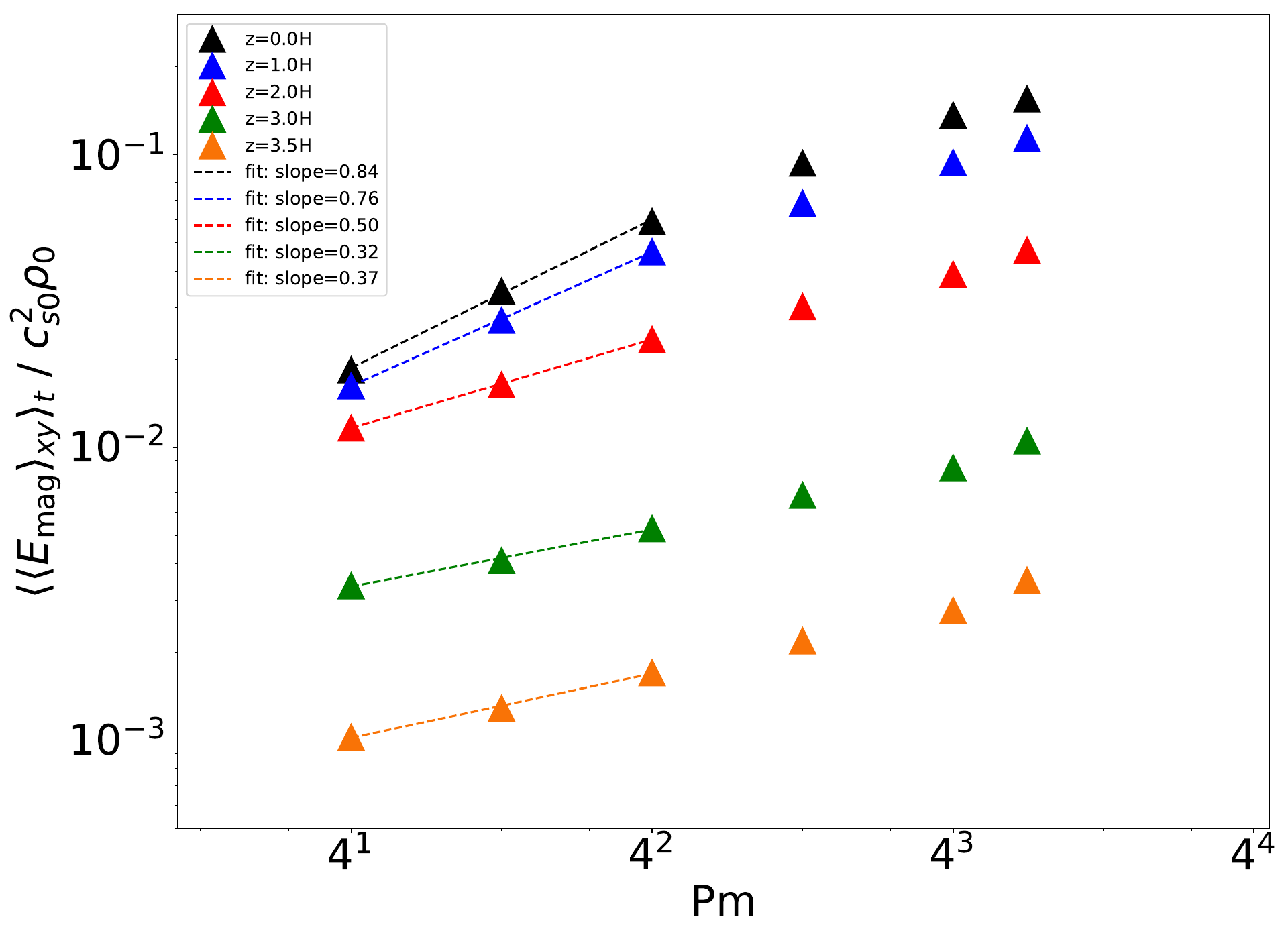}
\caption{Scaling of time- and horizontally- averaged magnetic energy density (in units of $c_{s0}^2 \rho_0$) as a function of magnetic Prandtl number Pm. The horizontal-averages are taken at different $z$-slices: $z=0$ (black), 1H (blue), 2H (red), 3H (green), and 3.5H (orange).}
\label{FIGURE_EmagPmScalingHeightDependence}
\end{figure}

\subsubsection{Magnetic buoyancy}
To determine whether (and where) the magnetic field becomes buoyantly unstable, we calculate the vertical profile of the square of the magnetic buoyancy frequency $N_m^2$, which we define as follows \citep{shi2009numerically}:

\begin{equation}
\langle \langle N_m^2 \rangle_{xy} \rangle_t \equiv \frac{1}{2} \frac{g}{\langle\langle\rho c_{s0}^2 +  B_y^2 \rangle_{xy} \rangle_t} \frac{d \langle \langle B_y^2 \rangle_{xy} \rangle_t}{d z},
\label{magneticbuoyancyfrequency}
\end{equation}
where $g=|g_{\text{eff},z}|=\Omega^2z$ is the vertical gravitational acceleration. Regions where $N_m^2 < 0$ are unstable to the Parker instability, which causes toroidal magnetic field lines to rise and escape from disc to the atmosphere.\footnote{Note that this expression characterizes the Parker instability for the \textit{total} toroidal magnetic field $B_y$, and hence in some sense gives the local instability condition for this toroidal field. Had we instead used $(\langle \langle B_y \rangle_{xy} \rangle_t)^2$ the expression would characterize the instability only of the \textit{mean} toroidal magnetic field.}

\section{Results in physical space}
\label{RESULTS_RealSpace}

In this section we present various diagnostics from our vertically stratified simulations of MRI-turbulence in the large magnetic Prandtl number regime (${\rm Pm} \geq 4$). We restrict our attention here to physical space, and defer a spectral analysis of the results to Section \ref{RESULTS_SpectralSpace}. Readers who are primarily interested in how turbulent transport and magnetic field strength scale with the magnetic Prandtl number may wish to focus their attention on \ref{RESULTS_PmComparisonAlphaPmAndEmagPmScalings} (in particular Figures \ref{FIGURE_EmagAndAlphaPmScaling} and \ref{FIGURE_EmagPmScalingHeightDependence}), only.

All the simulations described in this section have been run in a box of size $[L_x, L_y, L_z] = [4,4,8]H$, and at a resolution of $N_x\times N_y\times N_z = 512\times512\times1024$, corresponding to 128 cells per scale-height $H$ in all three directions. The magnetic Reynolds number $\text{Rm}$ is fixed at $\text{Rm}=18750$ across all runs (this value stems from a resolution study we carried out in our unstratified runs: it is high enough to achieve Reynolds numbers in the hundreds at large Pm, but low enough for the resistive scale to be well-resolved. See Appendix A of \cite{heldmamatsashvili2022}). We increase the magnetic Prandtl number from simulation to simulation by decreasing the Reynolds number $\text{Re}$.\footnote{In  \cite{heldmamatsashvili2022} we also considered the cases of keeping Re fixed and increasing Rm, and of keeping Pm fixed while simultaneously increasing Pm and Rm.} We run each simulation for 200 orbits, which gives the turbulence ample time to settle into quasi-steady state. All time-averages are taken between orbit 100 and orbit 200. For the $\text{Pm}=4$ case, we have checked that the key results are qualitatively similar regardless of choice of vertical boundary conditions or of the vertical box size (see Appendix \ref{APPENDIX_VerticalBoundaryConditions} and \ref{APPENDIX_VerticalBoxSize}, respectively). All the simulations are listed in Table \ref{TABLE_PmComparison} in Appendix \ref{APPENDIX_TablesOfSimulations}.

\begin{figure*}
\centering
\includegraphics[scale=0.4]{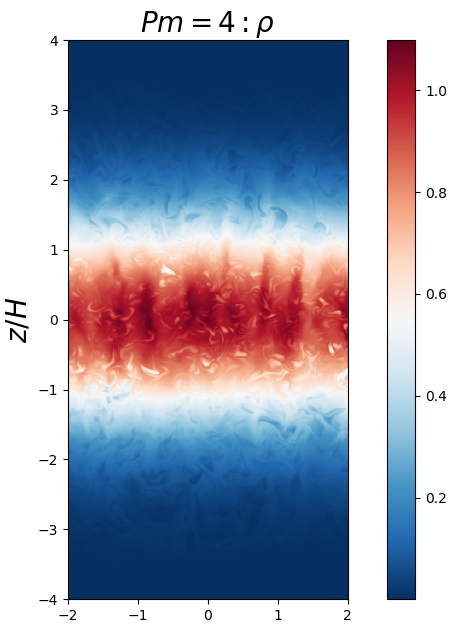}
\includegraphics[scale=0.4]{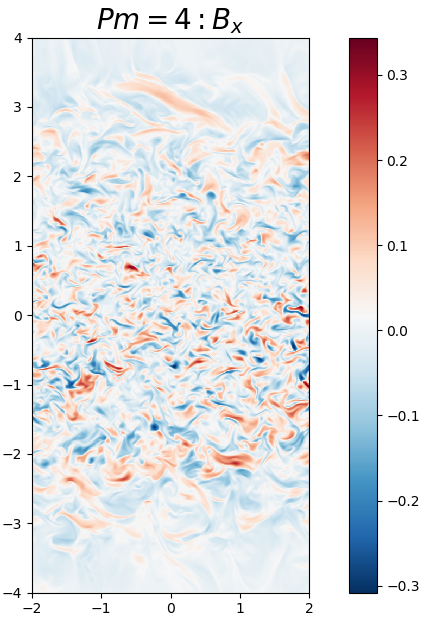}
\includegraphics[scale=0.4]{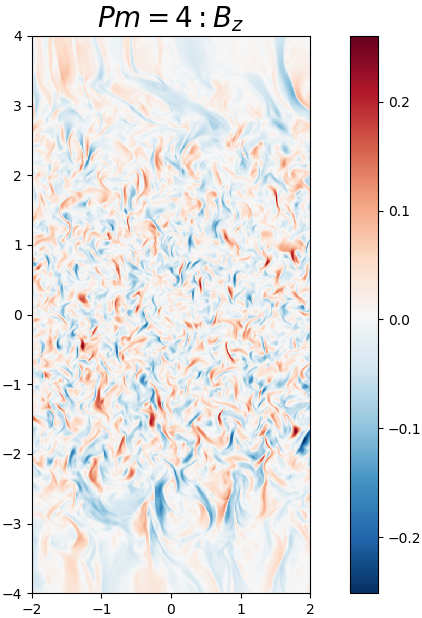}
\includegraphics[scale=0.4]{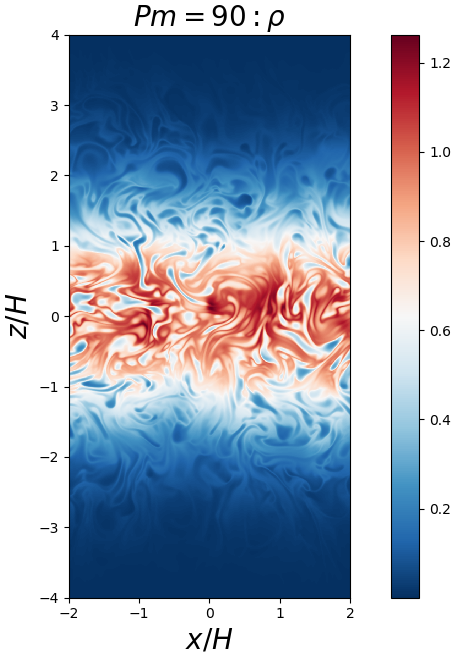}
\includegraphics[scale=0.4]{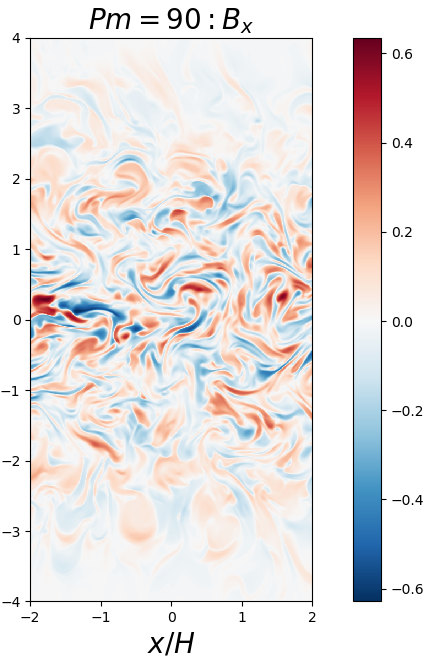}
\includegraphics[scale=0.4]{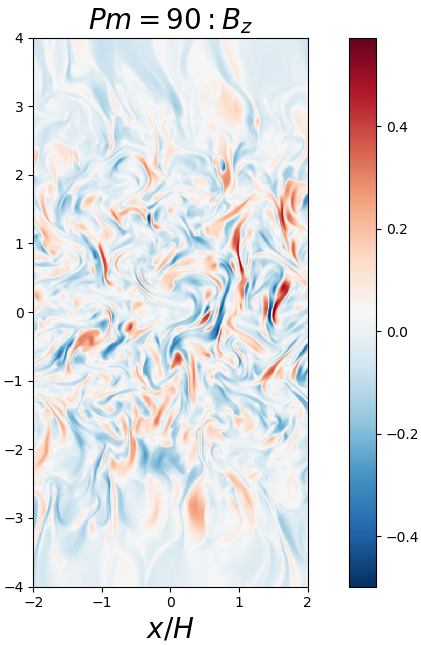}
\caption{Structures of density $\rho$ (left column) and magnetic field components $B_x$ (middle column) and $B_z$ (right column) in the $xz$-plane from two simulations at $\text{Pm = 4}$ (top row) and $\text{Pm} = 90$ (bottom row). The magnetic Reynolds number is fixed at $\text{Rm} = 18750$ in all simulations.}
\label{FIGURE_FlowFieldComparisonPm4Pm90Rm18750}
\end{figure*}

\subsection{Scalings of $\alpha$ and $E_{mag}$ with Pm}
\label{RESULTS_PmComparisonAlphaPmAndEmagPmScalings}
The key results are plotted in Figure \ref{FIGURE_EmagAndAlphaPmScaling}, where we show the dependence of the (time- and volume-averaged) total saturated magnetic energy $E_\text{mag} = (1/2)B^2$ and turbulent transport $\alpha$ parameter on magnetic Prandtl number. The green stars show the results of our previous \textit{unstratified} simulations \citep{heldmamatsashvili2022} at fixed $\text{Rm}=18750$ in boxes of size $4H\times4H\times4H$ and identical resolution $128$ cells per scale-height. All other data points are from our new, vertically stratified, simulations. Note that for the stratified simulations (denoted by triangles) we have restricted volume-averages to the bulk of the disc ($|z| < 2H$). In the bottom panel of this figure, for comparison we also show the two variants of $\alpha$ defined in Section \ref{METHODS_ReynoldsAndMagneticStressesAndAlpha}: empty triangles represent $\alpha$ (the total stress normalized by the \textit{volume-averaged} pressure $\langle P \rangle$; cf. Equation \ref{alpha1}), while solid triangles represent $\alpha_0$ (the total stress normalized by the \textit{mid-plane pressure at initialization} $P_0$; cf. Equation \ref{alpha2}). Due to the stratification (vertical pressure gradient), $\alpha$ is larger than $\alpha_0$ by a factor of around 1.70. Note that in unstratified boxes $\alpha$ is also sensitive to the box size and vertical-to-radial box aspect ratio, which we discuss in more detail in our unstratified work \citep{heldmamatsashvili2022}.

When we restrict our averages to the bulk of the disc ($|z| < 2H$), which contains $95\%$ of the total mass in the box, the results of the stratified simulations for both $E_{\text{mag}}$-Pm and $\alpha$-Pm scaling are in excellent agreement with those of our earlier unstratified simulations. At intermediate values of Pm we find power law scaling with slopes of $\delta \sim 0.74$ and $\delta \sim 0.71$ for each quantity, respectively, in agreement to two decimal places with the results of the unstratified runs, while for $\text{Pm}\gtrsim 32$ we observe the onset of a plateau. Thus, we find that in the bulk of the disc stratification does not appreciably change dependence of the turbulent dynamics on the magnetic Prandtl number. As we show below, this conclusion changes as we move away from the bulk of the disc where the MRI dominates the dynamics, and into into the atmosphere of the disc where another process due to magnetic buoyancy -- Parker instability -- comes into play.

\subsubsection{Vertical dependence of scaling}
\label{RESULTS_PmComparisonVerticalDependenceOfScaling}
A new result is that the scaling of turbulent transport and saturated magnetic energy with magnetic Prandtl number is stronger in the bulk of the disc than in the atmosphere, as seen in Figure \ref{FIGURE_EmagPmScalingHeightDependence}, where we plot the time- and \textit{horizontally}-averaged magnetic energy density $\langle \langle E_\text{mag} \rangle_{xy} \rangle_t$. Different colors correspond to horizontal-averages taken at fixed $|z|= \{0, 1, 2, 3, 3.5\}H$, respectively. The scaling clearly becomes weaker further away from the mid-plane: for example we measure a slope of $\delta \sim 0.84$ at the mid-plane ($|z|=0$), but a much more shallow slope of $\delta \sim 0.36$ at $|z|=3.5 H$. Thus our results foreshadow that different dynamics/dynamo processes (mainly MRI and Parker instability) are at play in different parts of the disc, as discussed in greater detail in Section \ref{RESULTS_MagneticPrandtlNumberDependence} below. Note that these processes have also been observed in earlier stratified simulations \citep{johansen2008high,blaes2007surface, shi2009numerically, blaes2011, kadowaki2018mhd}, although those simulations were ideal and initialized with relatively strong ($\beta_0 \sim 1-25$) net-toroidal-magnetic flux (which aids the development of the Parker instability), in contrast to our non-ideal, zero-net-flux runs with relatively weak initial magnetic field ($\beta_0  = 10^3$). We analyze the dynamics of these instabilities in more detail in Fourier space in Section \ref{RESULTS_SpectralSpace}.

\subsection{Dependence of results on Pm}
\label{RESULTS_MagneticPrandtlNumberDependence}
To determine whether the MRI is well-resolved we also measured the MRI quality-factor in the $\text{Pm}=4$ run. We define this diagnostic as $Q_z \equiv \lambda_{\text{MRI,z}}/\Delta z$, where $\Delta z$ is the grid size in the vertical direction.\footnote{The wavelength of fastest growing MRI mode in the $z$-direction is given by $\lambda_{\text{MRI,z}} \equiv 2\pi u_{A,z} / \Omega$, where $u_{A,z}$ is the vertical Alfv\'en speed.} At the mid-plane we find $Q_z \sim 26 \,(\lambda_z \sim 0.2H)$, while $z = 2H$ we find $Q_z \sim 68 \,(\lambda_z \sim 0.5H)$. Thus the critical MRI wavelength is very well resolved within the bulk of the disc, and also fits comfortably in the box.

\subsubsection{Structure of the flow}
\label{RESULTS_FiducialSim_StructureOfTheFlow}
Next we consider the flow field in the $xz$-plane, which is shown Figure \ref{FIGURE_FlowFieldComparisonPm4Pm90Rm18750} from a simulation characteristic of the power-law scaling region ($\text{Pm}=4$, top row) and also from a simulation characteristic of the plateau region ($\text{Pm}=90$, bottom row). The density is shown in the left-hand panels: due to the inclusion of stratification, the bulk of the disc is concentrated between $\pm 2H$, surrounded by a tenuous atmosphere. As in our unstratified runs, we observe density waves superimposed on the turbulence at lower Pm but not at higher Pm \citep{heldmamatsashvili2022, shi2016}. 

The magnetic field is shown  in the middle and right-hand panels of Figure \ref{FIGURE_FlowFieldComparisonPm4Pm90Rm18750} for $B_x$ and $B_z$, respectively. A small-scale, turbulent magnetic field dominates the flow within $\pm H$ of the mid-plane, while in the atmosphere ($|z| \gtrsim 2.2H$) the field is still turbulent, but characterized by noticeably larger structures. Further into the atmosphere still, the character of the magnetic field changes and we observe only large-scale structures (of size $\sim 0.8H$). In particular, $B_z$  is characterized by streaky, vertically-oriented structures. This indicates that in the disc atmosphere the magnetic field structure is not really characteristic of MRI turbulence, but rather an ordered, vertical field has formed. This is due to magnetic buoyancy, specifically the Parker instability, which dominates MRI in the atmosphere, as we show in Section \ref{RESULTS_ParkerInstability}. Finally, we observe larger structure in both the velocity (not shown) and magnetic fields as Pm is increased, as expected given the low Reynolds number ($\text{Re}=208$) at $\text{Pm}=90$.\footnote{The separation between the viscous scale and disk scale-height $H$ is characterized by the Reynolds number Re, where the viscous scale is given by $l_\nu = (2\pi/\sqrt{Re})H$ (see footnote 6 of \cite{heldmamatsashvili2022} for a derivation). Even in the the plateau region where Re reaches its minimum value ($\text{Re}=208$), $l_\nu \sim 0.4$H. Thus the viscous scale is less than $H$ in all our simulations.} Reassuringly, the flow field near the mid-plane at both values of Pm resembles that observed in our unstratified simulations (see Figure 3 of \citep{heldmamatsashvili2022})

\begin{figure}
\centering
\includegraphics[scale=0.25]{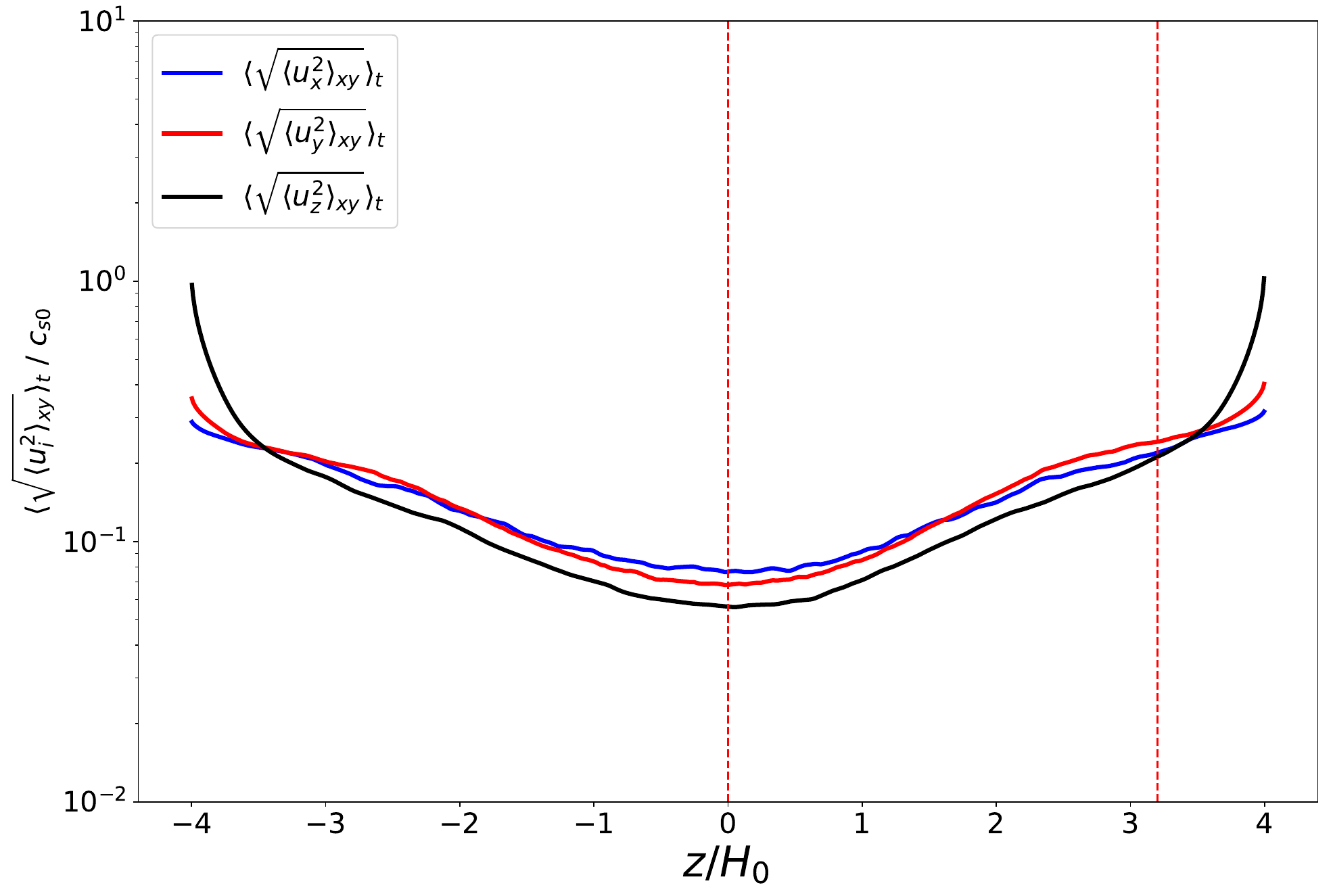}
\includegraphics[scale=0.25]{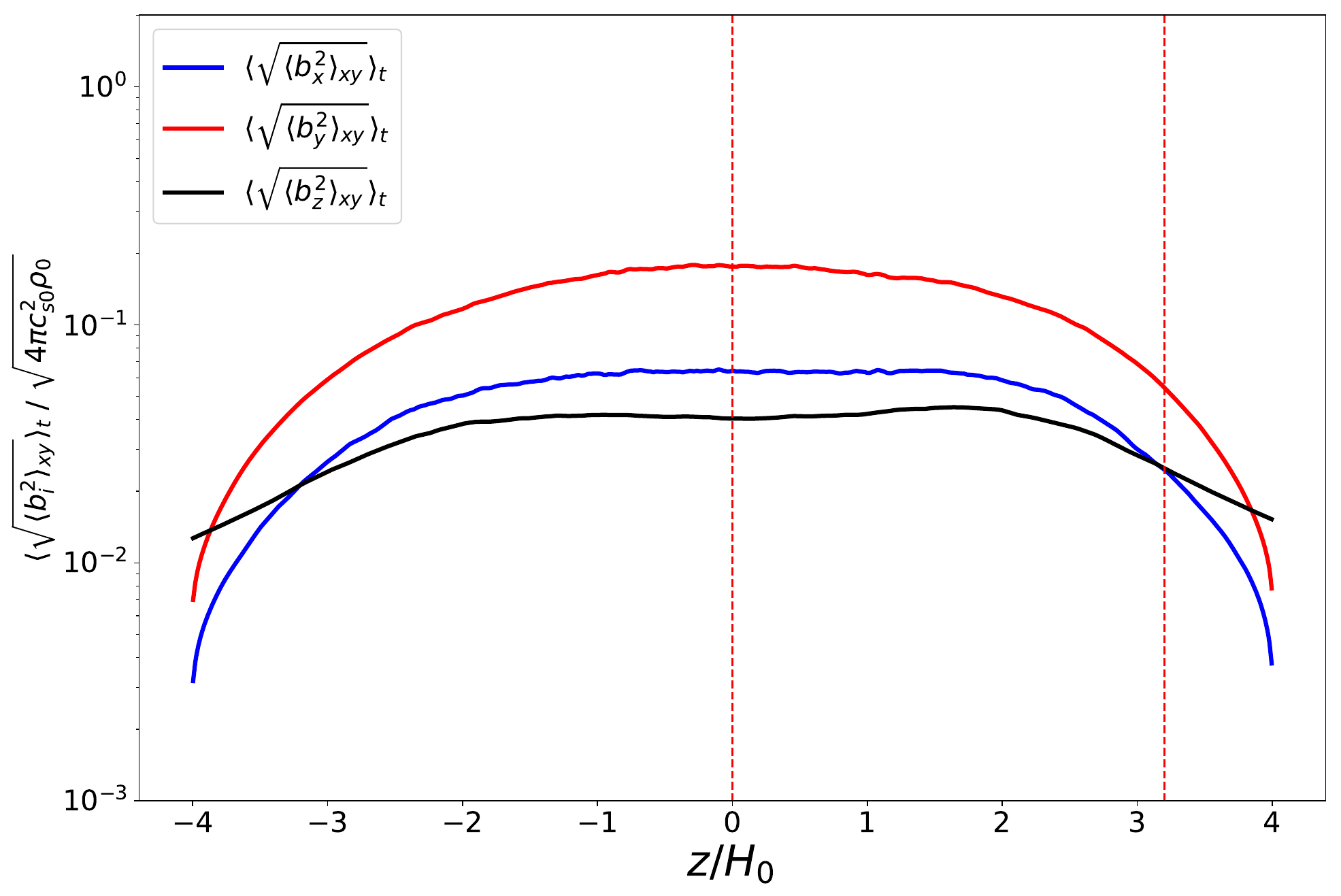}
\caption{Vertical disc structure (vertical profiles of time-and-horizontally averaged quantities) from the fiducial simulation at ${\rm Pm}=4$. Top: root-mean-square velocity components. Bottom: root-mean-square magnetic field components. The vertical dashed lines at $z = 0.0H$ and $z = 3.2H$ show locations at which we performed the spectral analysis (see Section \ref{RESULTS_SpectralFiducialSimulation}).}
\label{FIGURE_Pm4Verticaldiscstructure}
\end{figure}

\subsubsection{Vertical disc structure}
\label{RESULTS_FiducialSimVerticalProfiles}
To characterize the vertical structure of the disk, we look at diagnostics such as the time- and horizontally-averaged vertical profiles of density, plasma beta, and $\alpha$ at different Pm (not shown). As we increase Pm, we observe the largest changes to the disk structure within the bulk of the disk, as foreshadowed in Figure \ref{FIGURE_EmagPmScalingHeightDependence}. In the power-law scaling region at low Pm, the density remains nearly unchanged from its initial Gaussian profile, while the plasma beta parameter drops from $\beta \sim 55$ at the mid-plane to $\beta \sim 5$ at $|z| \sim 4 H$. The entire disc remains gas-pressure-dominated at all values of Pm that we investigated, although magnetic pressure contributes significantly to the total pressure as Pm is increased ($\beta \lesssim 10$ at the mid-plane at $\text{Pm}=90$).  Meanwhile, the $\alpha$ parameter is largest at the mid-plane and drops monotonically away from the mid-plane. The profiles of $\alpha$, plasma beta, and magnetic field are relatively uniform within the bulk of the disk at ${\rm Pm}=4$, but becomes increasingly peaked around $z = 0$ as ${\rm Pm}$ increases. The latter behavior is likely due to the increased importance of buoyancy in the disc region as ${\rm Pm}$ is increased, which results in the buoyantly unstable region extending closer to the mid-plane (see Figure \ref{FIGURE_ParkerInstabilityDirectEvidence}).

In Figure \ref{FIGURE_Pm4Verticaldiscstructure} we show vertical profiles of all horizontally-averaged root-mean-square (rms) velocity and magnetic field components at $\text{Pm}=4$. The profiles were time-averaged between around orbit 100 and orbit 200. At the mid-plane the dominant velocity components are the radial and azimuthal components (top panel). All velocity components increase away from the mid-plane (essentially scaling with the Alfv\'en speed $u_A \equiv B/\sqrt{\rho}$), and at $|z|\gtrsim 3H$ the vertical component $u_z$ (black curve) increases rapidly, quickly becoming the dominant component. This suggests that outflows dominate the dynamics in the upper reaches of the disc, probably due to the emergence of the large-scale vertical field $B_z$ in this region, as seen in its $xz$-slice in the top-right-hand panel of Figure \ref{FIGURE_FlowFieldComparisonPm4Pm90Rm18750}. Turning to the magnetic field vertical profiles (bottom panel of Figure \ref{FIGURE_Pm4Verticaldiscstructure}), the toroidal component $B_y$ (red curve) dominates over the radial (blue curve) and vertical (black curve) components at nearly all $z$. The toroidal field decreases monotonically with height away from the mid-plane, while the radial and vertical field are relatively constant within the bulk of the disc. In the atmosphere the toroidal and radial field drop rapidly, however, becoming comparable to the vertical field around $\pm 3H$. This suggests a magnetically-driven outflow dominates the dynamics away from the bulk of the disc even in the present case of zero-net-vertical magnetic flux, an outcome that otherwise generally requires sufficiently strong net-vertical-flux \citep{lesur2021magnetohydrodynamics}.

\subsubsection{Spacetime diagrams}
\label{RESULTS_FiducialSimSpacetimeDiagrams}
\begin{figure}
\centering
\includegraphics[scale=0.23]{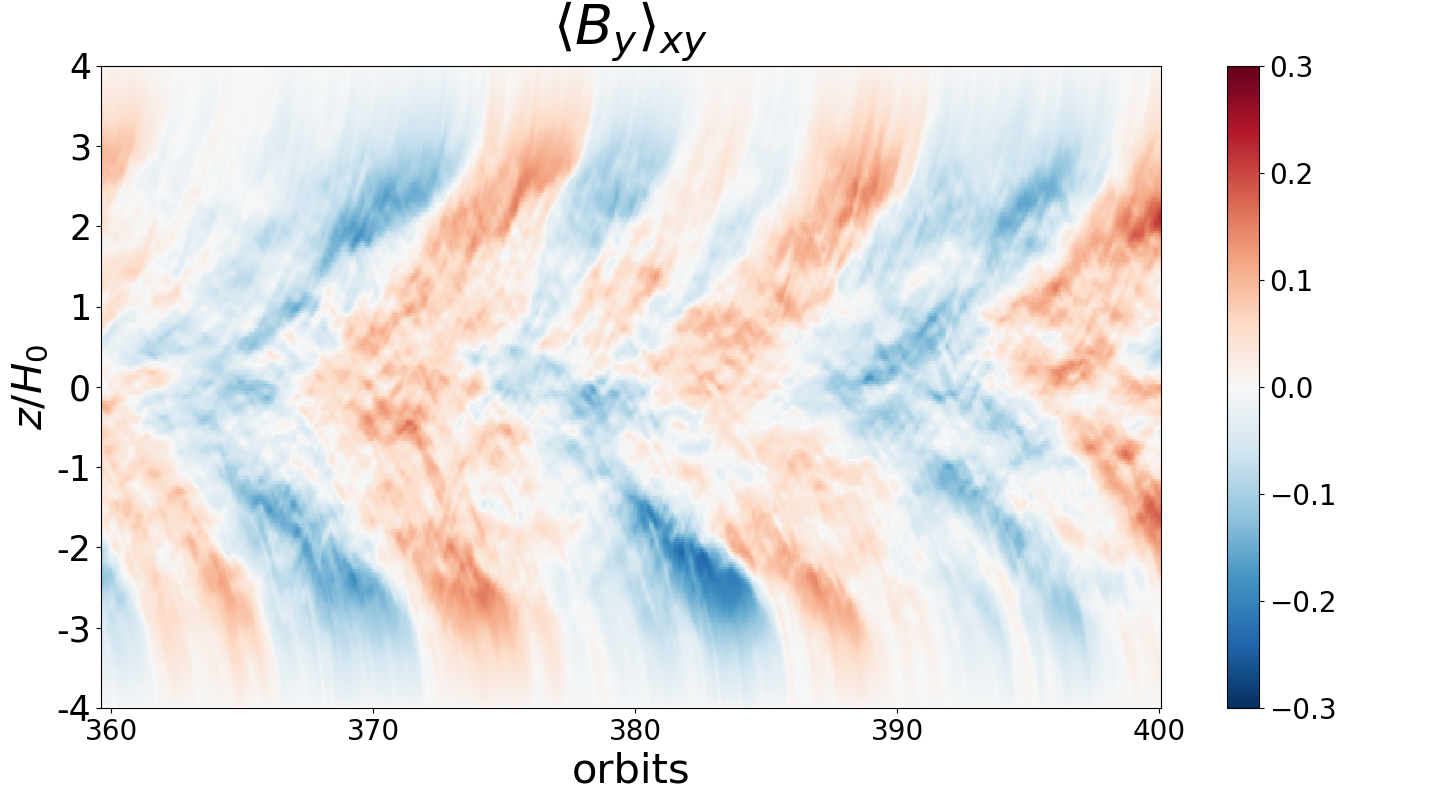}
\includegraphics[scale=0.23]{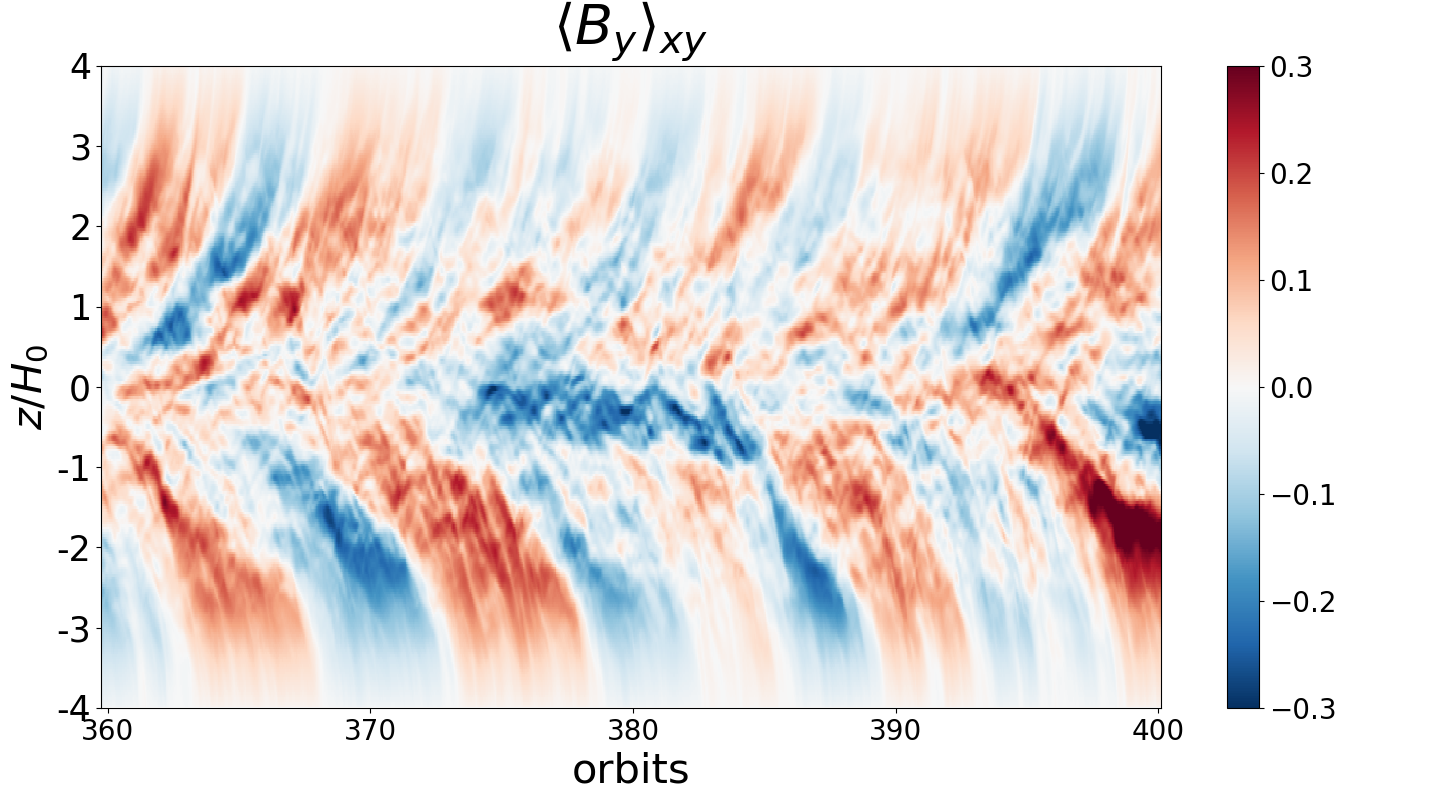}
\caption{Spacetime diagrams for the horizontally-averaged toroidal magnetic field $B_y$ taken from a simulation at the onset of the power-law scaling region at $\text{Pm}=4$ (top) and from a simulation in the plateau region at $\text{Pm}=90$ (bottom). The data are taken from low resolution runs (32 cells per scale-height $H$). Note, to facilitate easier comparison, we set the colorbar limits of the top panel to match those in the lower panel, where the toroidal field is somewhat stronger.}
\label{FIGURE_SpacetimeDiagrams}
\end{figure}

To examine the temporal behaviour of the magnetic field in more detail, we next turn to spacetime diagrams of toroidal field $B_y$ (see Figure \ref{FIGURE_SpacetimeDiagrams}), which are shown between orbit 360 and orbit 400 from our low resolution runs (32 cells per scale-height $H$). The top panel is from a simulation at $\text{Pm} = 4$, while the bottom panel is from a simulation at $\text{Pm} = 90$. We observe the characteristic `butterfly' pattern in $B_y$, which has been observed in previous stratified MRI-turbulence simulations \citep[e.g.,][]{davisstonepesssah2010,gressel2010,simon2011,bodo2014,salvesen2016a}. The toroidal field changes sign every $5$-$10$ orbits, and this period does not appear to be sensitive to the magnetic Prandtl number. The field reversals are most evident outside $\pm H$ of the mid-plane. Closer inspection reveals that field reversals are already present very close to the mid-plane at low Pm, where the disc is otherwise highly turbulent, but the butterfly pattern inside the disc $|z| < 2H$ is disrupted at $\text{Pm} = 90$. 

\subsubsection{Parker instability}
\label{RESULTS_ParkerInstability}
\begin{figure}
\centering
\includegraphics[scale=0.25]{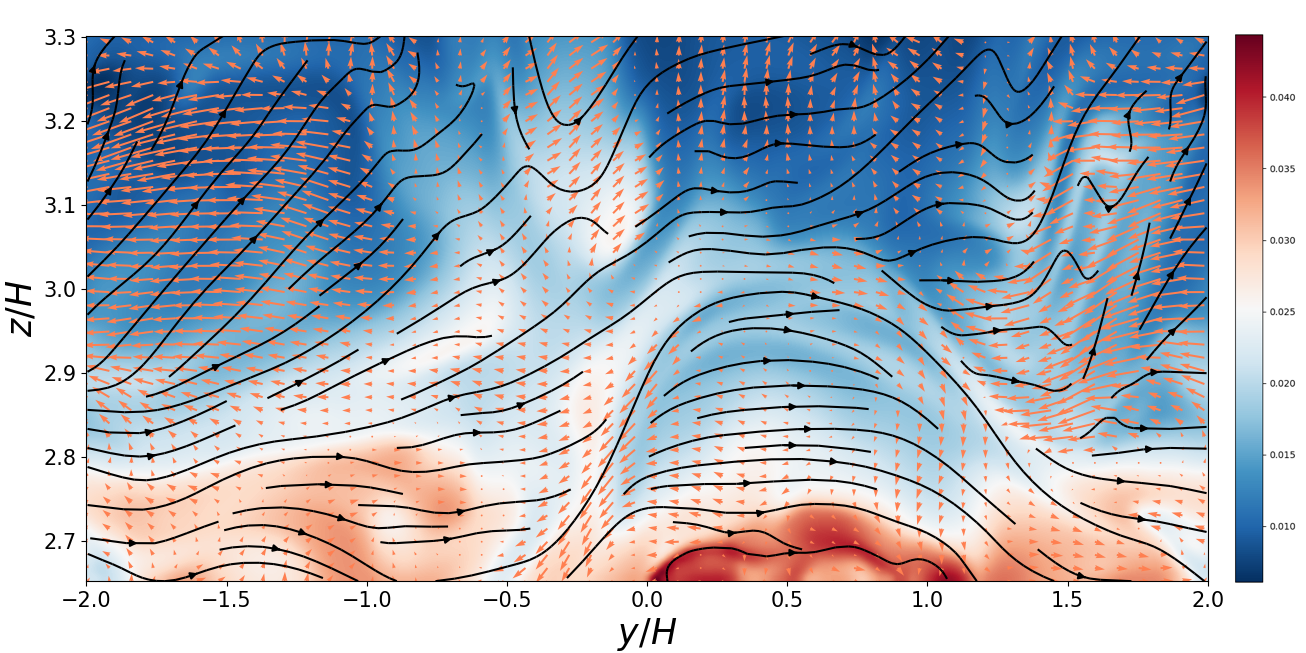}
\includegraphics[scale=0.25]{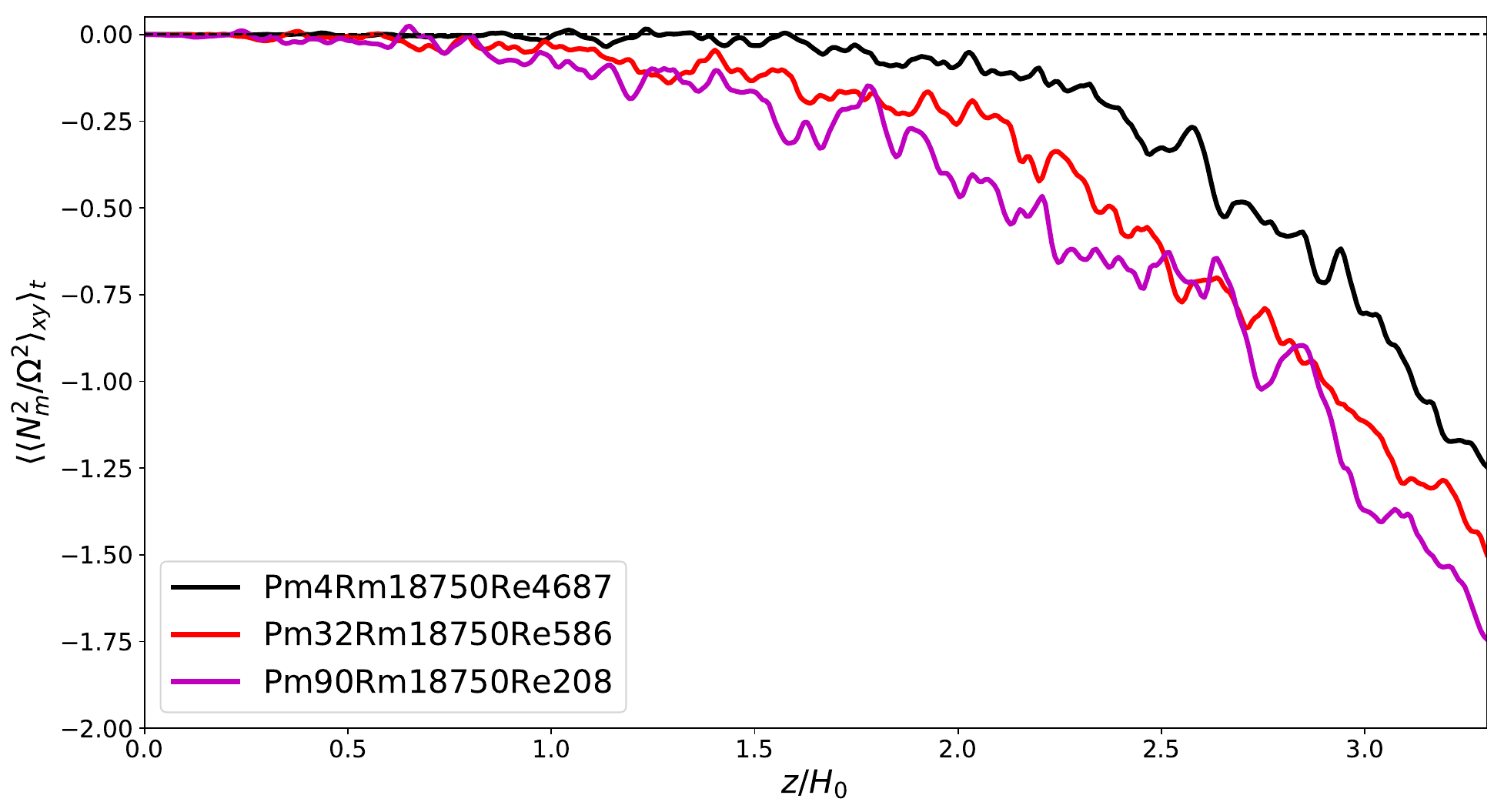}
\caption{Evidence of Parker instability. Top: flow field from a snapshot in the $yz$-plane at magnetic Prandtl number $\text{Pm}=4$. The colorplot shows the density, black lines show magnetic field lines, and orange arrows indicate the velocity field. The upwelling of magnetic field between $y=-0.5H$ and $y=2.0H$ with plasma flowing down from the crest of the upwelling along the magnetic field lines and into the troughs is characteristic of structures expected from the Parker instability. Bottom: vertical profiles of magnetic buoyancy frequency squared at different Pm in the upper half-plane. (Note that the upper $z$-cut-off of $z=3.3H$ on the $x$-axis of the bottom panel has been chosen to coincide with that on the $y$-axis of the top panel.)}
\label{FIGURE_ParkerInstabilityDirectEvidence}
\end{figure}

To investigate whether the disc is unstable to the Parker instability we plot the square of the magnetic buoyancy frequency $N_m^2$ (Equation \ref{magneticbuoyancyfrequency}). We find that, within the bulk of the disc ($|z| \lesssim 2H$), $N_m^2 \approx 0$  (bottom panel of Figure \ref{FIGURE_ParkerInstabilityDirectEvidence}), and thus the bulk of the disc is marginally stable to Parker instability. Outside this region $N_m^2 < 0$, and increases in absolute value with height. This shows that the atmosphere of the disc is magnetically buoyantly unstable. 

Another characteristic of the Parker instability is the bending (undulation) of magnetic field lines, and associated over- and under-densities in the gas density as plasma slides along the field lines away from the crests and collects in the troughs. See, for example, Figure 1 of \cite{johansen2008high} which is taken from a 2D simulation in an azimuthally extended box with strong net-toroidal-flux. In our 3D zero-net-magnetic-flux simulations, toroidal field is produced mostly by the shear and the flow field is much more complex than in 2D simulations. Thus such clear undulations of the field lines with over-densities in the troughs of the undulations is difficult to detect. Nevertheless, we have observed many examples of bending (and rising) field lines in our simulation. In Figure \ref{FIGURE_ParkerInstabilityDirectEvidence} we show an example from one snapshot at orbit 135. The colorplot shows the density in the $yz$-plane between $z=2.65H$ (upper edge of the disc) and $z=3.3H$. Note that in this part of the domain the square of the magnetic buoyancy frequency is negative, and so we expect the flow to be Parker unstable. The black lines are the magnetic field lines, while the yellow arrows indicate the velocity field. An upwelling of the magnetic field can clearly be seen between $y=-0.5H$ and $y=2H$. The velocity (yellow arrows) points along the magnetic field streamline on either side of the undulation, and two overdense regions can be seen in the troughs. Thus we have direct visual evidence of Parker instability in our vertically stratified zero-net-flux shearing box simulations.

\section{Spectral analysis}
\label{RESULTS_SpectralSpace}

To understand the energetics and dynamics that underpin the turbulence, following \cite{heldmamatsashvili2022}, we now present a detailed spectral analysis of the results first discussed above in physical space. We begin by transforming the induction equation in Fourier (wavenumber) space in Section \ref{RESULTS_SpectralGoverningEquations}. In Section \ref{RESULTS_SpectralFiducialSimulation} we present a spectral analysis at fixed $\text{Pm} = 4$, including energy spectra and the spectra of the dynamical terms governing magnetic field evolution at different heights in the disc. In Section \ref{RESULTS_SpectralPmDependence} we discuss how the results change with the magnetic Prandtl number. The reader who is only interested in how the main physical processes (MRI and Parker instability) sustain the turbulence/dynamo should jump to the summary in Section \ref{RESULTS_SummarySustenanceSchemesAtDifferentHeights}.

\subsection{Governing equations}
\label{RESULTS_SpectralGoverningEquations}
We start the spectral analysis by decomposing velocity and magnetic field perturbations into spatial Fourier modes in the radial $x-$ and toroidal $y-$coordinates, along which the perturbations are shear-periodic and periodic, respectively. Along the vertical $z$-coordinate, there is no periodicity due to the presence of stratification, so we do not Fourier transform in $z$. Thus, we have
\citep{riols2017gravitoturbulence}
\begin{equation}\label{eq:fourier}
f({\bf r},t)=\int \bar{f}(k_x,k_y,z,t)\exp\left({\rm
	i}k_xx+{\rm i}k_yy\right)dk_xdk_y,
\end{equation}
where $f\equiv ({\bf u},{\bf B})$ and $\bar{f}\equiv(\bar{\bf u},\bar{\bf B})$ denote the corresponding Fourier amplitudes. The grid in Fourier (${\bf k}$-)space is determined by the horizontal sizes of the flow domain $L_i$ and numerical resolution $N_i$, where $i\in \{x,y \}$, such that the cell sizes in Fourier space are given by $\Delta k_i = 2\pi /L_i$, and hence the wavenumbers run through the values $k_i=n_i\Delta k_i$, where $n_i = 0, \pm 1, \pm 2,..., \pm N_i/2$ is an integer.

Performing a Fourier decomposition in horizontal slices at each height $z$ allows us to explore different dynamical regimes discussed in Section \ref{RESULTS_RealSpace} --  MRI-turbulence dominating in the bulk of the disc at $|z| \lesssim 2H$ and a magnetic buoyancy, or Parker instability regime dominating in the upper layers $|z|\gtrsim 2H$.\footnote{Note, we refer here to which process \textit{dominates} the dynamics: the MRI is present at \textit{all} $z$, and is thus still active in the atmosphere.}  

Substituting Equation (\ref{eq:fourier}) into induction Equation (\ref{SB4}), we obtain the equations for the spectral magnetic field components,
\begin{multline}\label{eq:Bxk}
\frac{\partial \bar{B}_x}{\partial t}=-qk_y\frac{\partial \bar{B}_x}{\partial k_x}+ik_y(u_xB_y-u_yB_x)_{\bf k}+\frac{\partial}{\partial z}(u_xB_z)_{\bf k}\\-
\frac{\partial}{\partial z}(u_zB_x)_{\bf k}
+\frac{1}{Rm}\Delta_{\bf k}\bar{B}_x
\end{multline}
\begin{multline}\label{eq:Byk}
\frac{\partial \bar{B}_y}{\partial t}=-qk_y\frac{\partial \bar{B}_y}{\partial k_x}-q\bar{B}_x+ik_x(u_yB_x-u_xB_y)_{\bf k}
\\+\frac{\partial}{\partial z}(u_yB_z)_{\bf k}-\frac{\partial}{\partial z}(u_zB_y)_{\bf k}+\frac{1}{Rm}\Delta_{\bf k}\bar{B}_y
\end{multline}
\begin{multline}\label{eq:Bzk}
\frac{\partial \bar{B}_z}{\partial t}=-qk_y\frac{\partial \bar{B}_z}{\partial k_x}+ik_x(u_zB_x)_{\bf k}+ik_y(u_zB_y)_{\bf k}\\-ik_x(u_xB_z)_{\bf k}-ik_y(u_yB_z)_{\bf k}+\frac{1}{Rm}\Delta_{\bf k}\bar{B}_z,
\end{multline}
where in these and other spectral equations below we have chosen to denote Fourier transforms of the products of any two quantities with the subscript ${\bf k}$ instead of over-bars (which should not cause confusion) and $\Delta_{\bf k}=-k_x^2-k_y^2+\partial^2/\partial z^2$ is the 2D Laplace operator in the $(k_x,k_y)$-plane. The Fourier transform of the nonlinear terms -- the products of velocity and magnetic field components $(u_iB_j)_{\bf k}$, where $i,j\in\{x,y,z\}$, come from the Fourier transform of the electromotive force ${\bf u}\times{\bf B}$ in Equation (\ref{SB4}) and are given by convolutions in Fourier space \citep{mamatsashvili2020zero}
\[
(u_iB_j)_{\bf k}=\int d^2{\bf k}'\bar{u}_i({\bf k}',z,t)\bar{B}_j({\bf k}-{\bf k}',z,t).
\]
They describe the net effect of the nonlinear triadic interactions of a mode with ${\bf k}$ with two other modes ${\bf k}'$ and ${\bf k}-{\bf k}'$. 

Note that stratification does not explicitly appear in the induction equation (\ref{SB4}). However, it influences the vertical velocity $u_z$ in the momentum equation (\ref{SB2}) and, via the latter, the magnetic field dynamics. Thus, to characterize the effect of stratification, we have separated out the contributions of $u_z$ and $B_z$ in the nonlinear terms in the magnetic field Equations (\ref{eq:Bxk})-(\ref{eq:Bzk}). Multiplying both sides of these equations by the complex conjugates $\bar{B}_x^{\ast}$, $\bar{B}_y^{\ast}$, and $\bar{B}_z^{\ast}$, respectively, we obtain the equations for the squared moduli of the spectral magnetic field components,
\begin{equation}\label{eq:Bxk2}
\frac{\partial }{\partial t}\frac{|\bar{B}_x|^2}{2}=-qk_y\frac{\partial}{\partial k_x}\frac{|\bar{B}_x|^2}{2}+{\cal N}_x^{(h)}+{\cal N}_x^{(z)}+{\cal N}_{x}^{(va)}+{\cal D}_x,
\end{equation}
\begin{equation}\label{eq:Byk2}
\frac{\partial }{\partial t}\frac{|\bar{B}_y|^2}{2}=-qk_y\frac{\partial}{\partial k_x}\frac{|\bar{B}_y|^2}{2}+{\cal M}+{\cal N}_y^{(h)}+{\cal N}_y^{(z)}+{\cal N}_{y}^{(va)}+{\cal D}_y,
\end{equation}
\begin{equation}\label{eq:Bzk2}
\frac{\partial }{\partial t}\frac{|\bar{B}_z|^2}{2}=-qk_y\frac{\partial}{\partial k_x}\frac{|\bar{B}_z|^2}{2}+{\cal N}_{z}^{(h)}+{\cal N}_{z}^{(z)}+{\cal D}_z,
\end{equation}
which are central in the spectral analysis. Their right-hand sides contain terms of linear and nonlinear origin. The linear terms are:
\begin{enumerate}
\item
the shear-induced drift, $-qk_y\partial /\partial k_x$, which simply advects the field for non-axisymmetric $(k_y \neq 0)$ modes along the $k_x$-axis without any new energy production,
\item
The Maxwell stress multiplied by the shear parameter $q$,
\begin{equation}
{\cal M}=-\frac{q}{2}(\bar{B}_x\bar{B}_y^{\ast}+\bar{B}_x^{\ast}\bar{B}_y),
\end{equation}
which is responsible for the energy exchange between
(toroidal) magnetic field and the background disc flow. In this case, the Maxwell stress mediates the growth of toroidal field via stretching of radial field by Keplerian shear, which is a main part of the (non-modal) MRI process \citep{herault2011periodic,squire2014,mamatsashvili2013,gogichaishvili2017,riols2017magnetorotational}. It is the main term supplying (injecting) energy into turbulence.
\item
Ohmic dissipation terms,
\begin{equation}\label{eq:Dbi}
{\cal D}_i=-\frac{1}{\rm Rm}(k_x^2+k_y^2)|\bar{B}_i|^2+\frac{1}{\rm Rm}\bar{B}_i^{\ast}\frac{\partial^2 \bar{B}_i}{\partial z^2}, ~~~i\in{x,y,z}.
\end{equation}
\item
The nonlinear transfer terms are: ${\cal N}_i^{(h)}$, ${\cal N}_i^{(z)}$, ${\cal N}_{x}^{(va)}$, and ${\cal N}_{y}^{(va)}$, where $i\in{x,y,z}$, which correspond to the nonlinear terms in the original equations (Equations \ref{eq:Bxk}-\ref{eq:Bzk}). As mentioned above, they are classified into the terms explicitly containing the vertical velocity $u_z$ and magnetic field $B_z$, and the terms independent of these components. The explicit forms and physical meanings of these nonlinear terms are as follows: 
\end{enumerate}
The nonlinear induction-advection terms depending only on horizontal velocity, $(u_x, u_y)$, and magnetic field $(B_x, B_y)$ are:
\begin{equation}\label{eq:Nxh}
{\cal N}_x^{(h)}=ik_y\bar{B}_x^{\ast}(u_xB_y-u_yB_x)_{\bf k}+c.c.,
\end{equation}
\begin{equation}
{\cal N}_y^{(h)}=ik_x\bar{B}_y^{\ast}(u_yB_x-u_xB_y)_{\bf k}+c.c.,
\end{equation}
which describe, respectively, the production of $\bar{B}_x$ from $\bar{B}_y$ and vice versa due to stretching by horizontal velocity variation (shear) along $x$- and $y$-directions (the first terms in the brackets) and their horizontal advection/transport (second terms in the brackets). 

\begin{figure}
\centering
\includegraphics[scale=0.37]{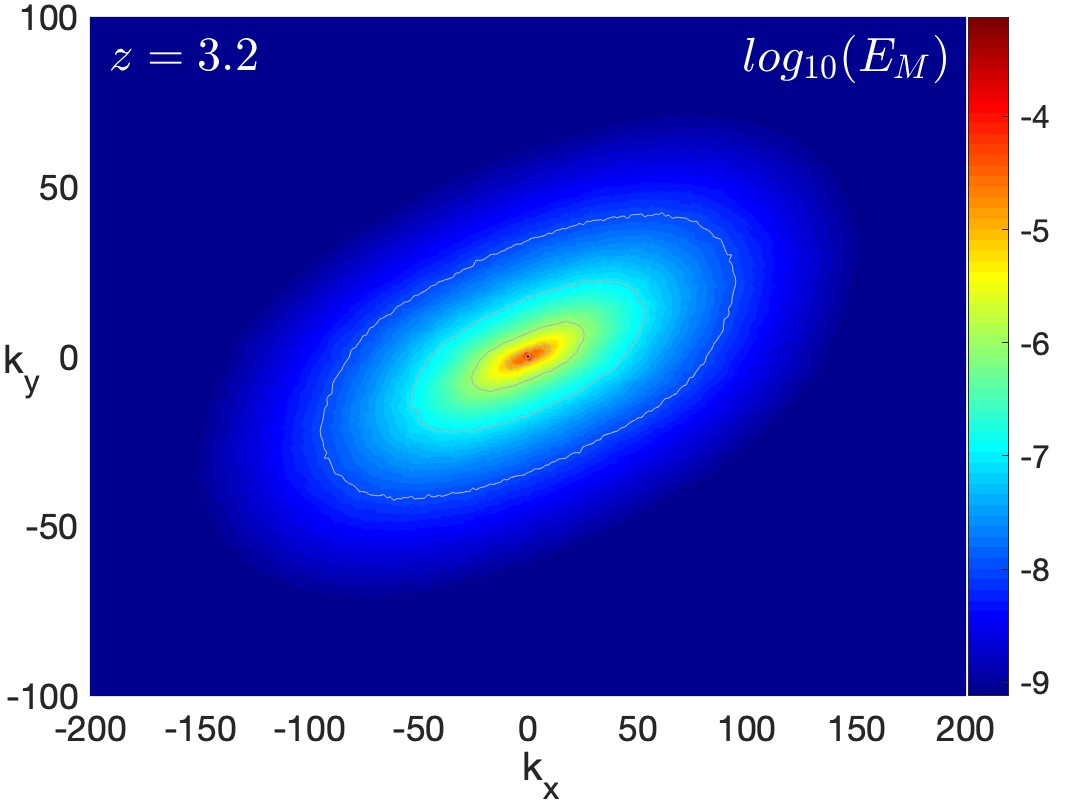}
\includegraphics[scale=0.37]{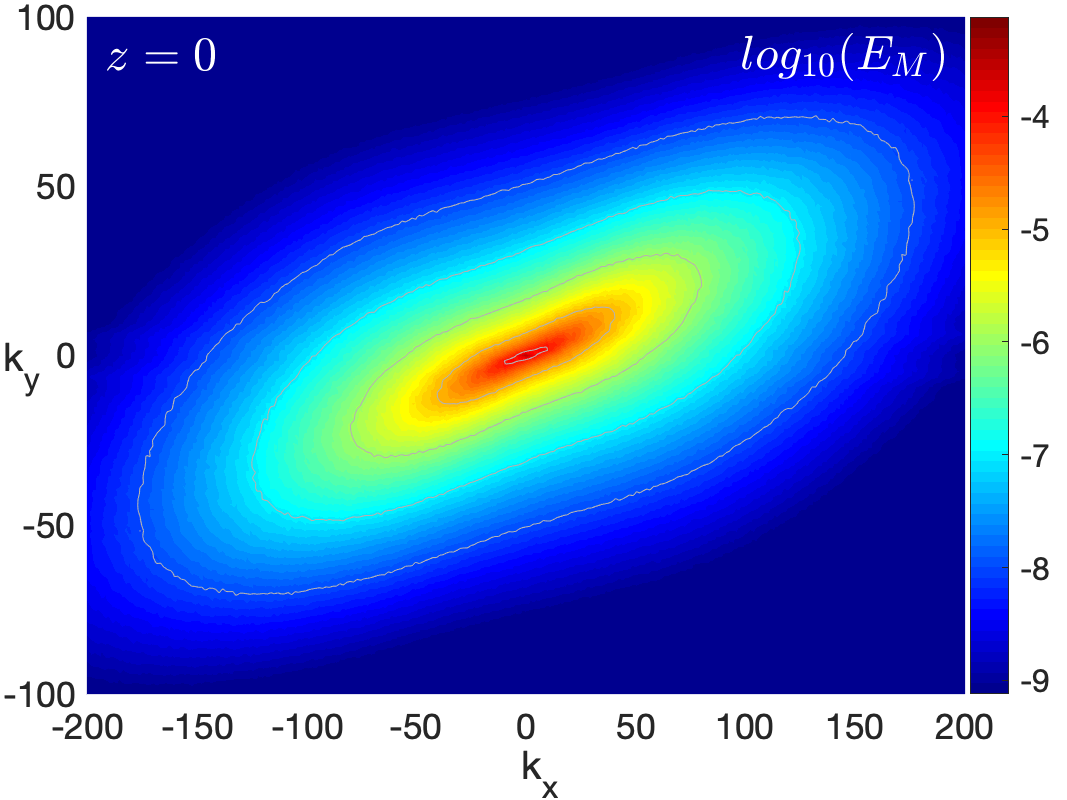}
\caption{Magnetic energy spectra in the $(k_x,k_y)$-plane near the mid-plane at $z=0$ (bottom panel) and in the atmosphere at $z=3.2$ (top panel) for ${\rm Pm}=4$, ${\rm Rm}=18750$ and resolution $128/H$. The spectra are anisotropic -- strongly inclined towards the $k_x$-axis -- due to Keplerian shear, and become concentrated towards smaller wavenumbers with increasing height.}\label{FIGURE_2Denergyspectra}
\end{figure}

The nonlinear terms  containing vertical field $B_z$ are:
\begin{equation}
{\cal N}_x^{(z)}=\bar{B}_x^{\ast}\frac{\partial}{\partial z}(u_xB_z)_{\bf k}+c.c.,
\end{equation}
\begin{equation}
{\cal N}_y^{(z)}=\bar{B}_y^{\ast}\frac{\partial}{\partial z}(u_yB_z)_{\bf k}+c.c.,
\end{equation}
which describe, respectively, the production of $\bar{B}_x$ and $\bar{B}_y$ from $\bar{B}_z$ due to stretching by the horizontal velocity variation (shear) along vertical $z$-axis, and  
\begin{equation}
{\cal N}_{z}^{(z)}=-\bar{B}_z^{\ast}\left[ik_x(u_xB_z)_{\bf k}+ik_y(u_yB_z)_{\bf k}\right]+c.c.,
\end{equation}
which describes the advection of vertical field by horizontal velocity.

Finally, the nonlinear terms explicitly depending on $u_z$ are:
\begin{equation}
{\cal N}_{x}^{(va)}=-\bar{B}_x^{\ast}\frac{\partial}{\partial z}(u_zB_x)_{\bf k} +c.c.,
\end{equation}
\begin{equation}
{\cal N}_{y}^{(va)}=-\bar{B}_y^{\ast}\frac{\partial}{\partial z}(u_zB_y)_{\bf k} +c.c.,
\end{equation}
which describe the vertical advection of the horizontal field components $\bar{B}_x$ and $\bar{B}_y$, respectively, by $u_z$, hence characterizing the effect of buoyancy on the magnetic field, and  
\begin{equation}
{\cal N}_z^{(h)}=\bar{B}_z^{\ast}\left[ik_x(u_zB_x)_{\bf k}+ik_y(u_zB_y)_{\bf k}\right]+c.c.,
\end{equation}
which describes production of the vertical field $\bar{B}_z$ from the horizontal field $(\bar{B}_x, \bar{B}_y)$ due to stretching of the latter by horizontal shear of $u_z$. The spectral equations (\ref{eq:Bxk2})-(\ref{eq:Bzk2}) are similar to those in the unstratified case in \cite{mamatsashvili2020zero} and \cite{heldmamatsashvili2022}, except that here we have isolated the nonlinear terms dependent on $u_z$ and $B_z$ and hence affected by stratification.

We also define spectral kinetic, $E_K$, and magnetic, $E_M$, energy densities \citep[e.g.,][]{simon2009},
\[
E_K=\frac{1}{2}\sum_{i=x,y,z}\frac{1}{2}|\overline{(\sqrt{\rho} u_i})|^2, ~~~~~E_M=\frac{1}{2}\sum_{i=x,y,z}|\bar{B}_i|^2.
\]

\begin{figure*}
\includegraphics[scale=0.32]{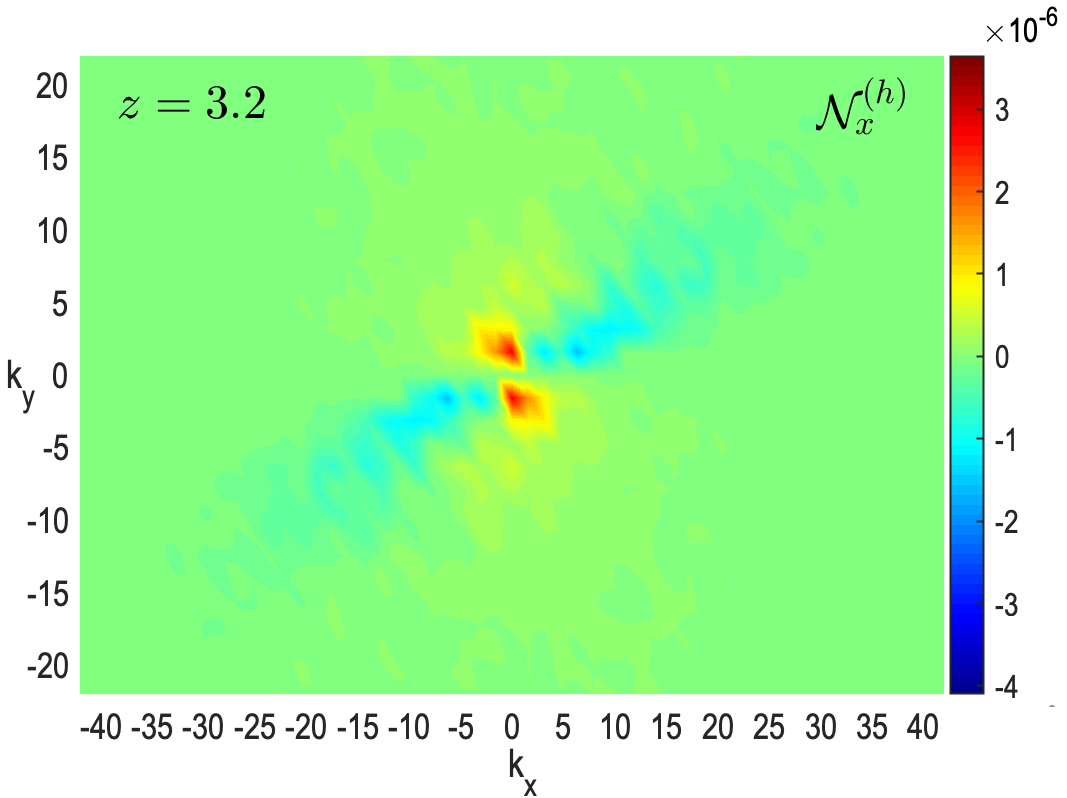}
\includegraphics[scale=0.32]{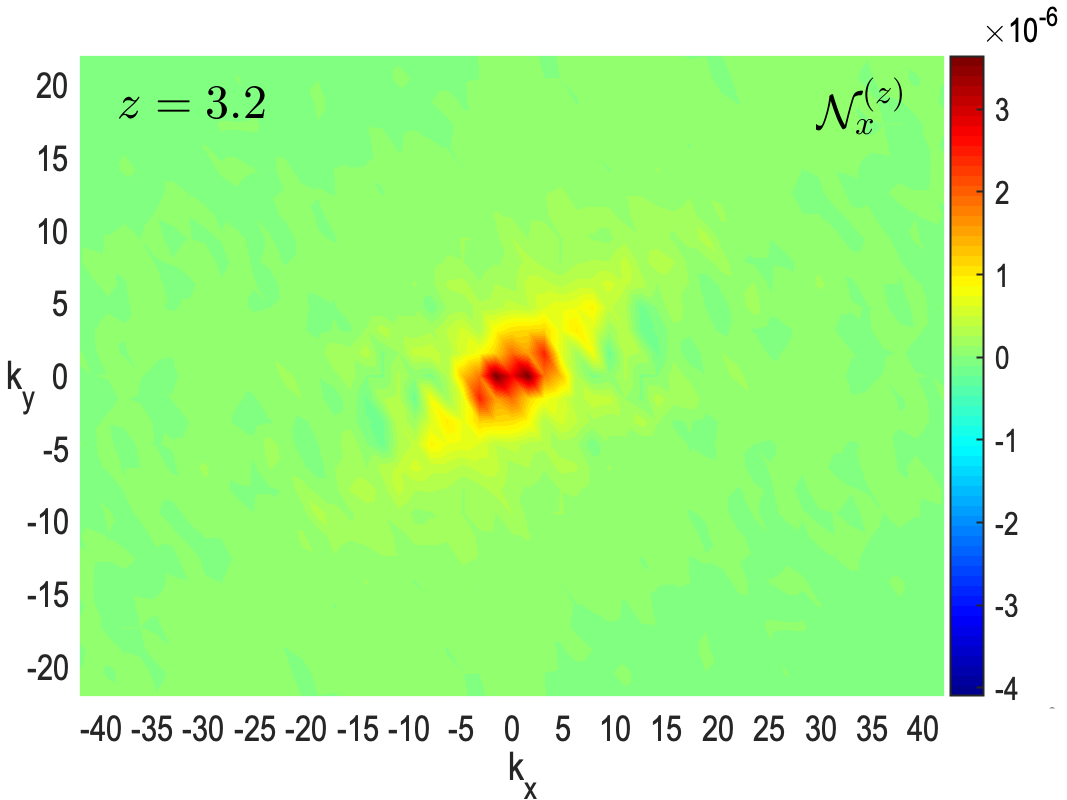}
\includegraphics[scale=0.32]{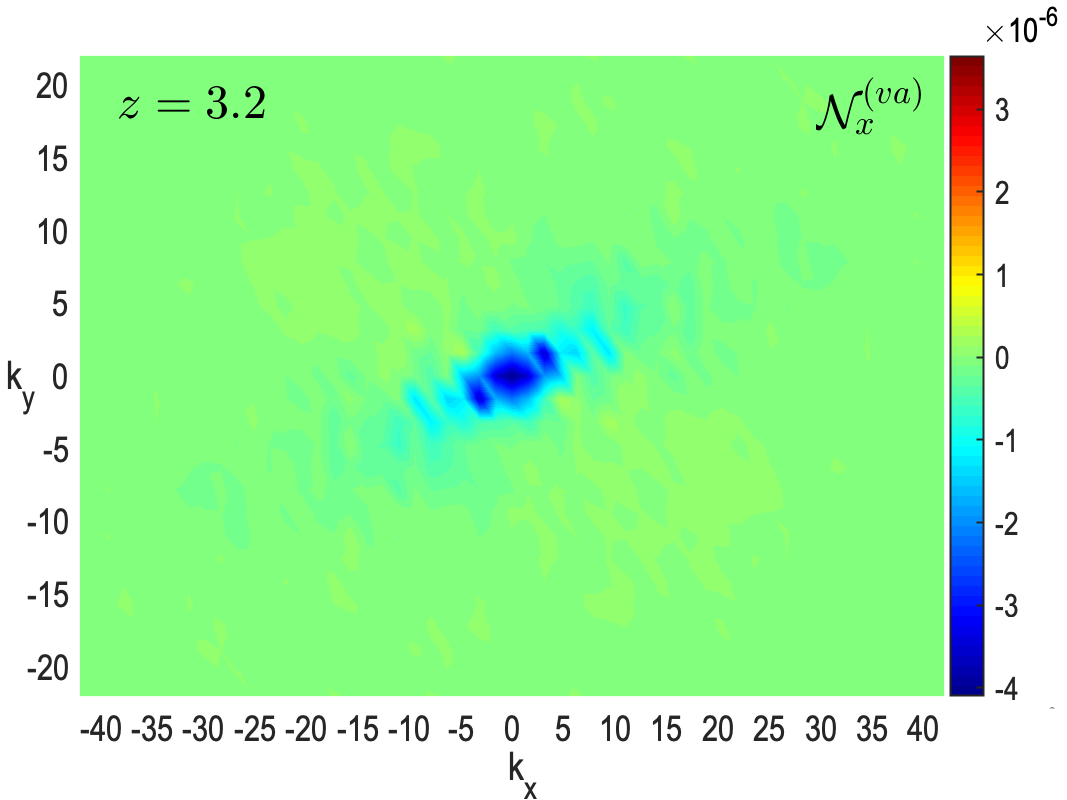}
\includegraphics[scale=0.32]{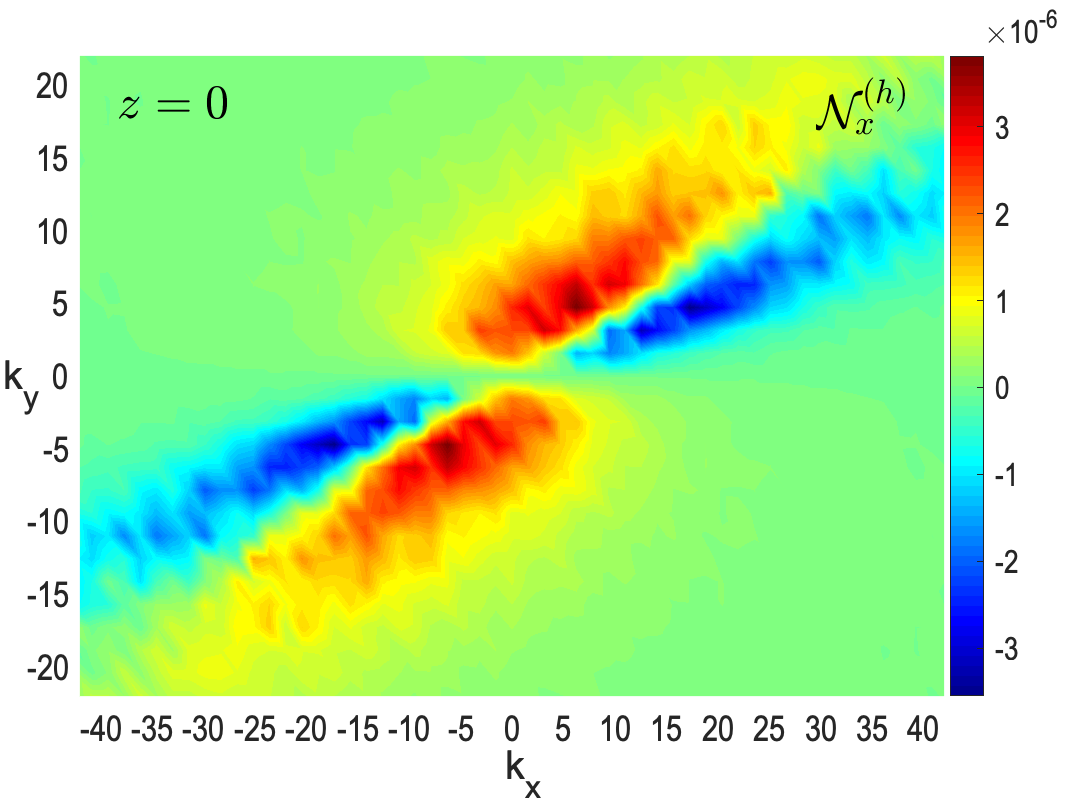}
\includegraphics[scale=0.32]{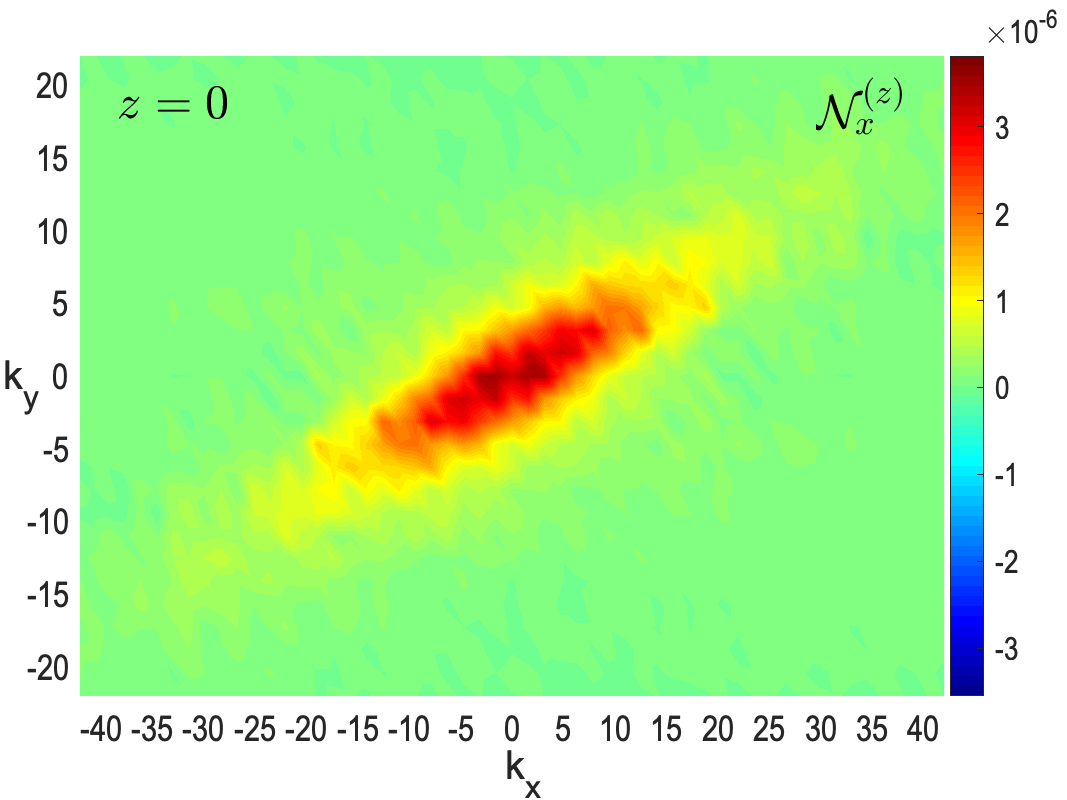}
\includegraphics[scale=0.32]{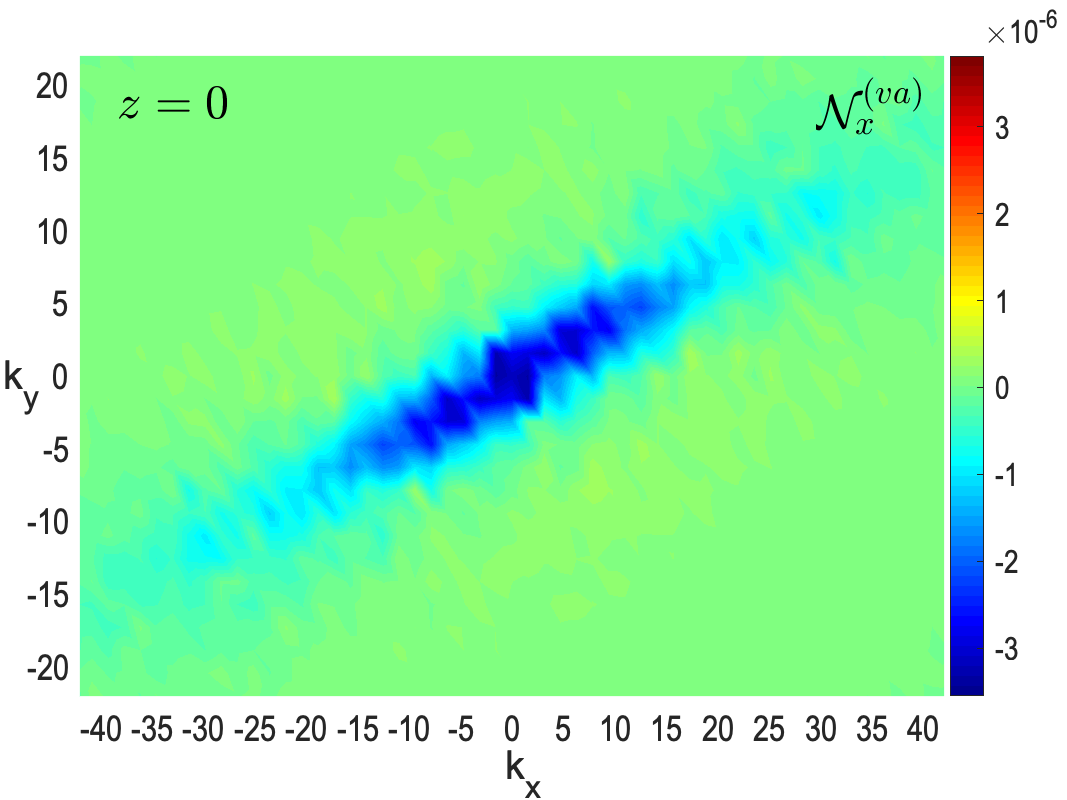}
\caption{Spectra of the nonlinear terms governing the \textit{radial} magnetic field $B_x$: induction-advection terms $N_x^{(h)}$ (left column) and $N_x^{(z)}$ (middle column) together with the vertical transport term $N_{x}^{(va)}$  due to buoyancy (right column) in the $(k_x,k_y)$-plane. The bottom row shows these terms at $z=0$ (the mid-plane), while the top shows the terms at $z=3.2$ (the disc atmosphere). The spectra were taken from a simulation run at ${\rm Pm}=4$ and ${\rm Rm}=18750$ with a resolution of $128/H$. At the mid-plane (bottom row), the dominant process is the production of the radial field from the toroidal field due to the non-linear transverse cascade, which is described by the positive values of $N_x^{(h)}>0$ (yellow/red areas in the left-hand column). The radial field is also produced from the vertical field, which
is described by the positive values of $N_x^{(z)}>0$ (yellow/red areas in the middle column). By contrast, the vertical advection term is negative, $N_x^{(va)}<0$, at small wavenumbers (blue areas in the right-hand column), implying removal of large-scale field by vertical buoyancy. In the atmosphere (top row), on the other hand, the efficiency of the transverse cascade is relatively small and the radial field is mainly produced from the vertical one.}
\label{FIGURE_2DSpectra_Bx}
\end{figure*}

\begin{figure*}
\centering
\includegraphics[scale=0.37]{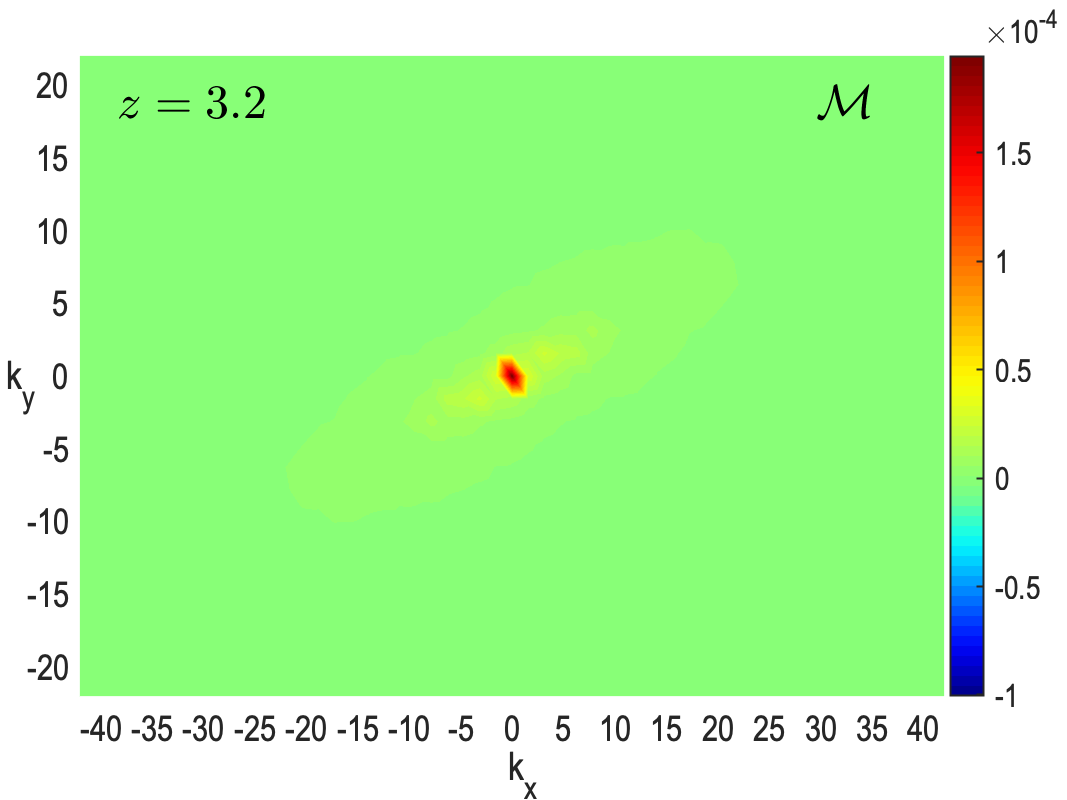}\hspace{0.6cm}
\includegraphics[scale=0.37]{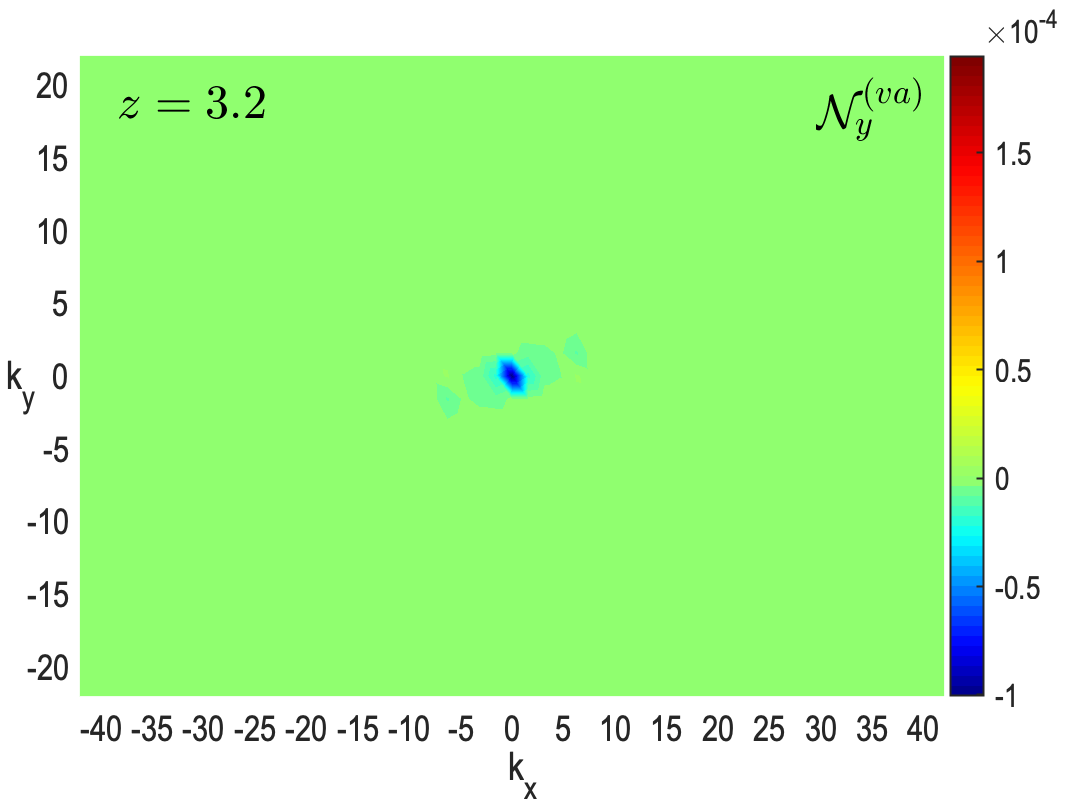}
\includegraphics[scale=0.37]{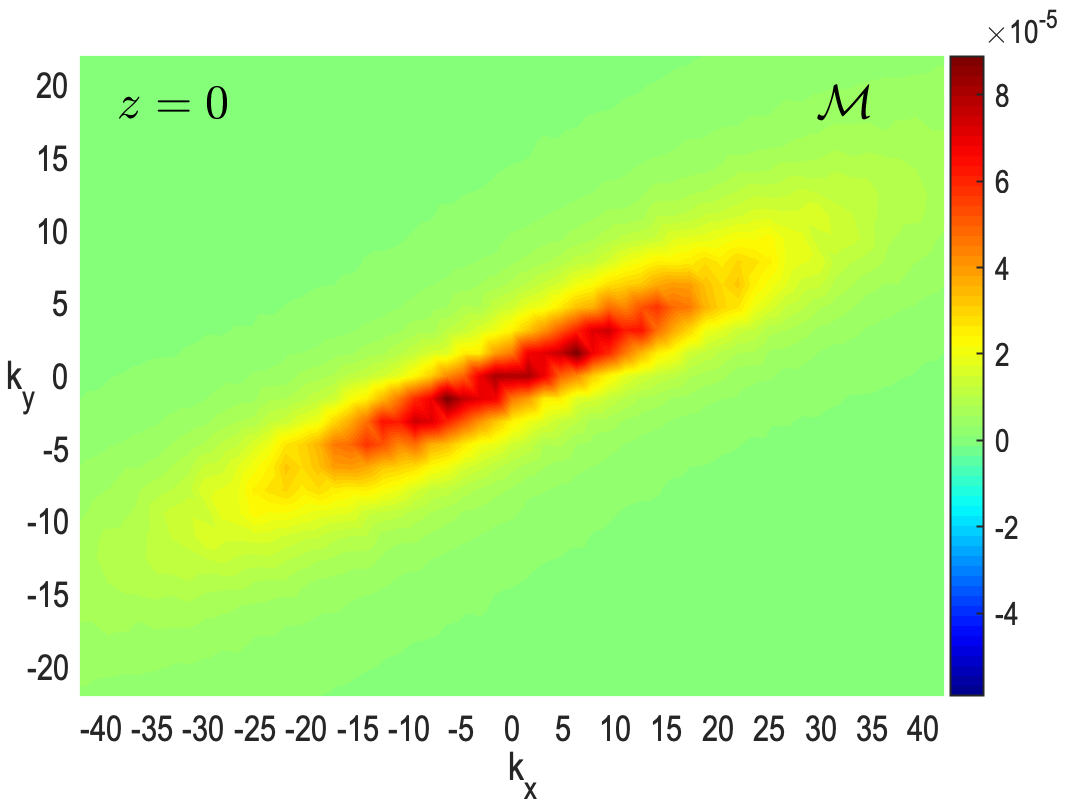} 
\hspace{0.6cm}
\includegraphics[scale=0.37]{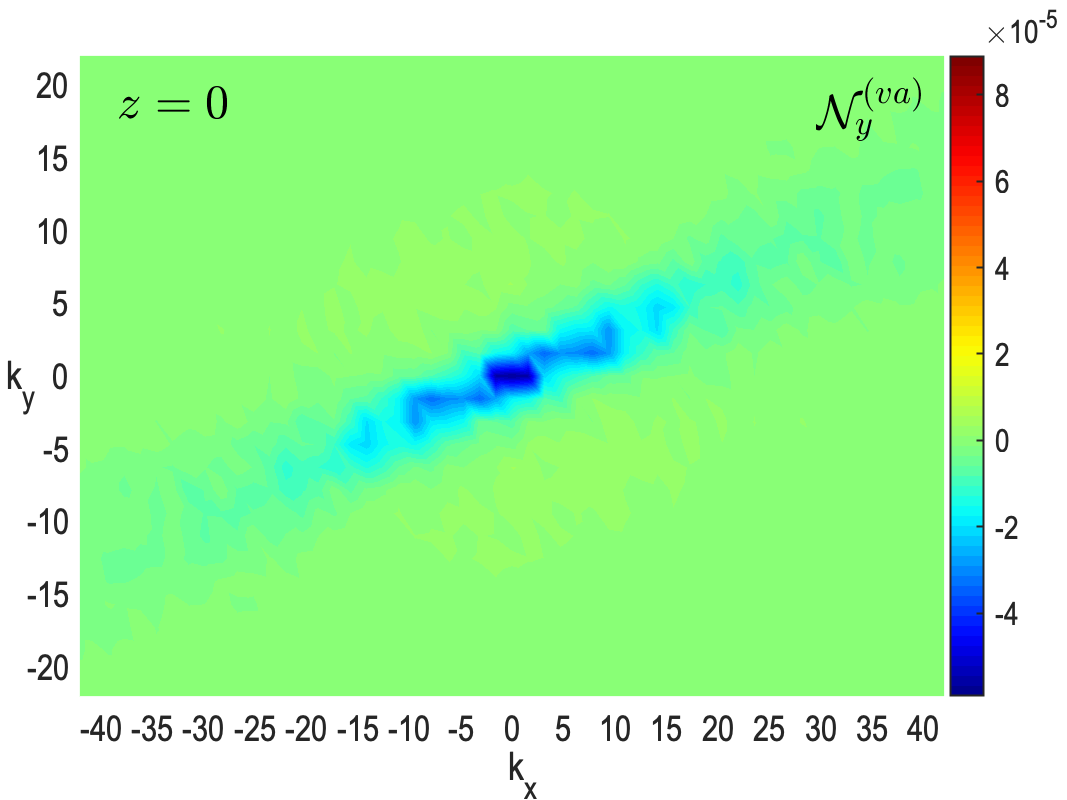} 
\caption{Spectra of the dynamical terms governing the \textit{toroidal} magnetic field $B_y$: Maxwell stress ${\cal M}$ (left column) and the vertical transport term $N_{y}^{(va)}$ (right column), which are the dominant terms for this field component, in the $(k_x,k_y)$-plane at $z=0$ (bottom) and $z=3.2$ (top) for the same run as in Figure \ref{FIGURE_2DSpectra_Bx}. The positive ${\cal M}$  (yellow/red) at small wavenumbers ensures amplification of toroidal field energy $|\bar{B_y}|^2/2$. This process at $z=0$ is a part of the non-modal MRI-growth and extends over a broader range of wavenumbers than at $z=3.2$, where it is localized about ${\bf k}=0$ and related to Parker instability. The term $N_{y}^{(va)}$ is negative at small wavenumbers near the mid-plane and in the atmosphere, thus acting as a sink for the toroidal field, which indicates removal of this field component due to buoyancy.}
\label{FIGURE_2DSpectra_By}
\end{figure*}

\begin{figure*}
\centering
\includegraphics[scale=0.37]{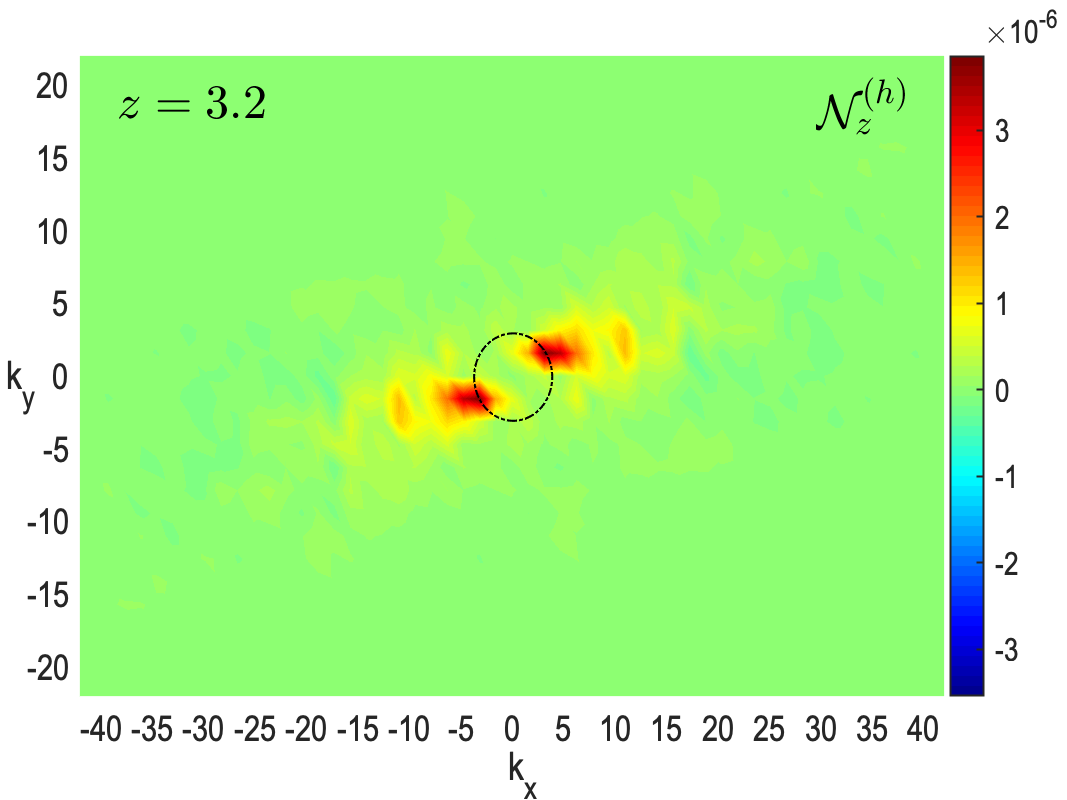}
\hspace{0.6cm}
\includegraphics[scale=0.37]{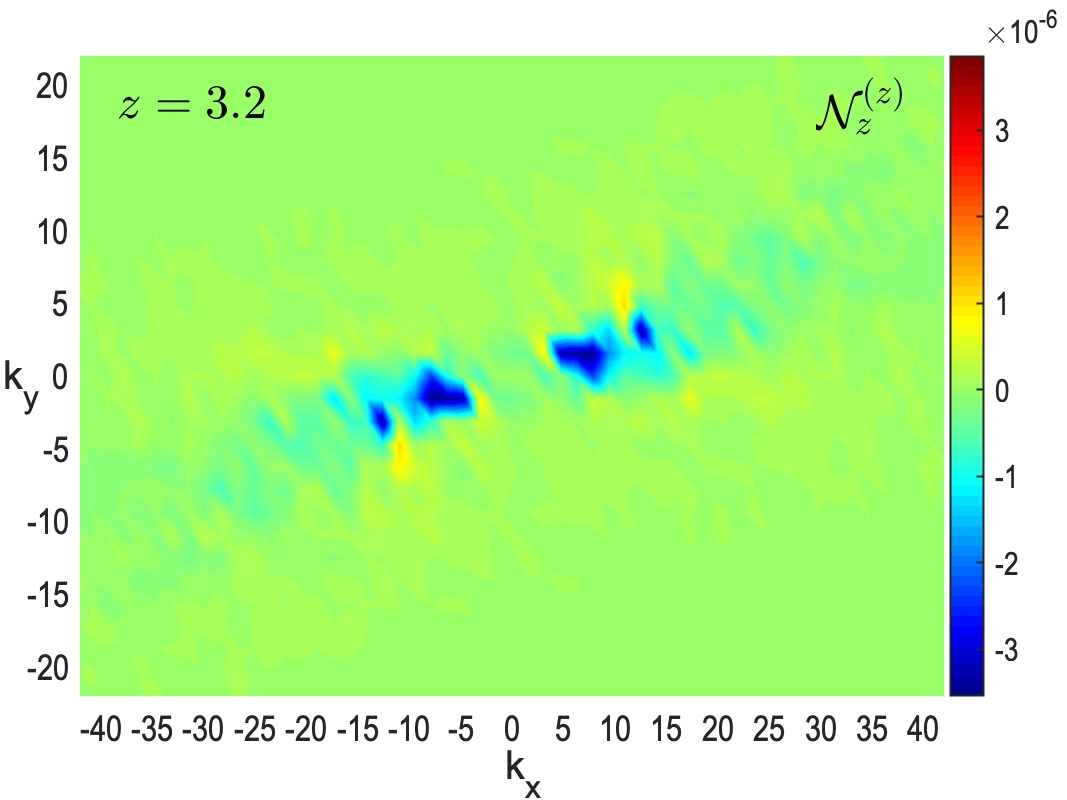}
\includegraphics[scale=0.37]{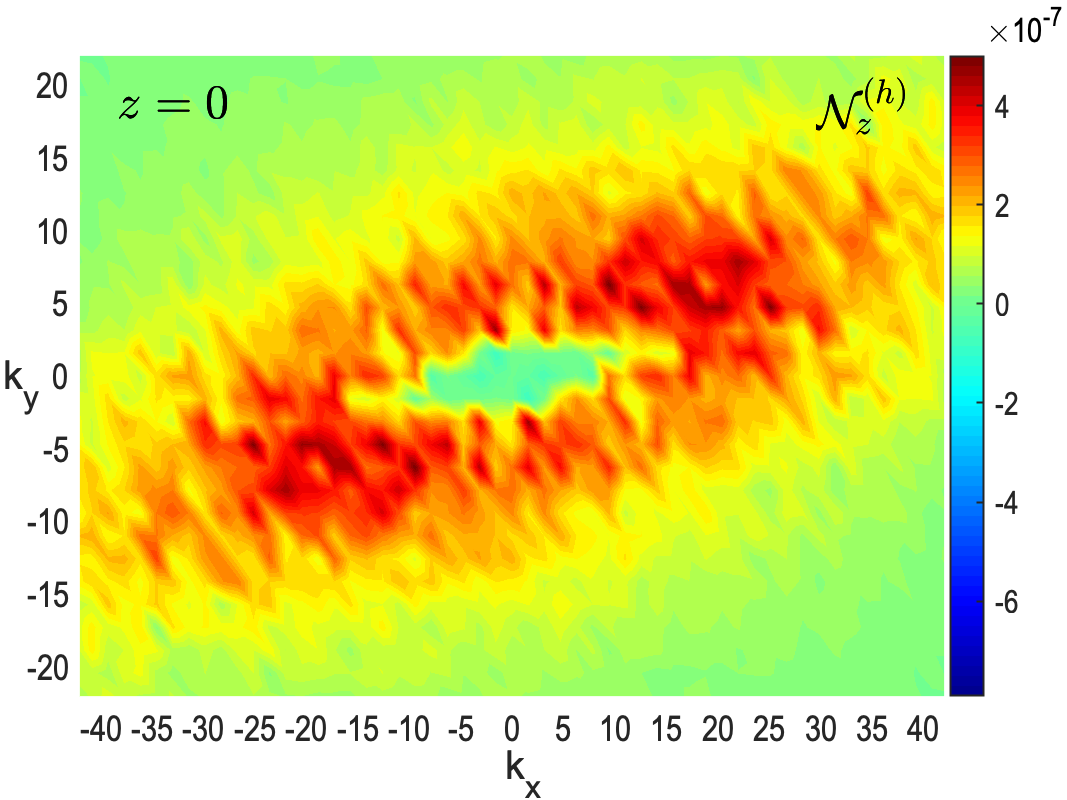}
\hspace{0.6cm}
\includegraphics[scale=0.37]{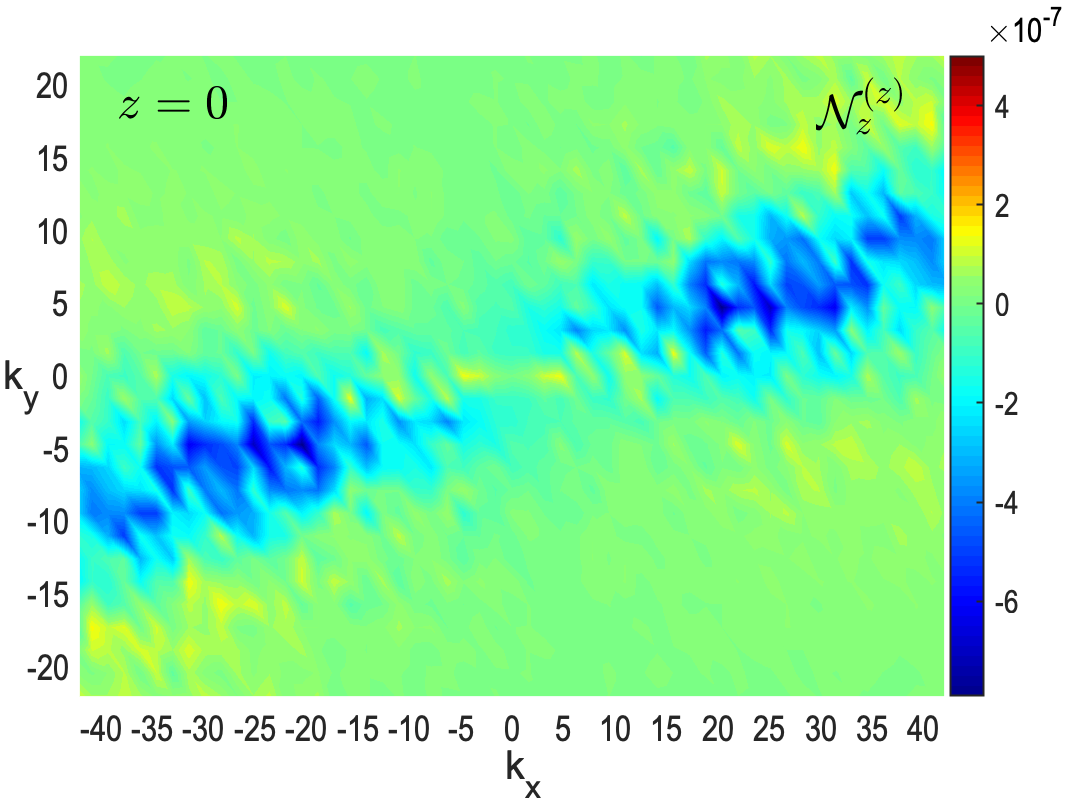}
\caption{Spectra of the dynamical terms for the \textit{vertical} magnetic field $B_z$: induction term $N_z^{(h)}$ (left-hand column) and the \textit{horizontal} advection term $N_z^{(z)}$ (right-hand column) in the $(k_x,k_y)$-plane at $z=0$ (bottom row) and $z=3.2$ (top row) for the same simulation as in Figure \ref{FIGURE_2DSpectra_Bx}. At small wavenumbers, the vertical field is produced from the horizontal (i.e. toroidal) field due to Parker instability at those wavenumbers where $N_z^{(h)}>0$ (yellow/red). These wavenumbers are comparable to the characteristic wavenumber of the Parker instability, $1/l_a\sim 3$ (see the text), as shown by a circle with this radius for reference (top-left hand panel). By contrast, the advection term $N_z^{(z)}<0$ acts as a sink (blue regions).}\label{FIGURE_2DSpectra_Bz}
\end{figure*}

\subsection{Fiducial simulation at fixed $\text{Pm} = 4$}
\label{RESULTS_SpectralFiducialSimulation}
Following our previous studies on the spectral dynamics of MRI-turbulence \citep{gogichaishvili2017,mamatsashvili2020zero, heldmamatsashvili2022}, we first focus on the fiducial simulation ${\rm Pm4Rm18750Re4687}$ with ${\rm Pm}=4$ and perform a detailed analysis of the energy spectra and the dynamical terms in the $(k_x,k_y)$-plane both near the disc mid-plane at $z=0$ and in the atmosphere at $z=3.2$ in order to understand the sustaining dynamics of turbulence/dynamo at different heights. Then, we explore how these spectra and the dynamical balances change with $\rm Pm$. Below, we average all the spectral quantities in time from 100 to 200 orbits when the turbulence is already well in a quasi-steady state. We also average over a small height interval $\Delta z=0.4$, around both $z=0$ and $z=3.2$, the two locations at which we perform a spectral analysis.

Using the spectral analysis we can study, apart from other modes, the dynamics of the special $k_x=k_y=0$ mode, referred to as the \textit{dynamo mode}, because in physical space it in fact represents a solution that is uniform in $x$ and $y$ and depends only on $z$. The dynamo mode gives rise to the large-scale horizontally-averaged magnetic field, which exhibits the well-known ``butterfly''-like variation in the $(t,z)$-plane in vertically stratified discs \citep[see Figure \ref{FIGURE_SpacetimeDiagrams} and, e.g.][]{davisstonepesssah2010, shi2009numerically, gressel2010, simon2011, guangammie2011, salvesen2016a}. This \text{dynamo field} has usually been studied in physical space using a conventional mean-field $\alpha-\Omega$ approach \citep[e.g.,][]{brandenburg1995dynamo,johansen2008high,gressel2010, gressel2015,shi2016,brandenburg2018,gressel2022}, \footnote{Note that \cite{gressel2015} studied the MRI-dynamo within the mean-field $\alpha-\Omega$ framework, but also characterized its dynamics in 1D Fourier space as a function of the vertical wavenumber $k_z$, though in the net vertical flux case.}  In this case, the details of non-linear mode interactions responsible for the sustenance and dynamics of the dynamo mode and the associated large-scale magnetic field are packed into $\alpha$ and $\eta$ tensorial parameters. On the other hand, the spectral approach that we use in this section, i.e. analysing in detail the nonlinear energy transfers in Fourier space, provides deeper insights into the self-sustenance mechanism of the dynamo in stratified discs.

\subsubsection{Energy spectra}
\label{RESULTS_SpectralFiducialEnergySpectral}
\begin{figure}
\centering
\includegraphics[scale=0.42]{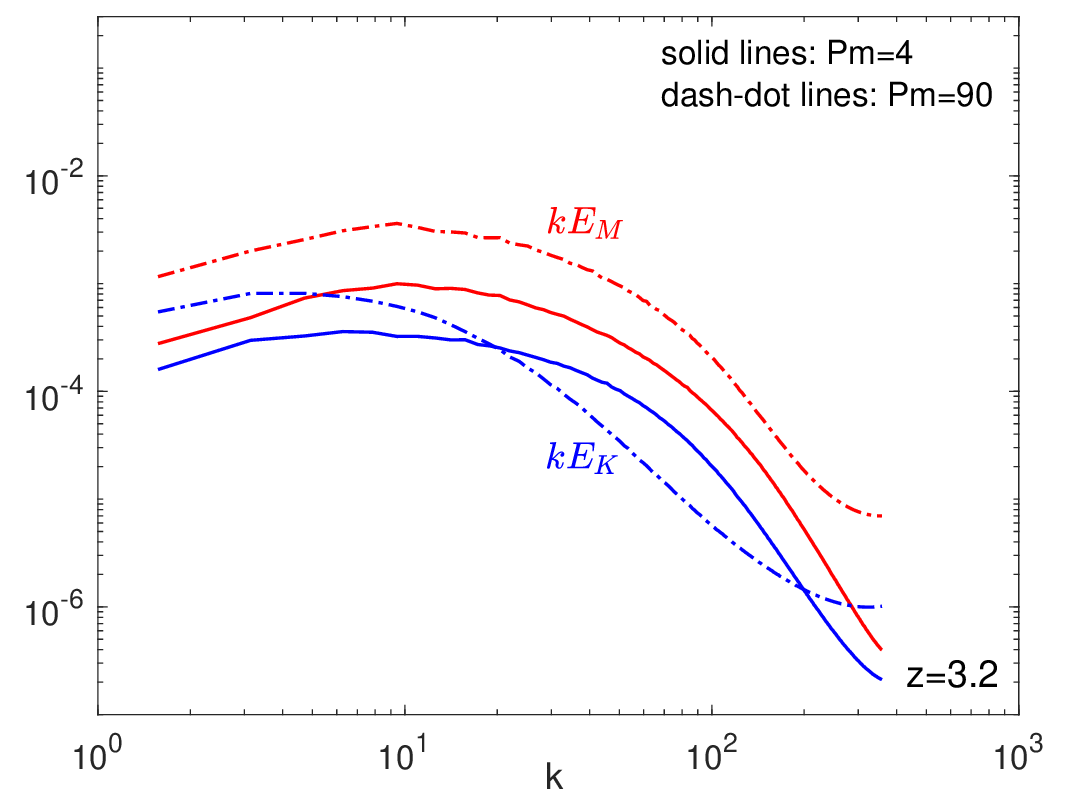}
\includegraphics[scale=0.42]{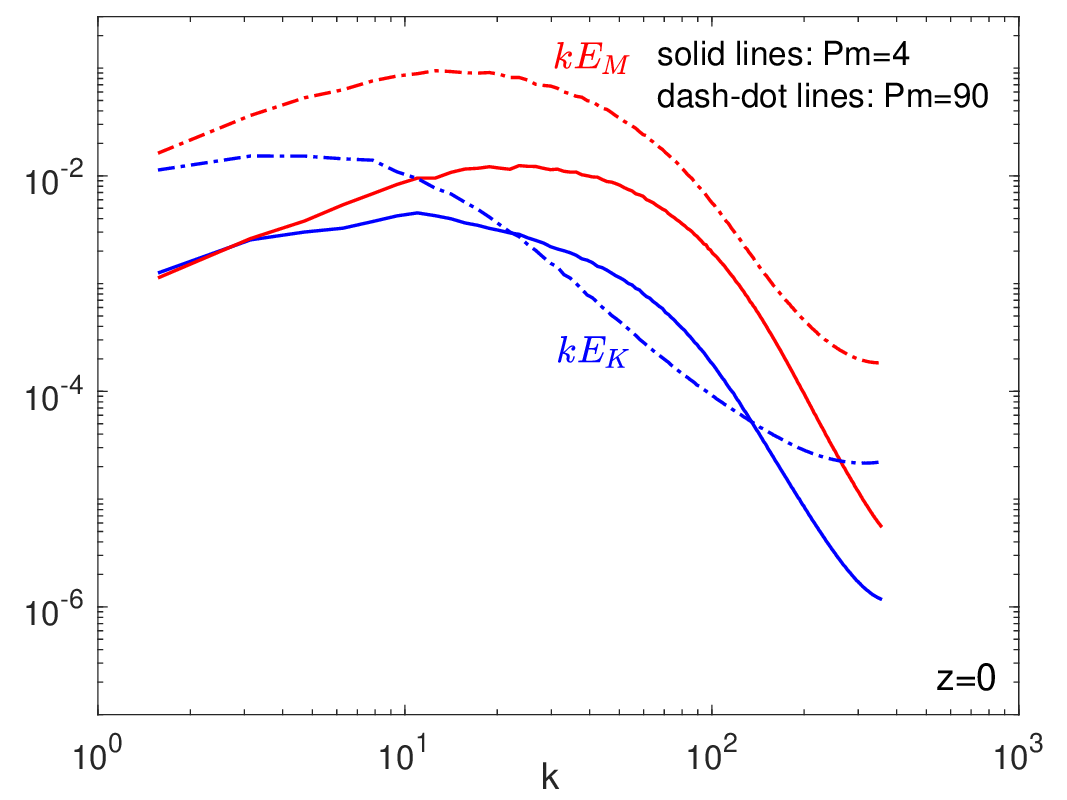}
\caption{Comparison of the ring-averaged compensated kinetic $kE_K$ (blue) and magnetic $kE_M$ (red) energy spectra at ${\rm Pm} = 4$ (solid) and ${\rm Pm}=90$ (dash-dot) at fixed ${\rm Rm}=18750$, both at the mid-plane $z = 0$ (bottom) and in the atmosphere $z=3.2$ (top). As ${\rm Pm}$ increases, the kinetic energy spectra increases mostly at small $k$ and becomes steeper due to increased viscosity. The spectral magnetic energy, on the other hand, increases at all $k$ with ${\rm Pm}$: at the mid-plane it shifts to lower $k$, while in the atmosphere it neither changes shape nor shifts, but increases only in magnitude by about the same factor at all $k$.}
\label{FIGURE_shell_averagedenergies}
\end{figure}

The time-averaged magnetic energy spectrum $E_M$ in the $(k_x,k_y)$-plane is shown in Figure \ref{FIGURE_2Denergyspectra} at $z=0$ and $3.2$ (the kinetic energy spectrum $E_K$ has a similar structure and is not shown here). It has a typical anisotropic structure in the $(k_x,k_y)$-plane due to the background shear, with nearly the same inclination towards the $k_x$-axis at both heights. Such an anisotropic spectrum is typical of MRI-turbulence and has also been observed in spectral analyses of unstratified MRI-turbulence \citep{Lesur_Longaretti2011, Murphy_Pessah2015, gogichaishvili2017, heldmamatsashvili2022}. With increasing $z$, the spectrum becomes more concentrated at smaller wavenumbers, i.e., larger-scale modes become dominant over smaller-scale ones (see also the 1D spectra in Figure \ref{FIGURE_shell_averagedenergies}). 

This behavior of $E_M$ with height is related to different characteristic (correlation) length-scales of the turbulence near the mid-plane and in the atmosphere. Near the mid-plane this length $l_m \sim u_A/\Omega_0$ \citep{Walker2016}, which is much less than the scale-height, since $u_{A} \ll c_s$ at $z=0$ and hence the spectrum extends to higher wavenumbers $\sim 1/l_m \gg 1/H$. On the other hand, in the buoyancy-dominated atmosphere, the dynamics is mainly governed by Parker instability, whose characteristic length-scale is set through the gravitational acceleration $g=\Omega_0^2z$, Alfv\'en $u_A$ and sound $c_s$ speeds, \citep{parker1967,blaes2007surface}
\begin{equation}\label{Parker_length}
l_a\sim \frac{c_s(2c_s^2+u_A^2)}{g(c_s^2+u_A^2)^{1/2}}\sim \frac{c_s^2}{g}\sim H,
\end{equation}
where it has been assumed that in the atmosphere $u_A \sim c_s$. Therefore, the energy spectra at $z=3.2$ extends to wavenumbers $\sim 1/l_a$ much smaller than $1/l_m$ at the midplane, since the ratio $l_a/l_m\sim c_s/u_{A,z=0}\gg 1$. This means that turbulent structures are of much smaller scale at the midplane than in the atmosphere consistent with the spatial distributions of the variables in Figure \ref{FIGURE_FlowFieldComparisonPm4Pm90Rm18750}     \citep[see also][]{blackman2009}.

\subsubsection{Spectra of the dynamical terms}
\label{RESULTS_2DSpectraDynamicalTermsPm4}
We now turn to the analysis of the spectral dynamics of turbulence and consider each magnetic field component separately:

\textit{The dynamics of the radial field $B_x$} is governed by the nonlinear terms ${\cal N}_x^{(h)}$, ${\cal N}_x^{(z)}$, and ${\cal N}_x^{(va)}$, which are shown in Figure \ref{FIGURE_2DSpectra_Bx} in the $(k_x, k_y)$-plane both at the disc mid-plane $z=0$ and in the atmosphere $z=3.2$. The radial component plays a central role in the self-sustaining process of MRI-turbulence since at all heights, the energy-carrying toroidal field can be produced only from the radial field via stretching by Keplerian shear (see below). At the mid-plane, like the energy spectra, all the nonlinear terms are strongly anisotropic in Fourier space due to the Keplerian shear, depending on the polar angle $\phi=arcsin(k_y/(k_x^2+k_y^2)^{1/2})$ of the horizontal wavevector ${\bf k}$. The main consequence of this anisotropy for ${\cal N}_x^{(h)}$ is the redistribution of power over wavevector orientation ($\phi$-angle) in the $(k_x,k_y)$-plane -- \textit{the nonlinear transverse cascade} \citep{mamatsashvili2014,gogichaishvili2017,mamatsashvili2020zero}, which transfers the radial field energy, $|\bar{B}_x|^2/2$, from ``giver'' wavenumbers for which ${\cal N}_x^{(h)} <0$ (blue) to ``receiver'' wavenumbers for which ${\cal N}_x^{(h)}>0$ (yellow and red), as seen in the left-hand column of Figure \ref{FIGURE_2DSpectra_Bx}. We refer to this region in spectral space encompassing small and intermediate wavenumbers (i.e. $|k_x|\lesssim 30,|k_y| \lesssim 20$, at $z=0$) where radial field is produced, as \textit{the vital area} of the turbulence. Since the toroidal field $B_y$ is the dominant field component, it gives the largest contribution to ${\cal N}_x^{(h)}$ (see equation \ref{eq:Nxh}), so the role of the nonlinear transverse cascade is in fact to continually replenish and amplify the radial field from the toroidal one at the ``receiver'' wavenumbers where ${\cal N}_x^{(h)}>0$. In the atmosphere, the vital area in the $(k_x,k_y)$-plane shrinks to smaller wavenumbers and the transverse cascade weakens. 

The radial field is also generated from the vertical field via ${\cal N}_x^{(z)}$ at those wavenumbers where it is positive, ${\cal N}_x^{(z)}>0$ (yellow/red areas in the middle column of Figure \ref{FIGURE_2DSpectra_Bx}). At the mid-plane, this process is less intensive but still comparable to the transverse cascade, whereas it dominates the latter in the atmosphere. However, the situation is different for the ${\bf k}=0$ dynamo mode, where ${\cal N}_x^{(h)}\approx 0$ and hence $\bar{B}_x$ is produced solely from $\bar{B}_z$ by positive ${\cal N}_x^{(z)}$. On the other hand, the vertical transport term is negative ${\cal N}_x^{(va)}<0$ in the vital area both at $z=0$ and $z=3.2$ (blue areas in the right-hand column), removing the large-scale radial field due to buoyancy at all $z$.

\textit{The dynamics of the toroidal field $B_y$} is governed by the Maxwell stress ${\cal M}$ and the non-linear terms ${\cal N}_y^{(h)}$, ${\cal N}_y^{(z)}$, ${\cal N}_y^{(va)}$. From these four terms the dominant ones are ${\cal M}$ and the vertical transport term ${\cal N}_y^{(va)}$ which are shown in Figure \ref{FIGURE_2DSpectra_By}. They exhibit a similar anisotropy due to the shear as do their counterparts for the radial field. In the vital area, only the Maxwell stress is positive, ${\cal M}>0$ (left-hand column) and hence amplifies (injects) toroidal field energy $|\bar{B}_y|^2/2$ by stretching the radial field due to shear. At the mid-plane, the amplification of $\bar{B}_y$ is due to the non-modal MRI-growth process and hence spans a much broader range of wavenumbers than in the atmosphere, where it is related to Parker instability and instead is concentrated at the smallest wavenumbers peaking around ${\bf k}=0$. In the vital area, ${\cal N}_y^{(va)}<0$ (right-hand column), draining the large-scale toroidal field energy due to buoyancy (see also \cite{blackman2004} and \cite{blackman2009}).

\textit{The dynamics of the vertical field $B_z$} is governed by the non-linear terms ${\cal N}_z^{(h)}$ and ${\cal N}_z^{(z)}$, which are shown in Figure \ref{FIGURE_2DSpectra_Bz}, having a similar anisotropic structure in the $(k_x,k_y)$-plane due to the shear. In the vital area, ${\cal N}_z^{(h)}$ is positive at intermediate wavenumbers (left-hand column), and generates the vertical field from the horizontal (mainly toroidal) one. This process is, however, due to two different sources: non-modal MRI process near the mid-plane at $z=0$ (where buoyancy is weak) and Parker instability in the atmosphere, where magnetic buoyancy, having  $N_m^2<0$, plays a major role (Figure \ref{FIGURE_ParkerInstabilityDirectEvidence}). Consequently, at $z=3.2$, the term ${\cal N}_z^{(h)}$ operates at wavenumbers comparable to $1/l_a$, which corresponds to the length-scale of the Parker instability $l_a\sim c_s^2/g$ as given by Equation (\ref{Parker_length}) above. We estimated this length to be $l_a\sim 0.3H$ from the simulations and showed for reference as a circle of radius $1/l_a\sim 3$ in the map of ${\cal N}_z^{(h)}$ in Figure \ref{FIGURE_2DSpectra_Bz}, which covers the location of the peaks of this term. At ${\bf k}\approx 0$, ${\cal N}_z^{(h)}=0$, implying that there is no production of the large-scale vertical field. By contrast, the horizontal advection term is negative ${\cal N}_z^{(z)}<0$ in the vital area at all $z$ (right-hand column), transferring the vertical field energy to larger wavenumbers.

Finally, the drift terms due to shear $q$ in equations (\ref{eq:Bxk2})-(\ref{eq:Bzk2}), do not produce (inject) energy into the modes, and serve only to balance the joint action of the dynamical terms in quasi-steady state, so we do not show the drift terms here. They cause the radial wavenumber $k_x$ of non-axisymmetric ($k_y\neq 0$) modes to change in time and cross the vital area. As a result, the growth of these modes acquires a transient, or non-modal character \citep{balbus1992,mamatsashvili2013,squire2014}.\footnote{Note that a transient growth of non-axisymmetric modes also occurs in the opposite case of negative shear $q<0$ due to a non-modal effect \citep{pessah2012}, which although MRI-stable, can be important in star-disc boundary layers.}

\begin{figure*}
\includegraphics[scale=0.32]{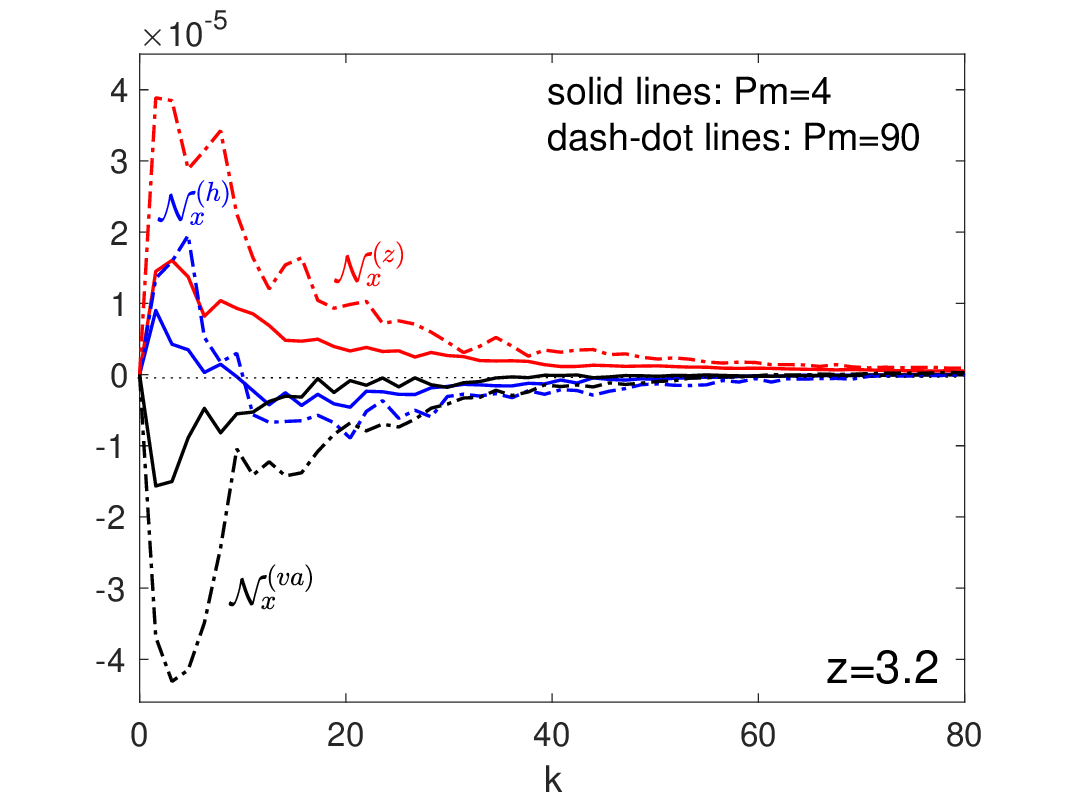}
\includegraphics[scale=0.32]
{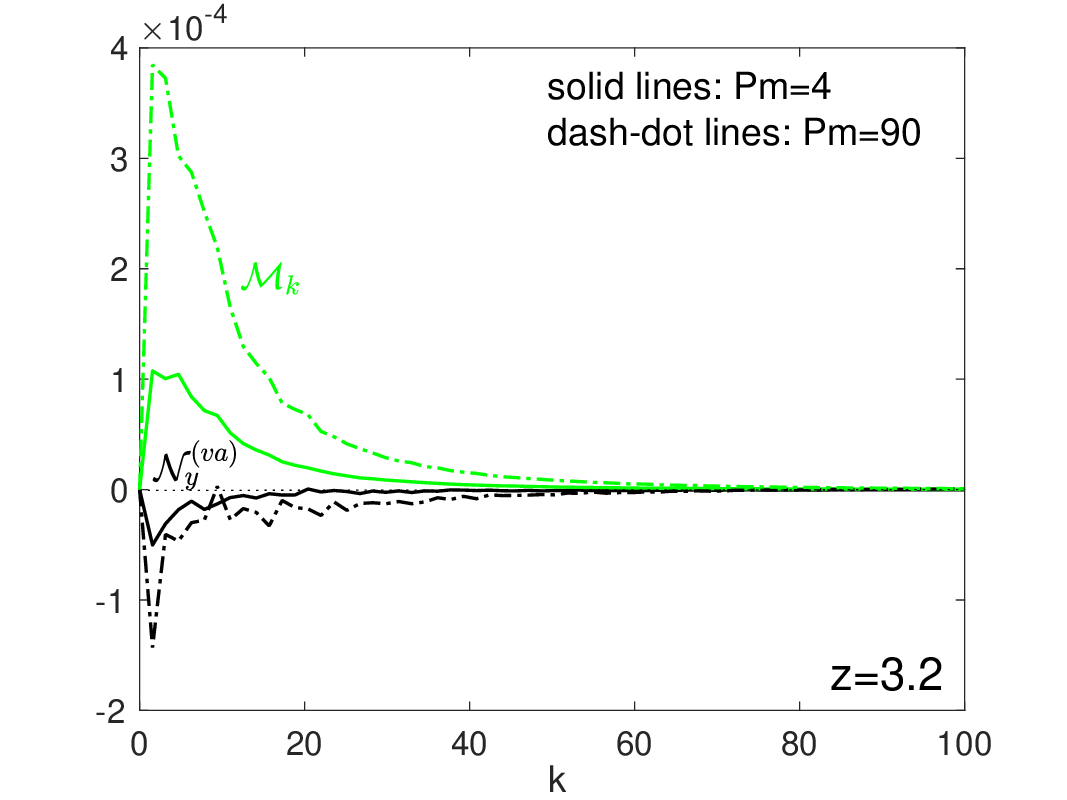}
\includegraphics[scale=0.32]
{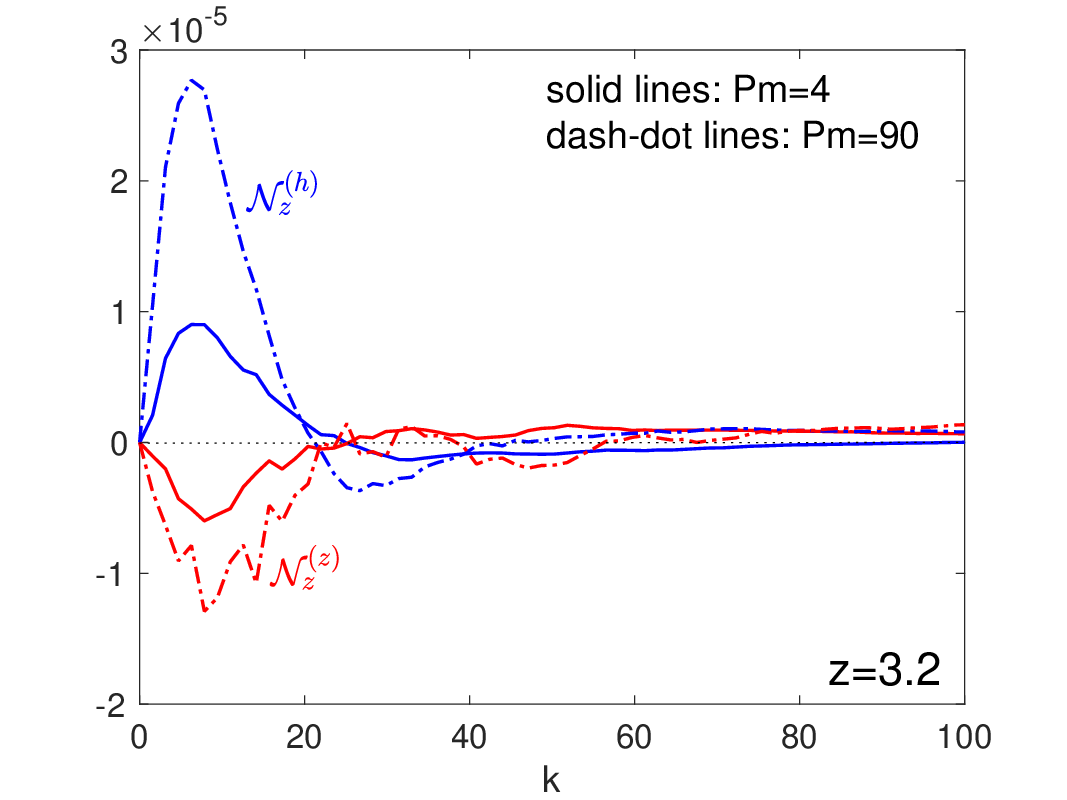}
\includegraphics[scale=0.32]{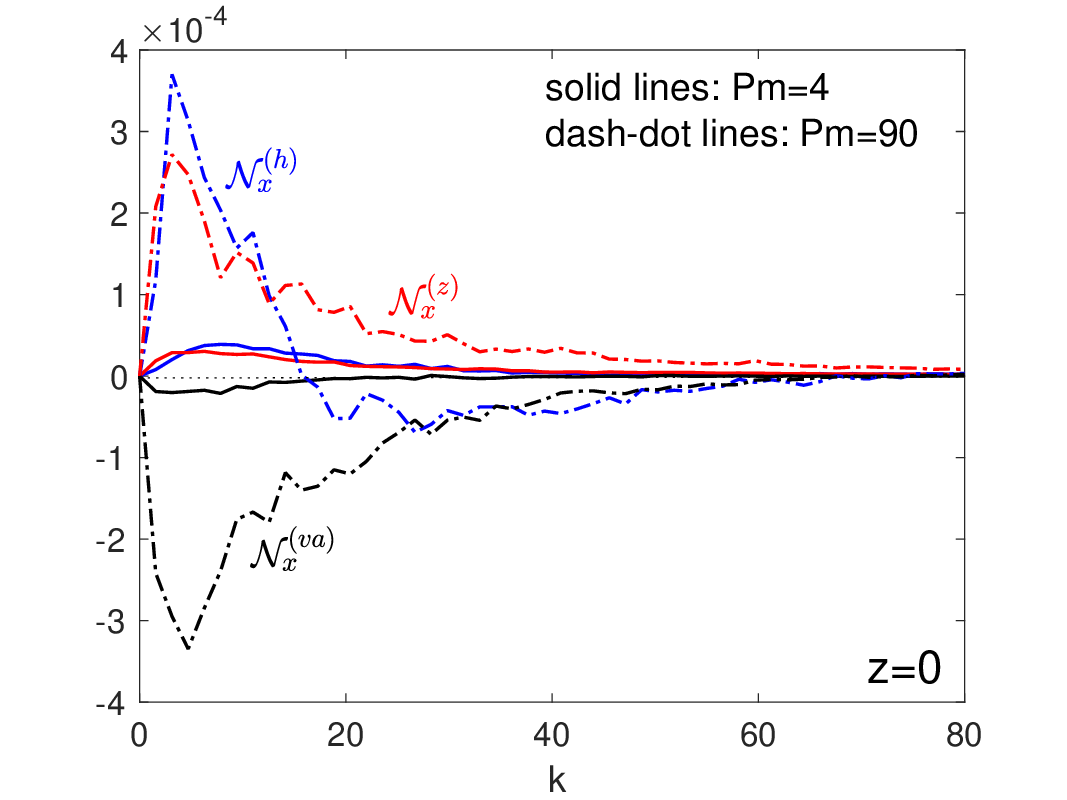}
\includegraphics[scale=0.32]
{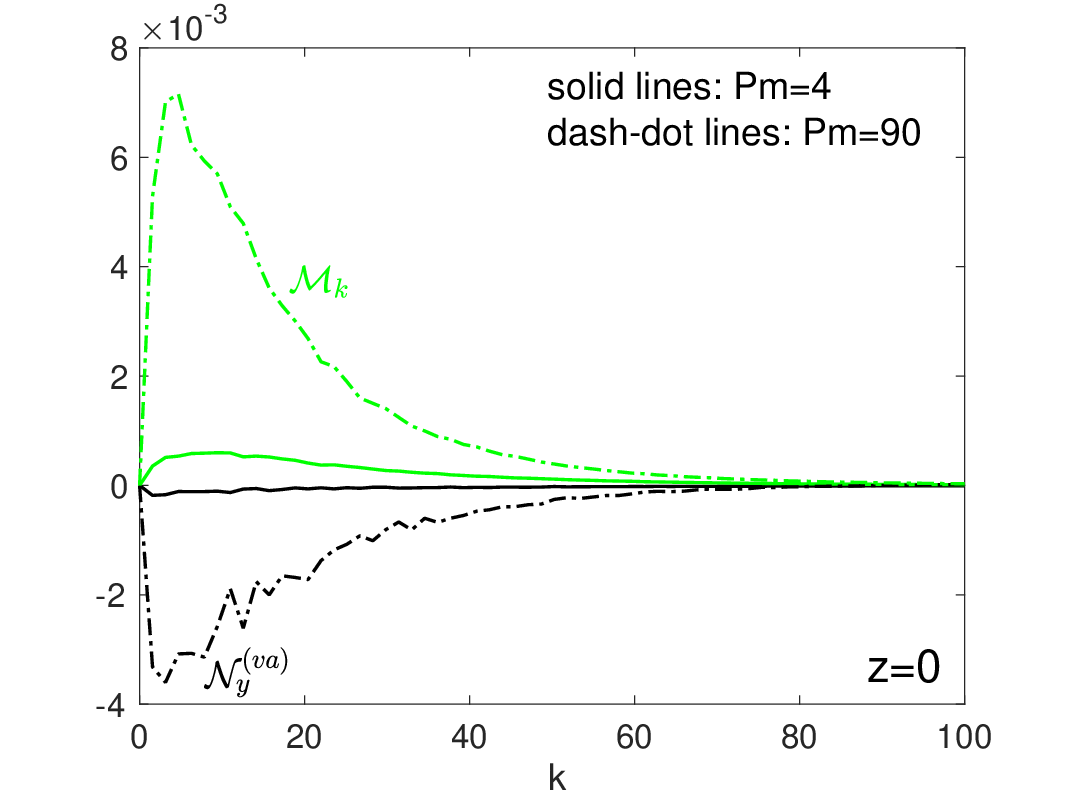}
\includegraphics[scale=0.32]
{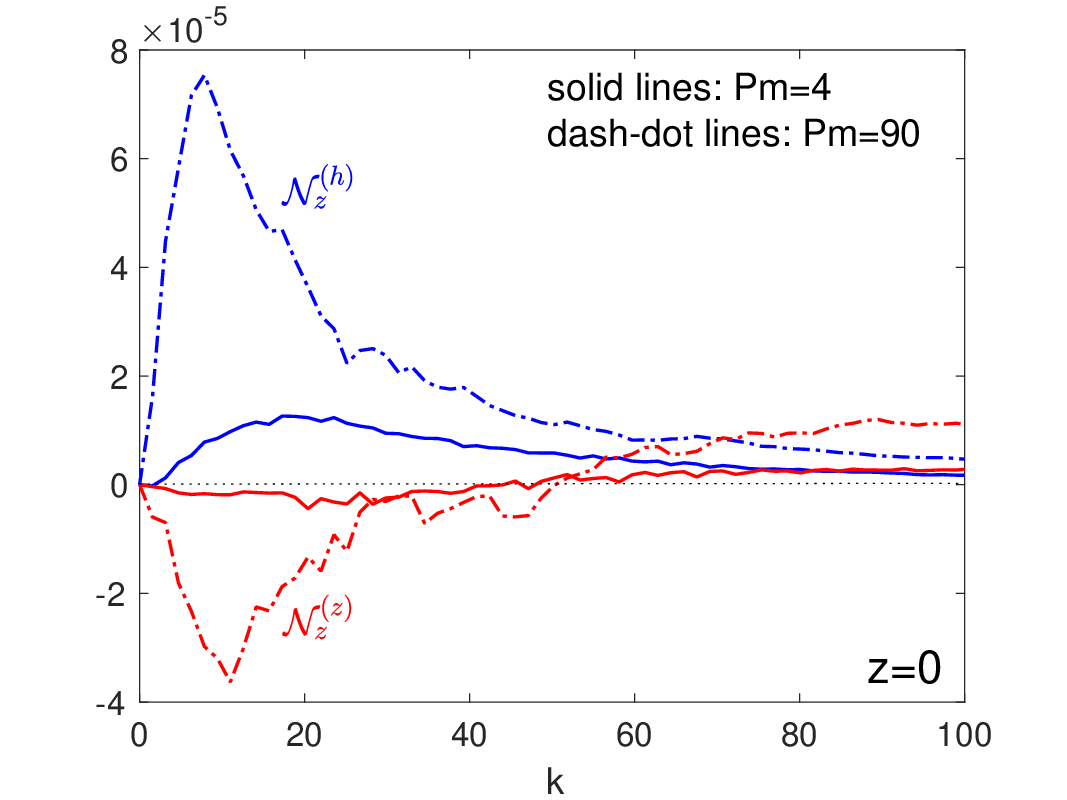}

\caption{Ring-averaged spectra of the dynamical terms ${\cal N}_x^{(h)}$, ${\cal N}_x^{(z)}$, and ${\cal N}_x^{(va)}$ (left-hand column), ${\cal M}$ and ${\cal N}_y^{(va)}$ (middle column) and ${\cal N}_z^{(h)}$ and ${\cal N}_z^{(z)}$ (right-hand column) at $\rm Pm=4$ (solid lines) and 90 (dash-dot lines) at the mid-plane $z=0$ (bottom row), and in the atmosphere $z=3.2$ (top row), for ${\rm Rm}=18750$, at a resolution of $128/H$. With increasing $\rm Pm$, all these terms generally increase in magnitude more significantly at the mid-plane than in the atmosphere. At the mid-plane they shift to smaller $k$, whereas in the atmosphere they retain their shape (peaks) as a function of $k$. The main nonlinear terms ${\cal N}_x^{(h)}$ and ${\cal N}_z^{(h)}$ sustaining the turbulence at $z=0$ and $z=3.2$, respectively, exhibit a similar trend: they become increasingly negative at $20\lesssim k \lesssim 50$ as $\rm Pm$ increases from 4 to 90, indicating a decrease in the direct cascade, i.e. a decrease in the transfer of energy to higher $k$ and hence loss by resistive dissipation, and this in turn results in an overall increase in the turbulence level with $\rm Pm$.}
\label{FIGURE_shellaveragedterms}
\end{figure*}

\subsection{Summary of dynamo self-sustenance mechanisms at different heights}
\label{RESULTS_SummarySustenanceSchemesAtDifferentHeights}
We now summarize the self-sustenance processes of the turbulence and dynamo at different vertical locations in the disc (at the mid-plane $z = 0$, and in the atmosphere $z = 3.2$). Let us briefly remind the reader of our key diagnostics: we have Fourier transformed the induction equation, and so obtained various terms governing the evolution of each spectral magnetic energy component (see Equations 
\ref{eq:Bxk2}-\ref{eq:Bzk2}). By plotting the 2D spectra of these dynamical terms at the disc mid-plane and in atmosphere we can construct a picture of the key processes sustaining each field component (and thus the dynamo and turbulence) in different parts of the disc, which we describe below. 

\subsubsection{Dynamical processes in the bulk of the disc.} 
The dynamical balances sustaining the turbulence near the mid-plane ($z = 0$) primarily involve the radial and toroidal field components, since these two carry the most energy (see bottom panel in Figure \ref{FIGURE_Pm4Verticaldiscstructure}). In this case, a kind of anisotropic inverse cascade induced by the shear -- more accurately known as a nonlinear \textit{transverse cascade} -- plays a key role, seeding the radial field $\bar{B}_x$ from the toroidal field $\bar{B}_y$ at small to intermediate wavenumbers (i.e. $|k_x|\lesssim 30,|k_y| \lesssim 20$, at $z=0$). This process is characterized by the nonlinear term ${\cal N}_x^{(h)}$ in the radial field equation (\ref{eq:Bxk2}), which is anisotropic in the $(k_x,k_y)$-plane due to shear, and which exhibits regions of ``giver'' wavenumbers where ${\cal N}_x^{(h)}<0$ (blue regions in left-hand column of Figure \ref{FIGURE_2DSpectra_Bx}) and ``receiver'' wavenumbers where ${\cal N}_x^{(h)}>0$ (yellow and red regions). Physically, this results in a transfer of radial field energy ($|\bar{B}_x|^2/2$) over angles (i.e., transversely) across the $(k_x,k_y)$-plane, from a select set of larger (``giver'') wavenumbers to another set of smaller (``receiver'') ones. Thus, the nonlinear transverse cascade replenishes the radial field at those wavenumbers where ${\cal N}_x^{(h)}>0$.   

At small wavenumbers, the radial field is also amplified through induction from the vertical field, since the nonlinear term responsible for this process is positive, ${\cal N}_x^{(z)}>0$ (see middle column of Figure \ref{FIGURE_2DSpectra_Bx}). The ``receiver'' modes then drift along the $k_x$-axis through the injection area, that is, the region in Fourier space where the Maxwell stress is positive (${\cal M}>0$) and hence injects magnetic energy into perturbations (see left-hand column of Figure \ref{FIGURE_2DSpectra_By}). In this case, the energy injection is mainly due to MRI, resulting in the amplification of the toroidal field energy, $|\bar{B}_y|^2/2$, during the time that the modes drift across the injection area.\footnote{The growth of perturbations due to MRI during some finite time interval is also known as the \textit{non-modal} MRI \citep{squire2014, gogichaishvili2017}.} This amplified toroidal field,  which provides the dominant contribution to ${\cal N}_x^{(h)}$ (see Equation \ref{eq:Nxh}), in turn produces the radial field via the transverse cascade process, as explained above, thereby closing the main self-sustenance cycle. (All the other non-linear terms in the toroidal field energy equation (Equation \ref{eq:Byk2}) are negative at small and intermediate wavenumbers, where most of the energy resides, as seen in Figure \ref{FIGURE_2DSpectra_By}, and thus act as sinks \textit {opposing} the self-sustenance.) This self-sustaining scheme at the mid-plane (where buoyancy is unimportant) is similar to that found in our unstratified studies \cite{mamatsashvili2020zero, heldmamatsashvili2022}. 

\subsubsection{Dynamical processes in the disc atmosphere.} 
In the disc atmosphere ($|z|\gtrsim 2$), the shear-modified Parker instability \citep{foglizzo1995, johansen2008high} dominates over the MRI, as seen in Figure \ref{FIGURE_ParkerInstabilityDirectEvidence} showing the increasingly negative magnetic buoyancy (squared), $N_m^2 <0$, and hence more intensive Parker instability with increasing $z$. Unlike at the mid-plane where the toroidal and radial fields dominates, in the atmosphere it is the  toroidal and \textit{vertical} field components that carry the most energy, and hence play the main role in the self-sustenance of the dynamo. The radial field is smaller, though still important (Figure \ref{FIGURE_Pm4Verticaldiscstructure}). In Figure \ref{FIGURE_2DSpectra_Bx} we show that at $z=3.2$ the dominant process producing the radial field is induction from the \textit{vertical} field described by ${\cal N}_x^{(z)}>0$, while the transverse cascade is subdominant. This radial field is subsequently stretched out by the Keplerian shear and thus produces and amplifies the toroidal field via the positive Maxwell stress ${\cal M}>0$ (Figure \ref{FIGURE_2DSpectra_By}). Finally, the toroidal field generates the vertical field for non-axisymmetric modes due to Parker instability (mediated by the positive nonlinear term ${\cal N}_z^{(h)}>0$ in the vital area, see Figure \ref{FIGURE_2DSpectra_Bz}), thereby closing  the self-sustenance cycle in the atmosphere. Note that despite the significance of the Parker instability in the atmosphere, the role of shear is still important, as it maintains the dominant toroidal field (which then becomes Parker-unstable) from the radial one.

Thus, we have identified two distinct but co-existing self-sustaining processes for the dynamo in vertically stratified discs. The first process operates primarily in the bulk of the disc ($|z| \lesssim 2H$), and is based on the interplay between the non-modal MRI, which amplifies the toroidal field $B_y$ from the radial field $B_x$, and the non-linear transverse cascade, which provides positive feedback, producing radial field from  toroidal field. The second process operates in the disc atmosphere ($|z| \gtrsim 2H$), and is based on the interplay between Parker instability (which produces  vertical field $B_z$ from toroidal field $B_y$), non-linear transfers (which generate radial field $B_x$ from the vertical field), and Keplerian shear (which stretches the radial field into the toroidal field). The self-sustaining process near the mid-plane extends over a much wider range of length-scales than self-sustaining process in the atmosphere: the latter process is localized to large scales (comparable to the size of the system). This picture is consistent with one of the first disc dynamo models proposed by \cite{Tout_Pringle1992} based on the interplay between MRI and Parker instability \citep[see also][]{blackman2004,gressel2010}. However, a key difference between our model and theirs is that here these two instabilities determine the dynamo action at \textit{different} heights: the MRI dynamo dominates near the mid-plane, whereas the shear-modified Parker instability dominates in the atmosphere.

\subsection{Dependence of the results on $\rm Pm$}
\label{RESULTS_SpectralPmDependence}
Having outlined the turbulence sustenance mechanisms at different heights in the disc, let us now analyse how the dynamical terms underlying these mechanisms  vary with \textit{magnetic Prandtl number} ${\rm Pm}$, both at the mid-plane and in the atmosphere. To capture the dynamics in the two regions of interest in the regime of large ${\rm Pm}$ (Figure \ref{FIGURE_EmagAndAlphaPmScaling}), we focus on ${\rm Pm} = 4$ (start of the power-law scaling region) and ${\rm Pm} = 90$ (plateau region). The ${\rm Pm}$-dependence of MRI-turbulence and clarification of its physical nature in \textit{unstratified} disc models were first elucidated in \cite{mamatsashvili2020zero} and \cite{heldmamatsashvili2022}. Here, we generalize these studies to the vertically stratified case. For comparison, we compute 1D ring-averaged spectra of the energies and the dynamical terms, i.e.  $\int_0^{2\pi}(..)kd\phi$, as a function of the horizontal wavenumber magnitude $k=(k_x^2+k_y^2)^{1/2}$, both at the mid-plane ($z = 0$) and in the atmosphere ($z=3.2$).

\subsubsection{Energy spectra}
\label{RESULTS_SpectralPmDependenceEnergySpectra}
Figure \ref{FIGURE_shell_averagedenergies} shows the ring-averaged kinetic and magnetic energy spectra (compensated by $k$) at ${\rm Pm}=4$ and 90. These spectra do not have a clear power-law form because of the strong anisotropy in the $(k_x,k_y)$-plane (Figure \ref{FIGURE_2Denergyspectra}), as also pointed out in \cite{Murphy_Pessah2015}. On a promising note, we do find pretty clear peaks in all the spectra (for the kinetic energy spectrum at ${\rm Pm}=90$ the peaks are shallow, but still visible), both at the mid-plane and in the atmosphere, whereas \cite{nauman2014characterizing} did not find such peaks in their vertically stratified ZNF simulations (they employed the same box size as we have, but much lower resolution). This led them to conclude that they were not capturing the outer scale of the turbulence in the atmosphere, whereas it follows from the energy spectra in Figure \ref{FIGURE_2Denergyspectra} that our box does encompass the outer scale (see below) in the atmosphere.

The ring-averaged kinetic energy spectra (blue curves) increase with increasing Pm at lower $k$ (large scales) both at the mid-plane and in the atmosphere, that is, the velocity of large-scale modes increases. The spectra reach peaks at $k_{m,d}=11$ at the mid-plane and $k_{m,a}=6.4$ in the atmosphere for ${\rm Pm}=4$, which move to smaller $k_{m,d}=3.1$ and $k_{m,a}=4.7$, respectively, for ${\rm Pm}=90$ (here the subscript $m$ stands for `maximum', and `d' and `a' denote mid-plane and atmosphere, respectively). After the peak, the kinetic energy decreases with $k$ more gently at ${\rm Pm}=90$ than at ${\rm Pm}=4$. Thus, the effect of increasing Pm is to ``stretch out'' the kinetic energy spectrum by extending the viscous range to smaller $k$: the viscous cut-off wavenumber $k_{\nu}\sim \sqrt{Re}$ \citep{heldmamatsashvili2022,Guilet2022} has shifted from $k_{\nu}\sim 68$ at ${\rm Pm}=4$ (${\rm Re}=4687$) to the left $k_{\nu}\sim 14$ at larger ${\rm Pm}=90$ (${\rm Re}=208$). As a result, the drop in the kinetic energy at intermediate wavenumbers  is sharper at $\rm Pm=90$ than at $\rm Pm=4$.  

The ring-averaged magnetic energy spectra (red curves), on the other hand, increase at all $k$ as ${\rm Pm}$ increases, however, they do so differently at the mid-plane and in the atmosphere. At the mid-plane, smaller $k$ (larger scales) gain more power than higher $k$ (smaller scales), so that the spectral peak shifts to lower $k$ with increasing ${\rm Pm}$. On the other hand, in the atmosphere, the spectrum retains its shape (i.e., does not shift in $k$) as $\rm Pm$ rises from 4 to 90 and mainly increases only in magnitude by about the same factor (of around 3.3) at all $k$. At ${\rm Rm}=18750$ used here, the resistive $k_{\eta}\sim \sqrt{\rm Rm}=137$ \citep{heldmamatsashvili2022} is larger than the viscous $k_{\nu}$. 

\subsubsection{Spectra of the dynamical terms}
\label{RESULTS_1DSpectraDynamicalTermsPmDependence}
Next, we examine the 1D ring-averaged dynamical terms and their dependence on $\rm Pm$, which are plotted in Figure \ref{FIGURE_shellaveragedterms} at ${\rm Pm} = 4$ and ${\rm Pm} = 90$ both at the mid-plane and in the atmosphere. It is seen in this figure that, like the energies, the dynamical terms generally increase in magnitude several times with increasing ${\rm Pm}$ (compare solid and dash-dot lines for ${\rm Pm}=4$ and ${\rm Pm}=90$, respectively), however, their behaviour with $\rm Pm$ differs at the mid-plane and in the atmosphere.  

\textit{Dynamics at the mid-plane of the disc.} At the mid-plane $z=0$, the ring-averaged spectra of the dynamical terms shift to smaller $k$ as Pm is increased (bottom row of Figure \ref{FIGURE_shellaveragedterms}), exhibiting a similar trend found in our unstratified simulations \citep{mamatsashvili2020zero, heldmamatsashvili2022}. In those studies, we interpreted this as a competition between the transverse and direct cascades. As discussed above, the central term in the sustaining dynamics near the mid-plane is ${\cal N}_x^{(h)}$, whose anisotropy (or angular dependence in the $(k_x,k_y)$-plane) describes the regeneration of the radial field due to the nonlinear transverse cascade (Figure \ref{FIGURE_2DSpectra_Bx}). On the other hand, the ring-average of ${\cal N}_x^{(h)}$ (blue lines in the bottom-left panel of Figure \ref{FIGURE_shellaveragedterms}), which involves integration over azimuthal angle, describes the action of the direct cascade along wavenumber ${\bf k}$ -- transfer of the radial field energy from the vital area to large wavenumbers where it is dissipated. As ${\rm Pm}$ increases, the latter quantity, which is positive at ${\rm Pm}=4$ at all $k$, decreases and becomes \textit{negative} at the higher wavenumbers $k \gtrsim 20$ at $\rm Pm=90$. This indicates that at higher $\rm Pm$, the efficiency of the direct cascade decreases: higher-$k$ modes gain \textit{less} power, while smaller-$k$ ones gain \textit{more} power, resulting in less resistive dissipation. Thus the production of $B_x$ by the transverse cascade in the vital area prevails over loss due to the direct cascade, and the turbulent radial field increases with $\rm Pm$. The ring-averages of ${\cal N}_x^{(z)}$ (red curves) and ${\cal N}_x^{(va)}$ (black curves) also increase and shift to lower $k$ with increasing ${\rm Pm}$.

The radial field in turn generates the toroidal field via the Maxwell stress, and the toroidal field generates the vertical field. As a result, the dependence of the ring-averages of the main dynamical terms governing these components -- the corresponding driver (positive) ${\cal M}$ and ${\cal N}_z^{(h)}$ as well as sink (negative) ${\cal N}_y^{(va)}$ and ${\cal N}_z^{(z)}$ -- on $\rm Pm$ actually stems from the dynamics of the radial field, similarly increasing in magnitude and shifting to lower $k$ with increasing $\rm Pm$, as seen in Figure \ref{FIGURE_shellaveragedterms}, but without changing sign, unlike ${\cal N}_x^{(h)}$.\footnote{The other sink terms for the toroidal field, ${\cal N}_y^{(h)}$ and ${\cal N}_y^{(z)}$, behave with $\rm Pm$ similarly to how  ${\cal N}_y^{(va)}$ behaves, so we do not show their ring-averages in the middle column of Figure \ref{FIGURE_shellaveragedterms} to avoid overcrowding the plots.} Note also that at $\rm Pm=90$, $N_y^{(va)}$ and $N_z^{(z)}$, attain values that are a significant fraction of $\mathcal{M}$ and $N_z^{(h)}$, respectively, so the physics these terms represent seems to play a more important role in the dynamics at large ${\rm Pm}$.

\textit{Dynamical terms in the atmosphere of the disc.} In contrast to the mid-plane, in the atmosphere at $z=3.2$ the 1D spectra of the dynamical terms do \textit{not} shift in $k$,  and retain their shapes (peaks) as a function of $k$ as $\rm Pm$ increases (top row of Figure \ref{FIGURE_shellaveragedterms}). 
This is because these 1D spectra are tied to the characteristic horizontal length-scale of the Parker instability, $l_a \sim c_s^2/g$ (Equation \ref{Parker_length}), which is independent of $\rm Pm$, and therefore they are concentrated around the corresponding wavenumber $1/l_a\sim 3$, as we have also seen above in the 2D spectral plots at $z=3.2$ in Figure \ref{FIGURE_2DSpectra_Bz}. 

As $\rm Pm$ is increased, the 1D spectra of the dynamical terms only increase in magnitude by about the same factors independent of $k$  (at least at lower and intermediate $k$), which is less than that observed near the mid-plane (bottom row of Figure \ref{FIGURE_shellaveragedterms}). Therefore, the dependence of the dynamics on $\rm Pm$ is \textit{weaker} in the atmosphere than in the bulk of the disc, consistent with the scaling laws found in these respective regions in Section \ref{RESULTS_PmComparisonVerticalDependenceOfScaling}. This can be explained by the reduced gain of magnetic energy at high $k$ as $\rm Pm$ increases. Indeed, the ring-average of ${\cal N}_z^{(h)}$, which is the central term supplying the vertical magnetic field in the atmosphere (which in turn produces the radial field and, through the shear, the toroidal field), increases at lower $k$ and at the same time becomes more negative at intermediate $20 \lesssim k\lesssim 50$ as $\rm Pm$ rises from 4 to 90, whereas the sink term ${\cal N}_z^{(z)}$ does not change much at these $k$. This implies that energy transfer from small to large $k$ decreases with ${\rm Pm}$, resulting in the increase in turbulent intensity.

\section{Conclusions}
\label{CONCLUSIONS}

We have carried out 3D shearing box simulations of MRI-driven turbulence and dynamo (zero-net-flux MRI) in the regime of large magnetic Prandtl numbers ($\text{Pm} \equiv \text{Rm}/\text{Re}$, equivalently, the ratio of viscosity to resistivity), a regime thought to be relevant to discs from binary neutron star and black-hole neutron star mergers, the inner regions of X-ray binaries and AGN, and the interiors of protoneutron stars. Our key aim was to investigate the self-sustenance mechanism of the turbulence and dynamo in the large $\text{Pm}$ regime, in both physical and Fourier space, the latter by means of spectral analysis of the induction equation. A new addition here is that we have included vertical stratification of the disc, thus building on our previous work \cite{heldmamatsashvili2022} on unstratified discs 

We carried out six high resolution simulations (128 cells per scale-height $H$) in boxes of size $[L_x, L_y, L_z] = [4,4,8]H$ between ${\rm Pm}=4$ and ${\rm Pm}=90$ at a fixed magnetic Reynolds number of ${\rm Rm}=18750$.\footnote{In Section 4.5 of \cite{heldmamatsashvili2022} we also investigated (i) increasing Pm by increasing Rm and keeping Re fixed, and (ii) keeping Pm constant and simultaneously increasing both Re and Rm.} A key result is that at intermediate values of ${\rm Pm}$ ($4 \lesssim {\rm Pm} \lesssim 32$), turbulent transport of angular momentum, as parameterized by the stress-to-thermal pressure ratio $\alpha$, exhibits power-law scaling within the bulk of the disc ($|z| \lesssim 2H$), with $\alpha \sim {\rm Pm}^\delta$ and $\delta \sim 0.71$, in excellent agreement with the results of our unstratified simulations. At very large values of $\rm Pm$, we observe the onset of a plateau in which turbulence depends only weakly (or not at all) on the magnetic Prandtl number, again in agreement with our unstratified simulations. We have checked that the results depend neither on the vertical boundary conditions, nor on the vertical box size (for $L_z \gtrsim 8H$).

A new result is that the scaling of turbulent transport with Pm is height-dependent, and becomes noticeably weaker as one moves away from the disc mid-plane and into the atmosphere of the disc ($|z| \gtrsim 2H$). In agreement with previous work \citep{blaes2007surface,davisstonepesssah2010, shi2009numerically, nauman2014characterizing}, we find that the addition of vertical stratification results in different dynamical processes dominating at different heights. Broadly speaking, the bulk of the disc is dominated by small-scale MRI turbulence, while in the disc atmosphere we find the emergence of a large-scale vertical magnetic field and evidence of Parker instability. These dynamics in the disc atmosphere could drive strong outflows/winds and support magnetised atmospheres/coronae, both of which have been observed in many disk systems (see e.g., \cite{Keek2016,Gallo2019,Kang2022,Masterson2022,Zhang2023} for AGN discs and a recent review by \cite{Pascucci2023} for protoplanetary discs).

To understand the interplay between these two processes -- MRI and Parker instability -- as well as their dependence on height and Pm, we have carried out a detailed spectral analysis of the non-linear terms governing the evolution of each magnetic field component. Our analysis shows that, at the mid-plane, the dynamo behaves similarly to what we found in the absence of stratification:  it primarily involves the interplay of radial and toroidal magnetic fields, with the toroidal magnetic field being generated mostly from shearing out of radial field, while radial field is reseeded primarily by means of a non-linear transverse cascade that serves to transfer energy from intermediate scales back to large-scales anisotropically in Fourier space. This transverse cascade becomes increasingly efficient at reseeding radial magnetic field as Pm is increased.

In the disc atmosphere, on the other hand, the dynamo primarily involves the interplay of toroidal and \textit{vertical} magnetic field components, and the dependence on Pm is weaker. The radial field is generated primarily from vertical field due to vertical gradients in the (perturbed) velocity. This field is then sheared out to produce toroidal field, and vertical field is generated primarily from toroidal field by means of magnetic buoyancy due to the Parker instability, thus closing the loop. This process is reminiscent of, though not entirely the same as, the idea put forward by \cite{Tout_Pringle1992} which involved the interplay between MRI and Parker instability, though our analysis shows that, while both instabilities play a role in the dynamo, they dominate in different parts of the disc.

By taking into account the effects of buoyancy and vertical stratification in this work (the second in a series of papers on the MRI in the regime of large magnetic Prandtl number), we have moved one step closer to modeling the conditions under which the MRI occurs in certain types of astrophysical objects, such as discs from binary neutron star mergers and proto-neutron stars. We have shown the turbulent transport is very sensitive to magnetic Prandtl number, at least at intermediate values of Pm, which could render the flow thermally unstable \citep{balbus2008, potter2017, kawanaka2019}. Thus future work should take into account the effects of thermodynamics, neutrino cooling, and temperature and and density-dependent diffusion coefficients, in order to probe the thermal stability of discs in the large-Pm regime.

\section*{Acknowledgements}
Simulations were run on the Sakura, Cobra, and Raven clusters at the Max Planck Computing and Data Facility (MPCDF) in Garching, Germany. The authors thank Masaru Shibata, Andrew Hillier, and Yosuke Mizuno for helpful discussions and clarifications. Finally, we would like to thank the referee for carefully reading the manuscript, and for helping us improve the clarity of the material. This work received funding from the European Union's Horizon 2020 research and innovation programme under the ERC Advanced Grant Agreement No. 787544 and from Shota Rustaveli National Science Foundation of Georgia (SRNSFG) [grant number FR-23-1277].
M.E.P. gratefully acknowledges support from the Independent Research Fund Denmark via grant ID 10.46540/3103-00205B.

\section*{Data availability}
The data underlying this article will be shared on a reasonable request to the corresponding author.




\bibliographystyle{mnras}
\bibliography{VstratAtHighPmBib} 




\appendix

\section{Effect of vertical boundary conditions and vertical box size}
\subsection{Effect of vertical boundary conditions}
\label{APPENDIX_VerticalBoundaryConditions}

\begin{figure}
\centering
\includegraphics[scale=0.23]{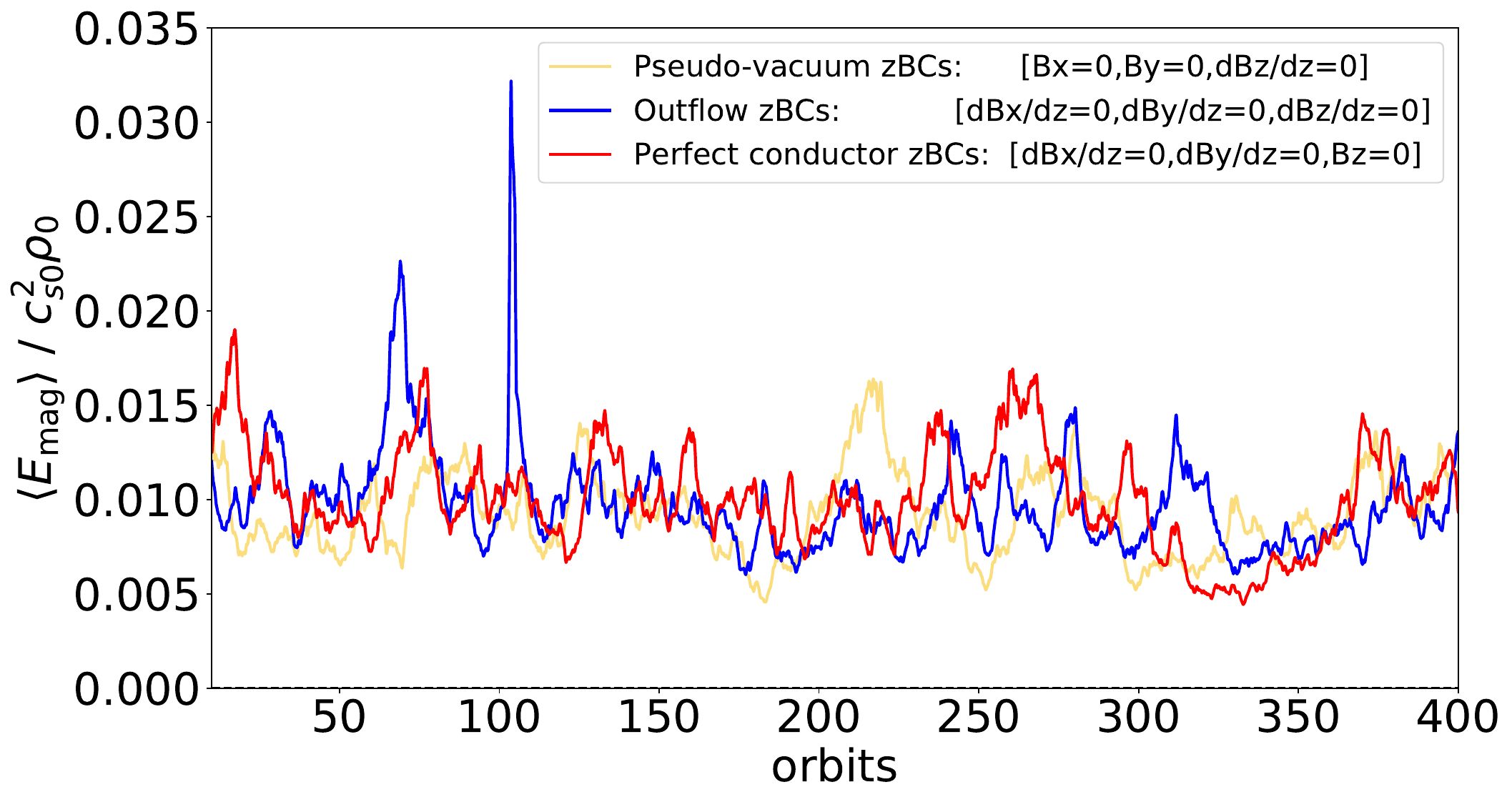}
\caption{Time-series of volume-averaged magnetic energy density from three simulations with different vertical boundary conditions (BCs): (i) pseudo-vacuum BCs (yellow), (ii) outflow BCs (blue), and (iii) perfect conductor BCs (red). For definitions see text. All simulations were run in boxes of size $[L_x,L_y,L_z]=[4,4,8]H$ at a resolution of $32$ cells per scale-height $H$, and at fixed magnetic Prandtl number of $\text{Pm}=4$.}
\label{FIGURE_zBCComparisonTimeSeries}
\end{figure}

\begin{figure}
\centering
\includegraphics[scale=0.24]{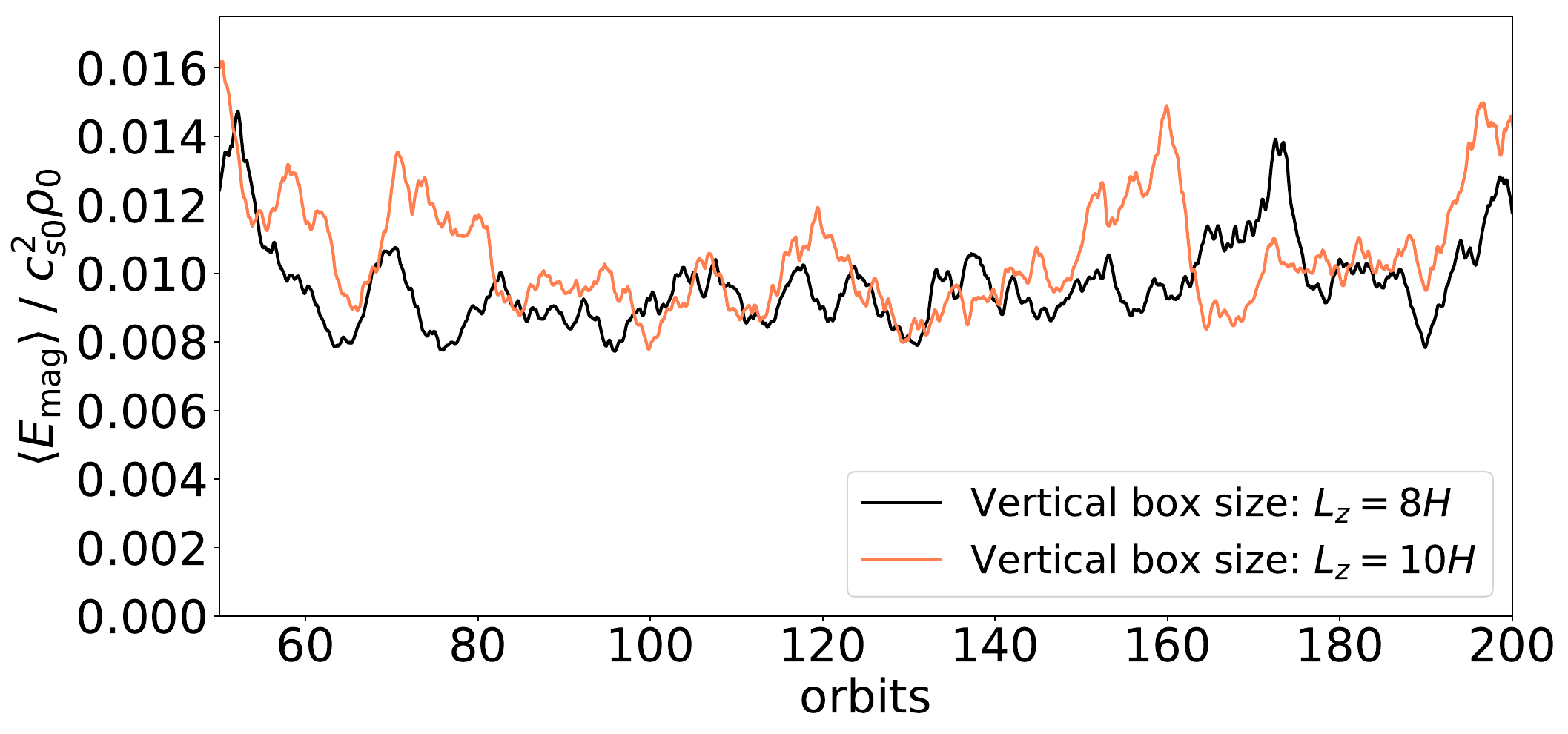}
\caption{Time-series of volume-averaged magnetic energy density from two simulations in boxes of different height: (i) $L_z = 8H$ (black), and (ii) $L_z = 10H$ (yellow). Both simulations were run at a resolution of $128$ cells per scale-height $H$ at fixed magnetic Prandtl number $\text{Pm}=4$, and employ outflow boundary conditions on the magnetic field in the vertical direction.}
\label{FIGURE_BoxHeightComparisonTimeSeries}
\end{figure}

In this section we investigate the effect of different vertical boundary conditions (zBCs) for the magnetic field on the results. All simulations were run at $\text{Pm} = 4$ at a resolution of $32$ cells per scale-height $H$ in a box of size $[L_x,L_y,L_z] = [4,4,8]H$ for 400 orbits. As in our fiducial simulations we employ a mass source term (see Section \ref{METHODS_BoundaryConditions}). All three simulations employed outflow boundary conditions on the velocity field in the vertical direction (the same BCs used in the simulations presented in the main text, i.e. $dv_x/dz = 0, dv_y/dz =0, dv_z/dz = 0$). For the magnetic field we tested three different vertical boundary conditions:

(i) \textit{pseudo-vacuum} (also known as \textit{vertical-field}): $B_x = 0, B_y = 0, dB_z/dz = 0$. See, for example, \cite{riols2018}.

(ii) \textit{outflow}: $dB_x/dz = 0, dB_y/dz = 0, dB_z/dz =0$. See, for example, \cite{simon2011, ryan2017}.

(iii) \textit{perfect conductor}: $dB_x/dz = 0, dB_y/dz = 0, B_z = 0$. See, for example, \cite{kapyla2010open}.

In Figure \ref{FIGURE_zBCComparisonTimeSeries} we plot time-series of the volume-averaged magnetic energy density from the three simulations. The time-series after orbit 100 are very similar in all three cases, indicating that the vertical boundary conditions do not affect bulk of the disc. Time-averaged values (from orbit 125 to orbit 400) are given in Table \ref{TABLE_zBCComparison} in Appendix \ref{APPENDIX_TablesOfSimulations}. The mass loss over 400 orbits is very similar in all three cases, and amounts to around $0.3\%$ of the total initial mass in the box per orbit. Finally, we observe little difference in the vertical profiles of plasma beta, Maxwell stress, and root-mean square magnetic field components, and in all three cases the disc is marginally stable to Parker stability for $z \lesssim 1.8H$, and unstable outside this region. We conclude that, for the box size we have chosen for most of our simulations ($[L_x,L_y,L_z] = [4,4,8]H$), the vertical boundary conditions on the magnetic field (pseudo-vacuum, outflow, and perfect conductor) do not significantly impact the result in the bulk of the disc ($|z| \lesssim 2H$), nor that in the disc atmosphere ($z| \gtrsim 2H$).

\subsection{Effect of vertical box size}
\label{APPENDIX_VerticalBoxSize}
We investigate the effect of changing the vertical box size by considering two high resolution simulations (128 cells per $H$) at fixed $\text{Pm}=4$. One simulation is in a box of vertical size $L_z=8H$ while the other is in a taller box of size $L_z=10H$. The simulations are otherwise identical. We employ pseudo-vacuum boundary conditions (see Appendix \ref{APPENDIX_VerticalBoundaryConditions} above) on the magnetic field at the vertical boundaries.

Figure \ref{FIGURE_BoxHeightComparisonTimeSeries} shows the time-series of magnetic energy from the two simulations. We find very little difference between the time-series of magnetic energy (and other diagnostics, such as stress) between these two runs. The mass lost per orbit is about an order of magnitude less in shorter box: $0.03\%$ of the initial mass in the box is lost (and added back in) per orbit in this run. The vertical disc structure differs at about the $10\%$ level (stress and magnetic energy being slightly larger in the taller box), likely on account of the smaller mass and magnetic flux loss rate in the taller box, but the profiles are qualitatively similar.

\section{Tables of Simulations}
\label{APPENDIX_TablesOfSimulations}

\begin{table*}
\centering
\caption{\textit{Magnetic Prandtl number comparison}: vertically stratified, isothermal, zero-net-magnetic-flux (ZNF) shearing box MHD simulations at various magnetic Prandtl numbers, $\text{Pm}$. All simulations were carried out in a box of size $[L_x,L_y,L_z] = [4,4,8]H$, a resolution of 128 cells per scale-height $H$, and with a shear parameter of $q=1.5$ (Keplerian shear). Here $\text{Rm}$ denotes the magnetic Reynolds number (which we fix at $\text{Rm}=18750$ in all runs), Pm the magnetic Prandtl number, and $\text{Re} \equiv \text{Rm}/\text{Pm}$ the Reynolds number (Re has been rounded down to the nearest whole number in the table). Other columns: $\langle\langle E_\text{mag}\rangle\rangle$,\,$\langle\langle\alpha\rangle\rangle$,\,$\langle\langle M_{xy} \rangle\rangle$,\,$\langle\langle R_{xy} \rangle \rangle$, and $R$ denote the time- and volume-averaged magnetic energy density, alpha (i.e. stress normalized by \textit{volume-averaged} pressure) , Maxwell stress, and Reynolds stress, respectively. Finally, $R \equiv \langle\langle M_{xy} \rangle\rangle / \langle\langle R_{xy} \rangle \rangle$. All simulations were run for 200 orbits ($1257\,\Omega^{-1}$). Time-averages were taken from orbit 100 to orbit 200. Volume-averages have been taken over the entire domain.}
\label{TABLE_PmComparison}
	\begin{tabular}{lcccccccccr}
		\hline
		Run	& Box Size & Resolution & Rm & Re & Pm &$\langle\langle E_\text{mag}\rangle\rangle$ &$\langle\langle\alpha\rangle\rangle$&$\langle\langle M_{xy} \rangle\rangle$&$\langle\langle R_{xy} \rangle \rangle$ &$R$ \\ 
		\hline
		VSTRMRIPm4Res128Re4687    & [4,4,8] & $128/H$ & $18750$ & $4687$ & $4$ & $0.009834$ & $0.016326$ & $0.004265$ & $0.000850$ & $5.019552$\\
		VSTRMRIPm8Res128Re2344    & [4,4,8] & $128/H$ & $18750$ & $2344$ & $8$ & $0.016211$ & $0.026406$ & $0.006990$ & $0.001282$ & $5.450631$\\
		VSTRMRIPm16Res128Re1172   & [4,4,8] & $128/H$ & $18750$ & $1172$ & $16$ & $0.026001$ & $0.041619$ & $0.011144$ & $0.001895$ & $5.880631$\\
		VSTRMRIPm32Res128Re586    & [4,4,8] & $128/H$ & $18750$ & $586$ & $32$ & $0.038039$ & $0.058550$ & $0.015880$ & $0.002464$ & $6.445064$\\
		VSTRMRIPm64Res128Re293    & [4,4,8] & $128/H$ & $18750$ & $293$ & $64$ & $0.053972$ & $0.078632$ & $0.021629$ & $0.003007$ & $7.192255$\\
		VSTRMRIPm90Res128Re208    & [4,4,8] & $128/H$ & $18750$ & $208$ & $90$ & $0.061722$ & $0.085051$ & $0.023609$ & $0.003038$ & $7.770238$\\
		\hline
	\end{tabular}
\end{table*}

\begin{table*}
\centering
\caption{\textit{Box size and resolution comparison}: all simulations carried out at a fixed magnetic Prandtl number $\text{Pm}=4$, Reynolds number $\text{Re}=4687$, and magnetic Reynolds number $\text{Rm}=18750$. Here, $[L_x,L_y,L_z]$ denotes the size of the box (in units of initial mid-plane scale-height $H$) in the $x$-, $y$-, and $z$-directions, respectively. All other quantities are the same as in Table \ref{TABLE_PmComparison}. Time-averages were taken from orbit 100 to orbit 200. (Note that simulation VSTRMRIPm4Res128Re4687 (Table 
\ref{TABLE_PmComparison}) and VSTRMRIPm4Res128Re4687H4 (below) are the same simulation.)}
\label{TABLE_BoxSizeComparison}
	\begin{tabular}{lcccccccccr}
		\hline
		Run	& Box Size & Resolution & Rm & Re & Pm &$\langle\langle E_\text{mag}\rangle\rangle$ &$\langle\langle\alpha\rangle\rangle$&$\langle\langle M_{xy} \rangle\rangle$&$\langle\langle R_{xy} \rangle \rangle$ &$R$ \\ 
		\hline
  		VSTRMRIPm4Res32Re4687H3    & [4,4,6] & $32$/H & $18750$ & $4687$ & $4$ & $0.006907$ & $0.007326$ & $0.002506$ & $0.000546$ & $4.592398$\\
		VSTRMRIPm4Res32Re4687H4    & [4,4,8] & $32$/H & $18750$ & $4687$ & $4$ & $0.008700$ & $0.012611$ & $0.003230$ & $0.000721$ & $4.476833$\\
		VSTRMRIPm4Res32Re4687H5    & [4,4,10] & $32$/H & $18750$ & $4687$ & $4$ & $0.008746$ & $0.015146$ & $0.003010$ & $0.000697$ & $4.450142$\\
              \hline
		VSTRMRIPm4Res64Re4687H4    & [4,4,8] & $64$/H & $18750$ & $4687$ & $4$ & $0.010512$ & $0.016942$ & $0.004365$ & $0.000943$ & $4.630342$\\
            \hline
		VSTRMRIPm4Res128Re4687H4   & [4,4,8] & $128$/H & $18750$ & $4687$ & $4$ & $0.009834$ & $0.016326$ & $0.004265$ & $0.000850$ & $5.019552$\\
		VSTRMRIPm4Res128Re4687H5   & [4,4,10] & $128$/H & $18750$ & $4687$ & $4$ & $0.010380$ & $0.020174$ & $0.004220$ & $0.000837$ & $5.043951$\\
		\hline
	\end{tabular}
\end{table*}

\begin{table*}
\centering
\caption{\textit{Vertical boundary condition comparison}: all simulations carried out  at fixed magnetic Prandtl number $\text{Pm}=4$, Reynolds number $\text{Re}=4687$, and magnetic Reynolds number $\text{Rm}=18750$ in boxes of size $[L_x,L_y,L_z]=[4,4,8]H$. In the first column, `VF' refers to vertical field boundary conditions (also known as `pseudo-vacuum boundary conditions'), `OF' refers to outflow BCs, and 'PC' refers to perfect conductor BCs (see Appendix \ref{APPENDIX_VerticalBoundaryConditions} for definitions.. All other quantities are the same as in Table \ref{TABLE_PmComparison}. All simulations were run for 400 orbits ($2514\,\Omega^{-1}$). Time-averages were taken from orbit 125 to orbit 400.}
\label{TABLE_zBCComparison}
	\begin{tabular}{lcccccccccr}
		\hline
		Run	& Box Size & Resolution & Rm & Re & Pm &$\langle\langle E_\text{mag}\rangle\rangle$ &$\langle\langle\alpha\rangle\rangle$&$\langle\langle M_{xy} \rangle\rangle$&$\langle\langle R_{xy} \rangle \rangle$ &$R$ \\ 
		\hline
  		VSTRMRIPm4Res32Re4687H4\_{VF}    & [4,4,8] & $32$/H & $18750$ & $4687$ & $4$ & $0.009206$ & $0.013370$ & $0.003424$ & $0.000765$ & $4.474018$\\
		VSTRMRIPm4Res32Re4687H4\_{OF}    & [4,4,8] & $32$/H & $18750$ & $4687$ & $4$ & 0.009026 & 0.013049 & 0.003341 & 0.000747 & 4.474206\\
		VSTRMRIPm4Res32Re4687H4\_{PC}    & [4,4,8] & $32$/H & $18750$ & $4687$ & $4$ & 0.009636 & 0.014034 & 0.003582 & 0.000808 & 4.430520\\
		\hline
	\end{tabular}
\end{table*}

\bsp	
\label{lastpage}
\end{document}